\documentclass[journal,onecolumn, 12pt]{IEEEtran}
\IEEEoverridecommandlockouts

\usepackage{tikz}
\usetikzlibrary{patterns}
\usepackage{bbm}
\usepackage{cite}
\usepackage{amsmath,amsfonts} 
\usepackage{algorithmic,algorithm}
\usepackage{graphicx}
\usepackage{caption}
\usepackage{subcaption}
\usepackage{textcomp}
\usepackage{xcolor}
\usepackage{color,soul}

\usepackage{pgfplots}
\pgfplotsset{compat=1.5}

\def\BibTeX{{\rm B\kern-.05em{\sc i\kern-.025em b}\kern-.08em
		T\kern-.1667em\lower.7ex\hbox{E}\kern-.125emX}}

\renewcommand{\baselinestretch}{1.2}
\newtheorem{theorem}{Theorem}
\newtheorem{lemma}{Lemma}

\newtheorem{definition}{Definition}

\newtheorem{remark}{Remark}
\newtheorem{example}{Example}
\newtheorem{assumption}{Assumption}
\begin{document}

\title{Scheduling to Minimize Age of Information with Multiple Sources}

\author{
	\IEEEauthorblockN{Kumar Saurav}
	\IEEEauthorblockA{\\\textit{School of Technology and Computer Science} \\
		\textit{Tata Institute of Fundamental Research}\\
		Mumbai, India. \\
		kumar.saurav@tifr.res.in}\\
	\IEEEauthorblockN{Rahul Vaze}
	\IEEEauthorblockA{\\\textit{School of Technology and Computer Science} \\
		\textit{Tata Institute of Fundamental Research}\\
		\textit{Mumbai, India}. \\
		rahul.vaze@gmail.com \\} 
}

\def\onehalf{\frac{1}{2}}
\def\etal{et.\/ al.\/}
\newcommand{\bydef}{\triangleq}
\newcommand{\tr}{{\it{tr}}}
\def\SNR{{\textsf{SNR}}}
\def\bydef{:=}
\def\bba{{\mathbb{a}}}
\def\bbb{{\mathbb{b}}}
\def\bbc{{\mathbb{c}}}
\def\bbd{{\mathbb{d}}}
\def\bbee{{\mathbb{e}}}
\def\bbff{{\mathbb{f}}}
\def\bbg{{\mathbb{g}}}
\def\bbh{{\mathbb{h}}}
\def\bbi{{\mathbb{i}}}
\def\bbj{{\mathbb{j}}}
\def\bbk{{\mathbb{k}}}
\def\bbl{{\mathbb{l}}}
\def\bbm{{\mathbb{m}}}
\def\bbn{{\mathbb{n}}}
\def\bbo{{\mathbb{o}}}
\def\bbp{{\mathbb{p}}}
\def\bbq{{\mathbb{q}}}
\def\bbr{{\mathbb{r}}}
\def\bbs{{\mathbb{s}}}
\def\bbt{{\mathbb{t}}}
\def\bbu{{\mathbb{u}}}
\def\bbv{{\mathbb{v}}}
\def\bbw{{\mathbb{w}}}
\def\bbx{{\mathbb{x}}}
\def\bby{{\mathbb{y}}}
\def\bbz{{\mathbb{z}}}
\def\bb0{{\mathbb{0}}}

\def\bydef{:=}
\def\ba{{\mathbf{a}}}
\def\bb{{\mathbf{b}}}
\def\bc{{\mathbf{c}}}
\def\bd{{\mathbf{d}}}
\def\bee{{\mathbf{e}}}
\def\bff{{\mathbf{f}}}
\def\bg{{\mathbf{g}}}
\def\bh{{\mathbf{h}}}
\def\bi{{\mathbf{i}}}
\def\bj{{\mathbf{j}}}
\def\bk{{\mathbf{k}}}
\def\bl{{\mathbf{l}}}
\def\bm{{\mathbf{m}}}
\def\bn{{\mathbf{n}}}
\def\bo{{\mathbf{o}}}
\def\bp{{\mathbf{p}}}
\def\bq{{\mathbf{q}}}
\def\br{{\mathbf{r}}}
\def\bs{{\mathbf{s}}}
\def\bt{{\mathbf{t}}}
\def\bu{{\mathbf{u}}}
\def\bv{{\mathbf{v}}}
\def\bw{{\mathbf{w}}}
\def\bx{{\mathbf{x}}}
\def\by{{\mathbf{y}}}
\def\bz{{\mathbf{z}}}
\def\b0{{\mathbf{0}}}
\def\opt{\mathsf{OPT}}
\def\on{\mathsf{ON}}
\def\off{\mathsf{OF}}
\def\bA{{\mathbf{A}}}
\def\bB{{\mathbf{B}}}
\def\bC{{\mathbf{C}}}
\def\bD{{\mathbf{D}}}
\def\bE{{\mathbf{E}}}
\def\bF{{\mathbf{F}}}
\def\bG{{\mathbf{G}}}
\def\bH{{\mathbf{H}}}
\def\bI{{\mathbf{I}}}
\def\bJ{{\mathbf{J}}}
\def\bK{{\mathbf{K}}}
\def\bL{{\mathbf{L}}}
\def\bM{{\mathbf{M}}}
\def\bN{{\mathbf{N}}}
\def\bO{{\mathbf{O}}}
\def\bP{{\mathbf{P}}}
\def\bQ{{\mathbf{Q}}}
\def\bR{{\mathbf{R}}}
\def\bS{{\mathbf{S}}}
\def\bT{{\mathbf{T}}}
\def\bU{{\mathbf{U}}}
\def\bV{{\mathbf{V}}}
\def\bW{{\mathbf{W}}}
\def\bX{{\mathbf{X}}}
\def\bY{{\mathbf{Y}}}
\def\bZ{{\mathbf{Z}}}
\def\b1{{\mathbf{1}}}

\def\bbA{{\mathbb{A}}}
\def\bbB{{\mathbb{B}}}
\def\bbC{{\mathbb{C}}}
\def\bbD{{\mathbb{D}}}
\def\bbE{{\mathbb{E}}}
\def\bbF{{\mathbb{F}}}
\def\bbG{{\mathbb{G}}}
\def\bbH{{\mathbb{H}}}
\def\bbI{{\mathbb{I}}}
\def\bbJ{{\mathbb{J}}}
\def\bbK{{\mathbb{K}}}
\def\bbL{{\mathbb{L}}}
\def\bbM{{\mathbb{M}}}
\def\bbN{{\mathbb{N}}}
\def\bbO{{\mathbb{O}}}
\def\bbP{{\mathbb{P}}}
\def\bbQ{{\mathbb{Q}}}
\def\bbR{{\mathbb{R}}}
\def\bbS{{\mathbb{S}}}
\def\bbT{{\mathbb{T}}}
\def\bbU{{\mathbb{U}}}
\def\bbV{{\mathbb{V}}}
\def\bbW{{\mathbb{W}}}
\def\bbX{{\mathbb{X}}}
\def\bbY{{\mathbb{Y}}}
\def\bbZ{{\mathbb{Z}}}

\def\cA{\mathcal{A}}
\def\cB{\mathcal{B}}
\def\cC{\mathcal{C}}
\def\cD{\mathcal{D}}
\def\cE{\mathcal{E}}
\def\cF{\mathcal{F}}
\def\cG{\mathcal{G}}
\def\cH{\mathcal{H}}
\def\cI{\mathcal{I}}
\def\cJ{\mathcal{J}}
\def\cK{\mathcal{K}}
\def\cL{\mathcal{L}}
\def\cM{\mathcal{M}}
\def\cN{\mathcal{N}}
\def\cO{\mathcal{O}}
\def\cP{\mathcal{P}}
\def\cQ{\mathcal{Q}}
\def\cR{\mathcal{R}}
\def\cS{\mathcal{S}}
\def\cT{\mathcal{T}}
\def\cU{\mathcal{U}}
\def\cV{\mathcal{V}}
\def\cW{\mathcal{W}}
\def\cX{\mathcal{X}}
\def\cY{\mathcal{Y}}
\def\cZ{\mathcal{Z}}

\def\sfA{\mathsf{A}}
\def\sfB{\mathsf{B}}
\def\sfC{\mathsf{C}}
\def\sfD{\mathsf{D}}
\def\sfE{\mathsf{E}}
\def\sfF{\mathsf{F}}
\def\sfG{\mathsf{G}}
\def\sfH{\mathsf{H}}
\def\sfI{\mathsf{I}}
\def\sfJ{\mathsf{J}}
\def\sfK{\mathsf{K}}
\def\sfL{\mathsf{L}}
\def\sfM{\mathsf{M}}
\def\sfN{\mathsf{N}}
\def\sfO{\mathsf{O}}
\def\sfP{\mathsf{P}}
\def\sfQ{\mathsf{Q}}
\def\sfR{\mathsf{R}}
\def\sfS{\mathsf{S}}
\def\sfT{\mathsf{T}}
\def\sfU{\mathsf{U}}
\def\sfV{\mathsf{V}}
\def\sfW{\mathsf{W}}
\def\sfX{\mathsf{X}}
\def\sfY{\mathsf{Y}}
\def\sfZ{\mathsf{Z}}

\def\bydef{:=}
\def\sfa{{\mathsf{a}}}
\def\sfb{{\mathsf{b}}}
\def\sfc{{\mathsf{c}}}
\def\sfd{{\mathsf{d}}}
\def\sfee{{\mathsf{e}}}
\def\sfff{{\mathsf{f}}}
\def\sfg{{\mathsf{g}}}
\def\sfh{{\mathsf{h}}}
\def\sfi{{\mathsf{i}}}
\def\sfj{{\mathsf{j}}}
\def\sfk{{\mathsf{k}}}
\def\sfl{{\mathsf{l}}}
\def\sfm{{\mathsf{m}}}
\def\sfn{{\mathsf{n}}}
\def\sfo{{\mathsf{o}}}
\def\sfp{{\mathsf{p}}}
\def\sfq{{\mathsf{q}}}
\def\sfr{{\mathsf{r}}}
\def\sfs{{\mathsf{s}}}
\def\sft{{\mathsf{t}}}
\def\sfu{{\mathsf{u}}}
\def\sfv{{\mathsf{v}}}
\def\sfw{{\mathsf{w}}}
\def\sfx{{\mathsf{x}}}
\def\sfy{{\mathsf{y}}}
\def\sfz{{\mathsf{z}}}
\def\sf0{{\mathsf{0}}}

\def\Nt{{N_t}}
\def\Nr{{N_r}}
\def\Ne{{N_e}}
\def\Ns{{N_s}}
\def\Es{{E_s}}
\def\No{{N_o}}
\def\sinc{\mathrm{sinc}}
\def\dmin{d^2_{\mathrm{min}}}
\def\vec{\mathrm{vec}~}
\def\kron{\otimes}
\def\Pe{{P_e}}
\newcommand{\expeq}{\stackrel{.}{=}}
\newcommand{\expg}{\stackrel{.}{\ge}}
\newcommand{\expl}{\stackrel{.}{\le}}
\def\SIR{{\mathsf{SIR}}}

\def\nn{\nonumber}

\maketitle
\begin{abstract}
  We consider a G/G/1 queueing system with a single server, where updates/packets arrive from different sources stochastically with possibly different update inter-generation time distributions. The server can transmit/serve at most one update at any time, with potentially different transmission/service times for updates belonging to distinct sources. The age of information (AoI) of any source is a function of the time difference between the departure time of successive updates of that source. Each fully/partially transmitted update incurs a fixed (energy) cost, and the goal of the scheduler is to minimize the linear combination of the sum of the age of information across all sources and the total energy cost. One distinguishing feature of the considered problem compared to the rich scheduling literature is that it is not necessary to transmit all updates, and transmitting only a subset (unknown) of updates is sufficient. We propose a simple non-preemptive randomized scheduling algorithm, that randomly marks arriving updates from a source to be eligible for transmission with a fixed probability and discards them otherwise. Every time the server becomes free, it chooses a source for transmission randomly with another fixed probability and begins to transmit the most recently marked update of the chosen source. Both the respective probabilities are chosen by solving a convex program. The competitive ratio of the proposed algorithm (against a non-preemptive offline optimal algorithm) is shown to be $3$ plus the maximum of the ratio of the variance and the mean of the inter-arrival time distribution of sources. For several common distributions such as exponential, uniform and Rayleigh, the competitive ratio is at most $4$.
  For preemptive policies, a G/M/1 system is considered and a non-preemptive policy is shown to have competitive ratio  (against a preemptive offline optimal algorithm) at most $5$ plus the maximum of the ratio of the variance and the mean of the inter-arrival time distribution of sources.
\end{abstract}

%



%
%
%
\section{Introduction}
With the advent of modern applications such as remote gaming, smart and connected cars, IoT, smart homes etc., information timeliness has become an important performance metric in addition to the traditional ones such as throughput and delay.  
Information timeliness refers to quick and periodic dissemination of information, for example, in networked cars, 
critical safety information needs to be updated quickly and often enough. 
In recent times, several metrics have been proposed for information timeliness that include the age of information (AoI) \cite{kaul2012real}, the age of incorrect information \cite{kam2020age}, the age of incorrect estimates \cite{Bhavya}. Because of its simplicity and elegance, AoI has become the de facto first choice for analysis,  where the {\it age} at time $t$ is 
defined as the time elapsed since the last received update was generated, and the AoI is the average of age across time.
In this paper, we consider a very general AoI scheduling problem, that we describe after presenting the following background. 

\subsection{Prior Work}
Research on analyzing the AoI, started with \cite{kaul2012real} that considered a M/M/1 system, that has a single server where updates from a single source arrive having exponential inter-generation
 times, and the server follows a FCFS policy with exponential service time. Subsequently, the same question was posed for the M/M/1 system following other scheduling policies such as LCFS, LCFS with preemption \cite{kaul2012status} etc. 
In addition to considering the  AoI, the peak AoI (maximum AoI over all times) \cite{peakAge} has also been studied for the M/M/1 system with a single source  under different scheduling policies. 
 
Extensions of this work to D/G/1 system for FCFS and LCFS has also been reported in \cite{champati2018statistical, inoue2019general}. An additional model called the generate-at-will has also been considered \cite{sun2017update}, where the source can generate the update at any time, however, the update is received at the monitor with a delay/service time that is stochastically distributed. 

Next step in this direction was to consider a M/M/1 system where updates arrive from multiple sources, and the considered objective was to characterize the mean, the distribution (equivalently the moment generating function) of the AoI or the related performance metrics for each of the sources \cite{multisourceyates, multisource}. 
Since AoI is a source based metric and depends on the relative delay between two consecutive departures corresponding to status updates received for any source at the monitor, this was more challenging than studying a single source case. Similar to the single source case, with multiple sources also, primarily the analysis was restricted for fixed scheduling policies such as FCFS, LCFS, FCFS with preemption etc \cite{multisourceyates, multisource}. Analyzing the distributional properties of AoI when the energy used to transmit updates is sourced from renewable sources called energy harvesting has also been considered in \cite{abd2021age}, and references therein.

An alternate AoI research direction has been the discrete-time model \cite{kadotascheduling, kadota2019minimizing}, where time is slotted, and there are multiple sources who want to update their information to the monitor.  
To model the delay/service time of the continuous-time model, for each slot, the probability that an update sent by source $i$ is successfully received at the monitor is assumed to be $p_i$, independent across slots. 
Under a natural constraint  that at most one source can transmit its update in any given slot, a centralized scheduling problem has been widely considered \cite{kadotascheduling, kadota2019minimizing}: in each slot which source should update its information to the monitor so as to minimize the long-term weighted sum of the AoI across all sources. 

Most of the research  in the discrete-time model considered the generate-at-will model,  where for example \cite{kadotascheduling} showed that a simple randomized policy is $2$-competitive. \footnote{Ratio of the cost of this policy and an optimal offline policy in the worst input case.}  Similar results have been extended for richer models, e.g. fading channel \cite{bhat2020throughput}, multiple access channel \cite{bhat2021minimization} etc. 
In the discrete-time model, when the updates arrive intermittently at the sources is somewhat less well investigated. As far as we know, only \cite{kadota2019minimizing} has considered this problem, and derived a $4$-competitive policy, when the inter-generation time distribution is geometric (corresponding to exponential distribution in continuous-time) with possibly different parameters for each source.

Most of the research on AoI related problems has ignored the cost of transmission, e.g. energy/power, and allows frequent 
 transmissions to lower AoI without accounting for the increased cost of transmission. There is, however, an inherent tradeoff between AoI and transmission cost, as studied in \cite{saurav2021minimizing} to minimize the sum of the AoI and the cost of transmission or minimize AoI subject to an average power constraint \cite{bhat2021minimization}. In addition, when partial packets are transmittable by accounting for distortion cost, \cite{rajaraman2021not} proposed a $2$-competitive policy for minimizing the sum of the distortion, the AoI and the energy cost.

\subsection{Considered Problem}
In this paper, we consider a scheduling problem for the continuous-time model with multiple sources. 
In particular, we consider a G/G/1 system, where updates from different sources arrive to a single queue. The inter-generation time of updates for source $i$ is assumed to be stochastic with distribution $\cG_i$, while the service time distribution of an update from source $i$ is $\cD_i$. At any time, only one update from any source can be under service. 
In a major departure from most of the prior work in the continuous-time model \cite{kaul2012real, kaul2012status, peakAge, champati2018statistical, inoue2019general, multisourceyates, multisource, abd2021age} that found the distribution of  AoI for a fixed scheduling policy, our focus in this work is on finding optimal scheduling policies. Moreover, we also consider a general metric compared to prior work, that is a linear combination of the weighted AoI and the total cost of transmission (service), where a constant cost 
(which may be different for each source) 
is counted for each partial/complete transmission (service).
For the G/G/1 case, we restrict our attention to finding an optimal 
non-preemptive policy,\footnote{The technical reasons for not considering preemptive policies are summarized in Remark \ref{rem:preempt} (in Section \ref{sec:SRP_details}).} while for the G/M/1 case, we 
consider finding an optimal preemptive policy. 

As far as we know, the considered problem remains unsolved even for a single source and zero transmission cost. 
For a single source, the problem of minimizing the sum of the AoI and the transmission cost has been considered in  \cite{saurav2021minimizing}, however, without any transmission delay, where an optimal threshold based policy has been derived when $G$ is exponential.
Similarly, for a single source, without accounting for the transmission cost, an always preempt policy (always transmit the newly arrived update while discarding the update under transmission) has been studied \cite{kaul2012status, multisourceyates, multisource} for minimizing the AoI only. However, note that forcibly preempting updates is not necessarily optimal. The decision question on  
whether to preempt/discard an on-going update transmission if a new update arrives at the server or not, has been considered in \cite{kavitha2021controlling}, however, without any theoretical guarantee.
With multiple sources having different $\cG_i,\cD_i$ distributions, this problem is further compounded. It is worthwhile noting that the considered continuous-time model also generalizes the discrete-time model \cite{kadotascheduling, kadota2019minimizing}, where scheduling problems have been studied for the specific case when $\cG_i$ and $\cD_i$ are geometric.

It turns out that directly finding the optimal non-preemptive/preemptive scheduling policy is challenging, and we take recourse in finding scheduling policies with small competitive ratios, where the competitive ratio is defined as follows. 
For a scheduling policy, the competitive ratio is defined as the ratio of the expected sum of the AoI and the transmission cost achieved by the scheduling policy and the expected sum of the AoI and the transmission cost achieved by an optimal offline policy that is aware of the input in advance. 


\subsection{Comparison with other Scheduling Problems}
One of the most well-studied scheduling problems in literature is the flow-time (sum of job response times) minimization problem, where both the stochastic  as well as arbitrary input is considered. In the arbitrary input case, jobs/packets with arbitrary sizes arrive at arbitrary time to a queue, while in the stochastic case both the arrival times and job/packet sizes follow a distribution. With a single server, that can process at most one packet at any time with a fixed speed, the problem is to find the scheduling policy that minimizes the  flow time of each job. It turns out that the elegant shortest remaining processing time (SRPT) policy  that at any time processes the job with the least remaining processing time is optimal in both the stochastic as well as the arbitrary input case  \cite{schrage1968letter}. For the M/G/1 system, flow time minimization with uncertainty about job sizes has been considered in \cite{scully2020simple}, which showed that the Gittin's policy is `universally' optimal.

The flow time minimization problem becomes more challenging in the presence of multiple servers. With multiple servers, for the stochastic input, multi-server SRPT policy is optimal in the heavy-traffic limit \cite{grosof2018srpt}, while for the arbitrary input, multi-server SRPT policy is order-wise optimal \cite{leonardi2007approximating}. 
In the arbitrary input case, the weighted counterpart of the flow time minimization problem where the objective is the weighted average of the job flow-times is also well-studied \cite{bansal2009weighted}, and for which it is known that competitive online policies do not exist without resource augmentation. 

Other related scheduling problems include the completion time problem \cite{hall1997scheduling, khuller2019select}, the makespan problem \cite{graham1966bounds,goel1999stochastic}, and the co-flow problem \cite{shafiee, sincronia, twoapprox, bhimaraju2020non}. Scheduling to minimize multi-class weighted flow time in a M/G/1 system, where class $j$ jobs have weight $c_j$ and service rate $\mu_j$ has been considered, and for which the 
famous $c\mu$-rule that schedules the job with the largest $c_j\mu_j$ has been shown to be optimal \cite{cox1991queues, klugel2019aoi}. 
In addition, there is some analysis of multi-server priority queues.
where there are finitely many classes of jobs with exponential
or phase-type service requirement distributions \cite{mitrani1981multiprocessor, sleptchenko2005exact,harchol2005multi}. 

Compared to these well studied scheduling problems, the AoI scheduling problem is more involved primarily because all the updates for each of the sources need not be transmitted/serviced. This critical observation follows directly
from the definition of the AoI, since updates on their own have no meaning, all that matters is the time difference between two successive updates received at the monitor from the same source.
This additional combinatorial feature (which subset of updates should be transmitted) makes the considered problem fundamentally different than the well studied scheduling problems. 


\subsection{Our Contributions}
We first describe our contributions for the G/G/1 system, where the goal is to find a non-preemptive policy with small competitive ratio.
\subsubsection{Policy}\label{Intro:policy}
To solve the considered problem, we propose a randomized non-preemptive scheduling policy, that on arrival of each update for source $i$, considers it for transmission with probability $p_i$ and discards it immediately otherwise. Whenever an update under transmission is received at the monitor and the channel becomes free, 
source $\ell$ is picked for transmission with probability $\hat{p}_\ell$, where 
\begin{align*} 
	\hat{p}_\ell=\frac{(p_\ell/\mu_\ell)}{\sum_{i=1}^{N}(p_i/\mu_i)},
\end{align*} and $\mu_i$ is the mean of the inter-generation time of updates at source $i$. The policy thus only needs the knowledge of $\mu_i$ and not of  the complete distribution $\cG_i$.
The selected source then transmits its most recently arrived update. 
Critically, if the selected source $i$ has no (new) update to transmit, the policy 
waits/idles for a random period of time, drawn independently from the service time distribution $\cD_i$. 
The probabilities $p_i$'s are chosen by solving a convex optimization problem having constraints that are functions of the expected service time for each source. The considered convex program tries to minimize an upper bound on the sum of the AoI and the average transmission cost for the considered policy. 

\subsubsection{Guarantee}
 
For the proposed policy, we show that its competitive ratio (against a non-preemptive optimal offline algorithm) is upper bounded by 
  \begin{align} \label{intro:eq:CR-for-SR}
    	\max\{4,\ \ 3+\max_{\ell}\{\sigma_\ell^2/\mu_\ell^2\}\}.
    \end{align}
 where $\sigma_\ell^2$ is the  variance of the packet inter-generation time for source $\ell$.
 Notably, for many of the `nice' distributions, e.g. exponential, uniform, Rayleigh, etc., the ratio $\sigma_\ell^2/\mu_\ell^2\le1$ and the competitive ratio is upper bounded by $4$. It is worth noting that the competitive ratio upper bound \eqref{intro:eq:CR-for-SR} is independent of the service time distributions $\cD_\ell$'s. Moreover, we recover the $4$-competitive result of \cite{kadota2019minimizing} for the discrete-time model that considered geometrically distributed inter-generation times of updates since for that $\sigma_\ell^2/\mu_\ell^2=1$. 
 As far as we know, this is a first such result in the area of AoI scheduling, with general inter-generation time and service time distributions $\cG_i$ and $\cD_i$, respectively. 
 
We also construct a `tight' example with a single source to show that the competitive ratio of the proposed policy is at least $\frac{\sigma^2}{2\mu^2}$, and conclude that the dependence of the competitive ratio of the proposed policy on $\frac{\sigma^2}{\mu^2}$ is unavoidable.
 
\subsubsection{Remarks on the Policy and the Guarantee}
One feature of the proposed policy that is worth discussing is : why should the policy idle when a source is selected for transmission without having an update to transmit, and not move on to pick any of the sources that have an update to transmit. 
The primary answer to this is the ease of analysis, since it keeps the system symmetric without having to keep track of sources that have an update to transmit. Moreover,  it is not obvious whether a simple modification to the policy that picks only those sources for transmission that have an update to transmit,  
strictly decreases the AoI over the proposed policy. The intuition for this can be borrowed  from \cite{sun2017update}, where it has been shown that introducing non-zero delay between two successive update transmissions is in fact optimal to minimize the AoI 
for a single source under the generate-at-will model and stochastic service times. Essentially, introducing idling time can help in transmitting `fresher' packets which in turn can potentially improve the AoI. 
Also, in Section \ref{sec:numerical-parametric} we show (using numerical simulation) that when the cost per transmission is large, idling indeed helps in minimizing the sum of the AoI and the average transmission cost. 

Next, we describe the reason for the competitive ratio of the proposed policy to be a function of  $\sigma_\ell^2/\mu_\ell^2$. To derive an upper bound on the competitive ratio, the first critical step is to lower bound the cost of the optimal offline policy ($\opt$). 
In particular, for the considered problem, we need to lower 
bound the AoI cost paid by the $\opt$, which depends on the variance of the time-difference between the generation time of two consecutive updates that 
are in fact transmitted by the $\opt$. As we show in Example \ref{ex:zero-var}, this variance can be made arbitrarily close to zero by the $\opt$, and hence we lower bound the variance of the $\opt$ by zero. In contrast, for the proposed policy, each arriving update at source $i$ is eligible to be transmitted with a fixed probability $p_i$, and consequently the mean and the variance of time-difference between the generation time of two consecutive updates that 
are  transmitted by the proposed policy for source $i$ are functions of $\mu_i/p_i$ and $\sigma_i^2/p_i$, respectively. 
Moreover, the eventual AoI cost of the proposed policy is a function of the ratio of the  variance and mean square, thus resulting 
in the bound  \eqref{intro:eq:CR-for-SR}. 

\subsubsection{Preemptive policies for G/M/1 system}
We also consider preemptive policies, however, only in a G/M/1 system. In particular, the offline optimal policy is allowed to be preemptive. 
We show that the competitive ratio of a non-preemptive policy that is conceptually identical to the policy defined in Section \ref{Intro:policy}, that uses only some different constants, has a competitive ratio of 
\begin{align} \label{intro:result:preempt}
    	\max\{6,\ \ 5+\max_{\ell}\{\sigma_\ell^2/\mu_\ell^2\}\}.
    \end{align}
    against an optimal offline algorithm that is allowed to be preemptive. 
    This is an interesting result, since even with the exponential service time distribution in the G/M/1 system, a preemptive algorithm can definitely improve its AoI performance by preempting packets since AoI improves by servicing fresher packets.
    However,  to prove this result we show that even with preemption, the total time between two consecutive completely transmitted/serviced packet for a source $i$ is still bounded by the expected service time $\gamma_i$ of $\cD_i$.




\section{System Model}
\label{sec:sysModel}
In this paper, we consider the G/G/1 system for non-preemptive policies, and a G/M/1 system for preemptive policies. To avoid notational inconvenience, we defer the 
discussion of preemptive policies to Section \ref{sec:preemption} that requires some added notations, and only consider non-preemptive policies for a G/G/1 system here.

Consider a system consisting of $N$ sources, where updates (henceforth, packets) are generated at source $\ell$ intermittently, and the inter-generation time between the $i^{th}$ and the $i+1^{st}$ packet is $X_{\ell i}$. We assume that $X_{\ell i}$  is 
independent and identically distributed according to some distribution $\cG_{\ell}$, with mean $\mu_\ell<\infty$ and variance $\sigma_\ell^2$. There is a single monitor, and all the $N$ sources wish to send their updates to the monitor as soon as possible. 
At any time, at most one source can transmit its packet to the monitor, and packet $i$'s transmission by source $\ell$ takes $d_{\ell i}$ time units (called transmission time) to complete (received at the monitor). We assume that transmission time ($d_{\ell i}$) experienced by packet $i$ of source $\ell$ is 
independent and identically distributed according to some general distribution $\cD_\ell$, with mean $\gamma_\ell<\infty$, independent of everything else. 
There is a single centralized scheduler, where at any time $t$, the scheduler has causal information of all the sources, and gets to decide which source should transmit when the channel becomes free (previous transmission is completed). With the non-preemptive restriction, packets can be transmitted in any order or discarded if their transmission has never started, however, a packet under transmission cannot be preempted or discarded.

This description is equivalent to a G/G/1 system without preemption, however, we are using the standard language found in the AoI literature. Also with abuse of notation, we will interchangeably use transmission or service time of a packet without any ambiguity.
The reasons for restricting to non-preemptive policies is  summarized in Remark \ref{rem:preempt}.

\begin{definition} \label{def:busy-channel}
	At any time $t$, the channel is said to be {\it busy}, if a packet is already under transmission by some source. Otherwise, the channel is {\it free}. A source can begin transmission of a packet only when the channel is free. 
\end{definition}

\begin{assumption}
	We assume that for each source $\ell$, the scheduler knows the transmission time distribution $\cD_\ell$. 
	However, regarding the packet inter-generation time distribution $\cG_\ell$, the scheduler only knows its mean $\mu_\ell$.
\end{assumption} 
\begin{remark}
	Note that for different sources, $\cG_\ell$'s and $\cD_\ell$'s may belong to different family of distributions. For example, we may have $\cG_1$ a uniform distribution, $\cG_2$ an exponential distribution, $\cD_1$ a Rayleigh distribution, and $\cD_2$ a log-normal distribution.
\end{remark}

At any time $t\ge 0$, age of source $\ell$ at the monitor is $\Delta_{\ell}(t)=t-\lambda_{\ell}(t)$, where $\lambda_{\ell}(t)$ denotes the generation time of the latest packet of source $\ell$ that has been received at the monitor (i.e., completely transmitted by source $\ell$) until time $t$. Thus, the \emph{age of information (AoI)} $\Delta_{\ell}^{av}(t)$ of source $\ell$ until time $t$ is  
\begin{align}\label{eq:avAge}
\Delta_{\ell}^{av}(t)=\frac{1}{t}\int_{0}^{t}\Delta_{\ell}(i)di.
\end{align}

Each time source $\ell$ transmits a packet, it incurs a transmission cost of $c_{\ell}$ units, where $c_\ell\ge 0$ is constant, and includes the cost for channel usage, as well as the energy required to transmit the packet. 
Hence, the average transmission cost incurred by source $\ell$ until time $t$ is given by
\begin{align}\label{eq:avTxCost}
C_{\ell}^{av}(t)=\frac{c_{\ell}R_\ell(t)}{t}, 
\end{align}
where, $R_\ell(t)$ denotes the number of packets transmitted by source $\ell$ until time $t$ (including the packet being transmitted at time $t$). 

\begin{definition} \label{def:causal-policy}
	A \textit{causal transmission policy} (in short, causal policy) refers to a centralized algorithm (employed by the scheduler) that at each time $t$ when the channel is free, based only on the causal information of all the sources available at time $t$, schedules at most one source to transmit its packet. For this section, all policies are non-preemptive.
\end{definition}

The objective is to find a causal policy (Definition \ref{def:causal-policy}) that minimizes a linear combination of the AoI and the average transmission cost of all the sources (henceforth, called the \emph{weighted sum cost}). Formally, the weighted sum cost of policy $\pi$ is given as   
\begin{align} \label{eq:tac}
		\Gamma(\pi)&=\lim_{t\rightarrow \infty}\frac{1}{N}\sum_{\ell=1}^{N}( C_{\ell,\pi}^{av}(t)+\rho_{\ell}\Delta_{\ell,\pi}^{av}(t)),
\end{align}
where $\rho_{\ell}\ge0$ is a constant (weight parameter) corresponding to each source $\ell$, and $C_{\ell,\pi}^{av}(t)$ and $\Delta_{\ell,\pi}^{av}(t)$ respectively denote the average transmission cost and the AoI of source $\ell$ until time $t$, under policy $\pi$. The objective is formulated as the optimization problem
\begin{equation} \label{eq:objective}
\underset{\pi\in\Pi}{\min} \ \ \Gamma(\pi),
\end{equation}
where $\Pi$ is the set of all causal policies $\pi$. 

\begin{remark}
	We consider the cost function \eqref{eq:tac} to be a linear combination of the AoI and the average transmission cost, as it naturally captures the tradeoff between the AoI and the transmission cost, specially in systems where there is no explicit constraint on the AoI/transmission cost, e.g. in systems with sufficient energy supply with cost per unit consumption.
\end{remark}
\begin{remark} \label{remark:rho}
	Any cost function of the form $\sum_{\ell=1}^{N}(\rho_{\ell,1} C_{\ell,\pi}^{av}(t)+\rho_{\ell,2} \Delta_{\ell,\pi}^{av}(t))/N$ (where $\rho_{\ell,1},\rho_{\ell,2}>0$ are constants $\forall \ell$) can be expressed as $\sum_{\ell=1}^{N}(C_{\ell,\pi}^{av}(t)+\rho_\ell\Delta_{\ell,\pi}^{av}(t))/N$ (for $\rho_\ell=\rho_{\ell,2}$), by subsuming $\rho_{\ell,1}$ inside cost per transmission $c_{\ell}$ (that appears in the expression for $C_{\ell,\pi}^{av}$). Therefore, the objective function \eqref{eq:tac} is equivalent to the general problem that considers weighted sum of the average transmission cost and the AoI across all the sources.  
\end{remark}

\begin{remark} \label{remark:finite-tx-cost}
	Without loss of generality, we assume that for each source $\ell$, $0<\rho_\ell<\infty$, and $0\le c_\ell<\infty$. When these conditions are not satisfied, the optimization problem \eqref{eq:objective} becomes trivial.
\end{remark}

\begin{remark} \label{remark:zero-transmission-cost}
	When $c_\ell=0$ for each source $\ell$, the objective \eqref{eq:objective} simplifies to an AoI minimization problem with multiple sources, where the packet inter-generation times, as well as the transmission time for packets may follow any general distribution. In prior work, such an AoI minimization problem has been considered under restricted settings, such as with single source \cite{saurav2021minimizing,sun2017update}, generate-at-will model \cite{kadota2018scheduling,sun2017update}, discrete-time model, \cite{kadota2019minimizing,kadota2018scheduling}, zero transmission time \cite{saurav2021minimizing}, etc.
\end{remark}

From prior work \cite{kadota2019minimizing,kadota2018scheduling,saurav2021minimizing,kadotascheduling}, it is known that for a multi-source setup, finding an optimal causal policy is challenging. Thus, in this paper, to quantify the performance of a causal policy, we compare it against  the performance of an optimal offline policy using the metric of competitive ratio (Definition \ref{def:CR-def}). Compared to a causal policy, an offline policy has access to more information, and that typically 
allows lower bounding of the cost of an optimal causal policy. However, more the extra information an offline policy has, larger is the gap between the 
performance of the optimal offline policy and a causal policy. Thus, ideally, we want the offline policy to have as little extra information as possible over the causal policy, while allowing analytical tractability. For the considered problem, we let the offline policy know the inter-generation 
time of updates non-causally, but the transmission time/delay experienced by any packet that is transmitted is revealed to it causally. 
We rigorously define the optimal offline algorithm that we consider next.

\begin{definition} \label{def:OPT}
	A policy $\pi_{OF}^\star$ is called an optimal offline non-preemptive policy, if its weighted sum cost $\Gamma(\pi_{OF}^\star)$ \eqref{eq:tac} is minimum among all non-preemptive policies that know the generation time of all the packets (at each source) in advance. 
	We assume that the transmission time for each packet is realized once the packet is transmitted, and it is not known to $\pi_{OF}^\star$ non-causally.
\end{definition}


\begin{definition} \label{def:CR-def}
	For a causal  non-preemptive policy $\pi$, its competitive ratio $\textsc{cr}_{\pi}$ is defined as the ratio of the expected weighted sum cost \eqref{eq:tac} for policy $\pi$ and the expected weighted sum cost \eqref{eq:tac} for an optimal offline non-preemptive policy $\pi_{OF}^\star$ (Definition \ref{def:OPT}), where the expectation $\bbE[\cdot]$ is jointly with respect to the distributions $\cG_\ell$ and $\cD_\ell$ (for each source $\ell$), 
	and the corresponding transmission policy (i.e., $\pi$ or $\pi_{OF}^\star$). 
	Mathematically,
	\begin{align} \label{eq:CR-def}
		\textsc{cr}_{\pi}=\frac{\bbE[\Gamma(\pi)]}{\bbE[\Gamma(\pi_{OF}^\star)]}.
	\end{align}
\end{definition}

In the following, we propose an easy to implement causal non-preemptive stationary randomized policy (in Section \ref{sec:SRP_details}), and derive an upper bound on its competitive ratio (Definition \ref{def:CR-def}), which is shown to be at most 4 for common distributions such as exponential, uniform and Rayleigh.  Before presenting these results, we need some preliminaries, which we discuss next. 

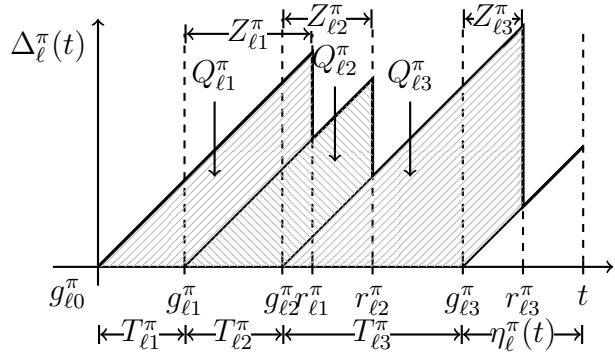
\begin{figure} 
	\begin{center}
		\begin{tikzpicture}[thick,scale=1, every node/.style={scale=1}]
		\draw[->] (-0.25,0) to (7.2,0); 
		\draw[->] (0.35,-0.25) to (0.35,3.3) node[below left]{$\Delta_\ell^\pi(t)$};
		\draw[very thick] (0.35,0) node[below left]{$g_{\ell 0}^\pi$} to (3.2,2.85) to (3.2,1.7) to (4,2.5) to (4,1.2) to (6,3.2) to (6,0.8) to (6.80,1.60); 
		
		
		
		\draw (1.5,-0.1) node[below]{$g_{\ell 1}^\pi$} to (1.5,0.1);
		\draw (2.8,-0.1) node[below]{$g_{\ell 2}^\pi$} to (2.8,0.1);
		\draw (5.2,-0.1) node[below]{$g_{\ell 3}^\pi$} to (5.2,0.1); 
		
		\draw (3.2,-0.1) node[below]{$r_{\ell 1}^\pi$} to (3.2,0.1);
		\draw (4,-0.1) node[below]{$r_{\ell 2}^\pi$} to (4,0.1);
		\draw (6,-0.1) node[below]{$r_{\ell 3}^\pi$} to (6,0.1); 
		\draw (6.8,-0.1) node[below]{$t$} to (6.8,0.1); 
		
		\draw[dashed] (1.5,0.1) to (1.5,3.1);
		\draw[dashed] (2.8,0.1) to (2.8,3.3); 
		\draw[dashed] (3.2,0.1) to (3.2,3.1); 
		\draw[dashed] (4,0.1) to (4,3.2);
		\draw[dashed] (5.2,0.1) to (5.2,3.2);
		\draw[dashed] (6,0.95) to (6,0.1);
		\draw[dashed] (6.8,3.1) to (6.8,0.1);
		
		
		\draw (1.5,0) to (3.2,1.7); 
		\draw (2.8,0) to (4,1.2);
		\draw (5.2,0) to (6,0.8); 
		
		\draw[|<->] (0.35,-0.9) -- (1.5,-0.9) node[rectangle,inner sep=-1pt,midway,fill=white]{$T_{\ell 1}^\pi$}; 
		\draw[|<->] (1.5,-0.9) -- (2.8,-0.9) node[rectangle,inner sep=-1pt,midway,fill=white]{$T_{\ell 2}^\pi$};
		\draw[|<->|] (2.8,-0.9) -- (5.2,-0.9) node[rectangle,inner sep=-1pt,midway,fill=white]{$T_{\ell 3}^\pi$};
		\draw[<->|] (5.2,-0.9) -- (6.8,-0.9) node[rectangle,inner sep=-1pt,midway,fill=white]{$\eta_\ell^\pi(t)$};
		
		\draw[|<->|] (1.5,3.1) -- (3.2,3.1) node[rectangle,inner sep=-1pt,midway,fill=white]{$Z_{\ell 1}^\pi$}; 
		\draw[|<->|] (2.8,3.3) -- (4,3.3) node[rectangle,inner sep=-1pt,midway,fill=white]{$Z_{\ell 2}^\pi$};
		\draw[|<->|] (5.2,3.3) -- (6,3.3) node[rectangle,inner sep=-1pt,midway,fill=white]{$Z_{\ell 3}^\pi$};
		
		
		
		\fill[pattern=north east lines, pattern color=lightgray] (0.35,0) to (3.2,2.85) to (3.2,1.7) to (1.5,0) to (0.35,0);
		\draw[->, thick] (1.9, 2.2) node[above]{$Q_{\ell 1}^\pi$} to (1.9,1.2);
		\fill[pattern=north west lines, pattern color=lightgray] (1.5,0) to (4,2.5) to (4,1.2) to (2.8,0) to (1.5,0);
		\draw[->, thick] (3.5, 2.4) node[above]{$Q_{\ell 2}^\pi$} to (3.5,1.4);
		\fill[pattern=north east lines, pattern color=lightgray] (2.8,0) to (6,3.2) to (6,0.8) to (5.2,0) to (2.8,0);
		\draw[->, thick] (4.5, 2.2) node[above]{$Q_{\ell 3}^\pi$} to (4.5,1.2);
		
		\end{tikzpicture}
	    \caption{Sample age plot of source $\ell$ in a multi-source system, where the time-averaged area under the solid black lines is the AoI. Packet $\ell_i^\pi$ generated at source $\ell$ at time $g_{\ell i}^\pi$ is received at the monitor at time $r_{\ell i}^\pi$.}
		\label{fig:multi-node-general-age} 
	\end{center}
\end{figure}

Let the packets generated at source $\ell$ be indexed as $\ell_1,\ell_2,...,\ell_i,...$ in increasing order of their generation time. Also, let $g_{\ell i}$ denote the generation time of packet $\ell_i$, with $g_{\ell i}-g_{\ell (i-1)}=X_{\ell i}\sim \cG_\ell$.
\begin{definition} \label{def:fresh}
	At any time $t$, a packet $\ell_i$ (generated until time $t$) is called {\bf fresh}, if the generation time of packet $\ell_i$ is greater than the generation time of the latest generated packet of source $\ell$ that has been received at the monitor until time $t$, i.e., $t\ge g_{\ell i}>\lambda_\ell(t)$. Note that at any time, a source can have multiple fresh packets. Also, for minimizing AoI, only fresh packets are useful.
\end{definition}
Among all the packets generated at source $\ell$, under a policy $\pi$, a subset of these packets get transmitted to the monitor. Let the packets of source $\ell$ that get transmitted to the monitor under a policy $\pi$ be indexed as $\ell_1^\pi,\ell_2^\pi,...$ in increasing order of their generation times. Also, let $g_{\ell i}^\pi$, $s_{\ell i}^\pi$, and $r_{\ell i}^\pi$ denote the generation time of packet $\ell_i^\pi $, the instant when the transmission of packet $\ell_i^\pi$ begins, and the time when packet $\ell_i^\pi$ is received at the monitor, respectively, under policy $\pi$. Therefore, $g_{\ell i}^\pi\le s_{\ell i}^\pi\le r_{\ell i}^\pi$.
\begin{definition} \label{def:system-time}
	The system time of packet $\ell_i^\pi$ (denoted by $Z_{\ell i}^\pi$) is defined as the difference between the generation time of packet $\ell_i^\pi$ and the time when it gets received at the monitor (i.e., when its transmission completes). Therefore, $Z_{\ell i}^\pi=r_{\ell i}^\pi-g_{\ell i}^\pi$, as shown in Figure \ref{fig:multi-node-general-age}. 
\end{definition}
	
	Note that the system time (Definition \ref{def:system-time}) of packet $\ell_i^\pi$ can also be written as 
	\begin{align} \label{eq:system-time}
		Z_{\ell i}^\pi=w_{\ell i}^\pi+d_{\ell i},
	\end{align}
	where $w_{\ell i}^\pi=s_{\ell i}^\pi-g_{\ell i}^\pi$ is the waiting time for packet $\ell_i^\pi$ (i.e., 	
	the duration of time when packet $\ell_i^\pi$ is available at source $\ell$, but is not being transmitted), and $d_{\ell i}=r_{\ell i}^\pi-s_{\ell i}^\pi$ denotes the total transmission time for packet $\ell_i^\pi$. 
	Note that  $d_{\ell i}$ is a realization from $\cD_\ell$ and hence independent of the transmission policy $\pi$. However, the waiting time $w_{\ell i}^\pi$ depends on $\pi$, which implies that the system time $Z_{\ell i}^\pi$ also depends on $\pi$. Moreover, $Z_{\ell i}^\pi\ge d_{\ell i}$ (since $w_{\ell i}^\pi \ge 0$, by definition). 

Further, we define a period as follows.
\begin{definition} \label{def:period}
	Under policy $\pi$, for each source $\ell$, we define a \textbf{period} as the time interval between the generation time of two consecutive packets of source $\ell$ that are received at the monitor.  Since the considered policies are non-preemptive, set of packets transmitted is the same as the set of packets received at the monitor.
As shown in Figure \ref{fig:multi-node-general-age}, the interval $\cP_{\ell i}^{\pi}=(g_{\ell (i-1)}^\pi,g_{\ell i}^\pi]$ represents the $i^{th}$ period of source $\ell$ under policy $\pi$, and the length (duration) of period $\cP_{\ell i}^\pi$ is $T_{\ell i}^\pi=g_{\ell i}^\pi-g_{\ell (i-1)}^\pi$.
	Note that a period starts as well as ends at the generation time of some packet. Hence, $T_{\ell i}^\pi$ depends on both the packet generation process, and the policy $\pi$. 
\end{definition} 

\begin{remark} \label{assume:init-AoI}
	Without loss of generality, we assume that the age of all the sources at time $t=0$ is $0$ (i.e., $\Delta_\ell(0)=0$, $\forall \ell\in\{1,\cdots,N\}$). In \eqref{eq:objective}, since we are interested in the weighted sum cost over infinite time horizon, this assumption does not affect the final solution of \eqref{eq:objective}, but simplifies the analysis, and  for each source $\ell$, allows us to assume $g_{\ell 0}^\pi=0$ (i.e., the first period of every source starts at time 0).
\end{remark}

As shown in Figure \ref{fig:multi-node-general-age}, for any policy $\pi$ and source $\ell$, when $g_{\ell 0}^\pi=0$, any time $t$ can be written as 
\begin{align} \label{eq:sumT=t-n}
t=\sum_{i=1}^{R_\ell^{\pi}(t)}T_{\ell i}^\pi+\eta_\ell^\pi(t),
\end{align}
where $R_\ell^\pi(t)$ denotes the number of packets transmitted by source $\ell$ until time $t$ under policy $\pi$, $\eta_\ell ^\pi(t)=t-\lambda_\ell(t)=t-g_{\ell R_\ell^\pi(t)}^\pi$ denotes the length of the ongoing period of source $\ell$ at time $t$, and $\lambda_\ell^\pi(t)$ denotes the generation time of the latest generated packet of source $\ell$ that has been received at the monitor until time $t$, under policy $\pi$.

Using this formalism, we next propose and analyze randomized scheduling policy to solve problem \eqref{eq:objective}.

\section{Stationary Randomized Causal Transmission Policy} \label{sec:SRP_details}
 
Consider a stationary randomized policy $\pi_{sr}$ (Algorithm \ref{algo:distribution-independent-policy}), that at any time $t$, $(i)$ if a packet is generated at source $\ell$, marks the generated packet with probability $p_{\ell}$ (and discards it otherwise), and $(ii)$ if the channel becomes free (latest transmission is complete/received at the monitor), chooses source $\ell$ (among all the $N$ sources) for transmission with probability 
\begin{align} \label{eq:sr-p-hat}
	\hat{p}_\ell=\frac{(p_\ell/\mu_\ell)}{\sum_{i=1}^{N}(p_i/\mu_i)}.
\end{align}
If the chosen source $\ell$ has at least one fresh marked packet, then its latest generated marked packet is transmitted. Else, the policy waits for a random time duration, independently sampled from distribution $\cD_\ell$. 
The probability vector $[p_1,p_2,...,p_N]$ is obtained by solving the following \emph{convex optimization problem}   
\begin{align} \label{eq:tx_prob}
	&\underset{[p_1,...,p_N]}{\arg\min}\sum_{\ell=1}^{N}\left(\frac{2\rho_{\ell}\mu_{\ell}}{p_\ell}+\frac{c_\ell p_\ell}{\mu_\ell}\right), \\
	\label{eq:interference-constraint}
	& s.t. \ \ \sum_{\ell=1}^{N}\frac{p_\ell \gamma_\ell}{\mu_\ell}\le 1, \\
	\label{eq:probability-constraint}
	& \hspace{5ex} p_\ell\in [0,1], \ \ \forall \ell\in\{1,\cdots,N\}. 
\end{align} 


\begin{remark}
	The rationale for choosing $p_\ell$'s as the solution of \eqref{eq:tx_prob}---\eqref{eq:probability-constraint} is that in Lemma \ref{lemma:gamma-pisr-ub-gen-dist},
	we show that 
the expression \eqref{eq:tx_prob}, with some additional constant terms, is an upper bound on the expected weighted sum cost for the proposed policy $\bbE[\Gamma(\pi_{sr})]$. 
Thus the proposed policy is choosing $p_\ell$'s to minimize an upper bound on its weighted sum cost. 
	Note that the constraint \eqref{eq:interference-constraint} is derived in Lemma \ref{lemma:actual-tac-with-t} (that relates the number of packets that each source may transmit). 
\end{remark}

\begin{remark} \label{remark:sr-non-wc}
	Under $\pi_{sr}$ (Algorithm \ref{algo:distribution-independent-policy}), if a source $\ell$ is chosen by SR-NSS which does not have a fresh marked packet, then it waits for a random amount of time as per distribution $\cD_\ell$ (instead of choosing a different source). This is to ensure that the time duration between two successive instants when a given source $\ell$ is chosen under SR-NSS (to transmit), does not depend on SR-PMS (i.e., when the packets are marked at source $\ell$). 
	Also, it has been shown in \cite{sun2017update,zou2019waiting} that when the transmission times are random, non-zero wait might help in minimizing the AoI for a source. In fact, in Section \ref{sec:numerical-parametric} using numerical simulation (Figure \ref{fig:WSC_vs_TxCost}), we show that when the cost per transmission is large, waiting helps in minimizing the weighted sum cost \eqref{eq:tac}. Thus, it is not trivially wasteful to wait. 
\end{remark}

\begin{remark}
	The probability $\hat{p}_\ell$ \eqref{eq:sr-p-hat} that source $\ell$ is chosen to transmit is a function of both the probability $p_\ell$ \eqref{eq:tx_prob} with which source $\ell$ marks a generated packet, and the mean packet inter-generation time $\mu_\ell$ at source $\ell$. Intuitively, this synchronizes SR-PMS and SR-NSS such that mostly when a source $\ell$ is chosen to transmit, it has a fresh marked packet to transmit (discussed in Section \ref{sec:numerical-parametric}). Note that the subroutine SR-PMS (in particular, the choice of $p_\ell$'s \eqref{eq:tx_prob}) used to mark/discard packets is critical as it prevents $\pi_{sr}$ from transmitting unnecessarily when the cost per transmission is large.
\end{remark}

\begin{algorithm}
	\caption{Stationary randomized policy $\pi_{sr}$.}
	\label{algo:distribution-independent-policy}
	\begin{algorithmic} [1]
		\STATE \emph{/* SR-Packet Management Subroutine (SR-PMS) */}  
		\FOR{each packet generated at source $\ell\in\{1,\cdots,N\}$}
		\STATE mark the packet with probability $p_\ell$, and discard it with probability $1-p_\ell$;
		\ENDFOR

\STATE \emph{/* SR-Node Scheduling Subroutine (SR-NSS) */}
\FOR{time $t\ge0$, if channel is free}
\STATE among the $N$ sources, choose 
source $\ell$ with probability $\hat{p}_\ell$ \eqref{eq:sr-p-hat}; 
\IF{source $\ell$ has at least one fresh marked packet}
    \STATE transmit the latest marked packet of source $\ell$;
\ELSE 
    \STATE wait for random time $d_\ell\sim\cD_\ell$;
\ENDIF

\ENDFOR
\end{algorithmic}
\end{algorithm}


\begin{remark}
	In past, the concept of using two independent subroutines for minimizing AoI has also been used in \cite{sun2019age} in the design of separation principle policy (SPP). In \cite{sun2019age}, a multi-source slotted-time model has been considered, where at each source $\ell$, packets are generated at rate $q_\ell$, 
	and in each slot $t$, at most one source $\ell$ can transmit, with transmission success probability $u_\ell$. 
	Further, each source has a FCFS constraint, i.e., in any slot, a source can only transmit its earliest generated packet that has not been received at the monitor. The objective is to minimize the weighted sum AoI of the sources by designing packet generation rate control, and a scheduling policy for the sources.\footnote{
	Due to FCFS constraint, if packet generation rate is high, the waiting time for packets will be large, thereby increasing AoI.} To achieve this objective, SPP is proposed, that considers packet generation rate control and scheduling of sources, independently, and is shown to be near optimal.
\end{remark}

{\bf The first main result of this paper is as follows.}
\begin{theorem} \label{thm:dist-ind-algo-bound-cr}
	The stationary randomized policy $\pi_{sr}$ (Algorithm \ref{algo:distribution-independent-policy}) has competitive ratio (against non-preemptive optimal offline algorithm)
    \begin{align} \label{eq:CR-for-SR}
    	\textsc{cr}_{\pi_{sr}}\le\max\{4,\ \ 3+\max_{\ell}\{\sigma_\ell^2/\mu_\ell^2\}\},
    \end{align}
    where $\sigma_\ell^2$ and $\mu_\ell$ respectively denote the variance and the mean of packet inter-generation times at source $\ell$.
\end{theorem}

Theorem \ref{thm:dist-ind-algo-bound-cr} shows that the competitive ratio for $\pi_{sr}$ is independent of the transmission time distribution $\cD_\ell$. Intuitively, this is because the optimal offline policy $\pi_{OF}^\star$ also does not know the realization of the transmission times of packets non-causally (Definition \ref{def:OPT}), and hence, the impact of random transmission time on $\pi_{OF}^\star$ is similar to that on $\pi_{sr}$. 

However,  the competitive ratio for $\pi_{sr}$ depends on the distribution of inter-generation time of packets (in particular, on $\sigma_\ell^2/\mu_\ell^2$). This is because the optimal offline policy knows the generation time of packets in advance ($\pi_{sr}$ only knows the expected inter-generation time of packets, not even the distribution), and for certain distributions, $\pi_{OF}^\star$ can use this information and minimize the variance of its period lengths to 0 (irrespective of $\sigma_\ell^2$), whereas the AoI of $\pi_{sr}$ always depends on $\sigma_\ell^2$.  Next, we make this concrete via constructing a tight example, and show that   the dependence of the competitive ratio on $\pi_{sr}$ on $\sigma_\ell^2/\mu_\ell^2$ for $\pi_{sr}$ is unavoidable. 

\begin{example} \label{ex:zero-var}
	Consider a system with a single source ($N=1$), where the packet inter-generation time is distributed as 
	\begin{align}
	X = \begin{cases} \alpha, & \text{with probability} \ 0.5, \\
	 \epsilon & \text{with probability} \ 0.5, \end{cases}
	\end{align}
	 and $\epsilon\to 0^+$, while $\alpha$ is a large (but finite) positive constant. Thus, the mean and the variance of the packet inter-generation time $X$ are $\mu=(\alpha+\epsilon)/2\approx \alpha/2$ and $\sigma^2=(\alpha-\epsilon)^2/4\approx \alpha^2/4$, respectively. Also, let the cost per transmission $c=0$, and the transmission time $d_i=0$ for each packet (hence the expected transmission time $\gamma=0$). 
	
	Consider a threshold policy $\pi_{tp}$, that on generation of a packet at time $t$, transmits it immediately if $t-\lambda(t)\ge\alpha$ (where $\lambda(t)$ denotes the generation time of the latest packet that got transmitted until time $t$), and discards it otherwise. Let $g_i^{tp}$ denote the generation time of the $i^{th}$ packet transmitted under policy $\pi_{tp}$.
	Then the period lengths under policy $\pi_{tp}$ are $T_i^{tp}=g_i^{tp}-g_{i-1}^{tp}=m_i\alpha+n_i\epsilon$, where $m_i$ and $n_i$ denote the number of packets generated in the $i^{th}$ period with inter-generation time $\alpha$ and $\epsilon$, respectively. 
	Since the threshold for transmission of packets is $\alpha$, any packet $i$ with inter-generation time $X_i=\alpha$ is always transmitted. Therefore, either $m_i=0$ and $n_i=\alpha/\epsilon$, or $m_i=1$ and $n_i<\alpha/\epsilon$. 
	
	However, since $\alpha/\epsilon\to\infty$ (because $\epsilon\to 0^+$), and the inter-generation time of packets take values $\epsilon$ or $\alpha$, each with probability 0.5, the probability that $m_i=0$ and $n_i=\alpha/\epsilon\to\infty$ is 0. Hence, with probability 1, $m_i=1$, and $n_i$ is finite. Also, $n_i<\infty$ implies that as $\epsilon\to 0^+$, $n_i\epsilon\to 0$ as well. 
	Therefore, the period lengths $T_i^{tp}=m_i\alpha+n_i\epsilon\approx \alpha=2\mu$ (a constant). 
	Thus, substituting $N=1$, $c=0$, $T_i^{tp}= 2\mu$ and $Z_i^{tp}=0$ (for $\pi_{tp}$ waiting time is 0, and transmission time $d_i=0$ is given),  in \eqref{eq:actual-tac-with-t} an equivalent expression to \eqref{eq:tac} for the weighted sum-cost, $\forall i\in\bbN$, we get $\Gamma(\pi_{tp})\to \rho\mu$.
	
	Next, consider the performance of the stationary randomized policy $\pi_{sr}$ (Algorithm \ref{algo:distribution-independent-policy}) for the same input. 
	Since $\gamma=0$, $c=0$ and $N=1$, from the convex optimization problem \eqref{eq:tx_prob}, it is immediate that $p=1$ (i.e., $\pi_{sr}$ marks every generated packet). Also, since the transmission time $d_i=0$ for every packet, the channel is always free, and every marked packet gets immediately transmitted. Therefore, $T_i^{sr}=X_i$ and $Z_i^{sr}=0$.
	Thus, substituting $N=1$, $c=0$, $T_i^{sr}=X_i$ and $Z_i^{sr}=0$ ($\forall i\in\bbN$) in \eqref{eq:actual-tac-with-t},\footnote{Since $\bbE[X_i]=\mu<\infty$, $T_{i}^{sr}=X_i$ is finite with probability 1. Hence, for the setup in the considered example, $\pi_{sr}\in\Pi_S$.} and using the renewal reward theorem \cite{ross2014introduction}, we get $\Gamma(\pi_{sr})=\rho\bbE[X^2]/(2\bbE[X])=(\rho/2)(\sigma^2+\mu^2)/\mu$. Hence,
	\begin{align}
		\frac{\Gamma(\pi_{sr})}{\Gamma(\pi_{tp})}=\frac{1}{2}\left(\frac{\sigma^2}{\mu^2}+1\right).
	\end{align} 
	Since $\pi_{OF}^\star$ is at least as good as $\pi_{tp}$, we have the result that the 
	competitive ratio of $\pi_{sr}$ is proportional to $\sigma^2/\mu^2$. 
\end{example}

	Although $\textsc{cr}_{\pi_{sr}}$ \eqref{eq:CR-for-SR} depends on $\sigma_\ell^2/\mu_\ell^2$, for several common distributions, it is upper bounded by a constant. Some examples are as follows. 
	\begin{paragraph}
		{Exponential Distribution} For exponential distribution, the ratio $\sigma_\ell^2/\mu_\ell^2=1$. Therefore, if packet inter-generation time at all the sources is exponentially distributed (packet generation rates may be different for each source), then $\textsc{cr}_{\pi_{sr}}\le 4$.
	\end{paragraph}
	\begin{paragraph}
		{Uniform distribution} Let the support of the uniform distribution for source $\ell$ be over interval $[a_\ell,b_\ell]$ ($0\le a_\ell\le b_\ell$). Then, $\mu_\ell=(b_\ell+a_\ell)/2$, and $\sigma_\ell^2=(b_\ell-a_\ell)^2/12$. Therefore, $\sigma_\ell^2/\mu_\ell^2\le 1/3<1$. Hence, $\textsc{cr}_{\pi_{sr}}\le 4$.
	\end{paragraph}
	\begin{paragraph}
		{Rayleigh Distribution} Let the scale parameter of Rayleigh distribution for source $\ell$ be $\nu_\ell$. Then $\mu_\ell=\nu_\ell\sqrt{\pi/2}$, and $\sigma_\ell^2=\nu_\ell^2(4-\pi)/2$. Therefore, $\sigma_\ell^2/\mu_\ell^2=(4/\pi)-1<1$. Hence, $\textsc{cr}_{\pi_{sr}}\le 4$.
	\end{paragraph}


\vspace{0.1in}

Next, we present the proof of Theorem \ref{thm:dist-ind-algo-bound-cr} in two steps. First, we derive a lower bound on the weighted sum cost for an optimal offline policy $\pi^\star_{OF}$ (Definition \ref{def:OPT}), as follows. 
\begin{lemma} \label{lemma:Gopt-lb-gen-dist}
	Let $h_\ell(t)$ denote the number of packets generated at source $\ell$ until time $t$, and an optimal offline policy $\pi_{OF}^\star$ transmits $R_\ell^\star(t)$ number of these packets. 
	Then, the expected weighted sum cost for policy $\pi_{OF}^\star$ is 
	\begin{align} \label{eq:final-lb-Goff}
		\bbE[\Gamma(\pi_{OF}^\star)]&\ge 
		\frac{1}{N}\sum_{\ell=1}^{N}\left(\frac{\rho_{\ell}\mu_{\ell}}{2f_\ell^\star}+\rho_\ell \gamma_\ell+\frac{c_\ell f_\ell^\star}{\mu_\ell}\right),
	\end{align}
	where $f_\ell^\star=\lim_{t\to\infty}R_\ell^\star(t)/h_\ell(t)\in[0,1]$. Further, $\sum_{\ell=1}^{N}\gamma_\ell f_\ell^\star/\mu_\ell\le 1$.
\end{lemma}

\begin{remark}\label{rem:preempt} We briefly discuss why we limit ourselves to non-preemptive scheduling policies for a G/G/1 system. One unique aspect of the considered scheduling problem is that all packets need not be 
transmitted to the monitor. This aspect can be exploited by a preemptive algorithm as follows. Assume that the rate of packet arrivals is very high, and a new packet is available very often. 
In this case, an algorithm can potentially choose to preempt and discard the packet under transmission as soon as the delay experienced by it exceeds a threshold and move on to transmit the newly arrived packet. This essentially skews the distribution of the delay seen by the packet that is completely transmitted and received by the monitor compared to $\cD_\ell$, and is the primary reason why our lower bound (Lemma \ref{lemma:Gopt-lb-gen-dist}) on the cost of an offline optimal algorithm does not extend to preemptive policies. 
\end{remark}

Next, we compute an upper bound on the expected weighted sum cost of policy $\pi_{sr}$ (Algorithm \ref{algo:distribution-independent-policy}), described as follows. 
\begin{lemma} \label{lemma:gamma-pisr-ub-gen-dist}
	The expected weighted sum cost for policy $\pi_{sr}$ (Algorithm \ref{algo:distribution-independent-policy}) is 
	\begin{align} \label{eq:dummy1}
		\bbE[\Gamma(\pi_{sr})]\le \frac{1}{N}\sum_{\ell=1}^{N}\left(\left(\frac{2\rho_{\ell}\mu_{\ell}}{p_\ell}+\frac{c_\ell p_\ell}{\mu_\ell}\right)+\left( \rho_\ell\gamma_\ell-\frac{\rho_\ell\mu_\ell \theta_\ell}{2}\right)\right), 
	\end{align}
where $\theta_\ell=1-\sigma_\ell^2/\mu_\ell^2$ ($\sigma_\ell^2$ and $\mu_\ell$ respectively denotes the variance and the mean of packet inter-generation times at source $\ell$), and $p_\ell$ is as defined in for $\pi_{sr}$. 
\end{lemma}
\begin{IEEEproof}[Proof of Theorem \ref{thm:dist-ind-algo-bound-cr}]
Using Lemma \ref{lemma:Gopt-lb-gen-dist} and \ref{lemma:gamma-pisr-ub-gen-dist}, we complete the proof of Theorem \ref{thm:dist-ind-algo-bound-cr} as follows.
Recall that $f_\ell^\star$'s (defined in Lemma \ref{lemma:Gopt-lb-gen-dist}) satisfy the constraints \eqref{eq:interference-constraint} and \eqref{eq:probability-constraint}. Also, under the same constraints, $p_\ell's$ minimize \eqref{eq:tx_prob}. Therefore,
\begin{align} \label{eq:dummy2}
	\sum_{\ell=1}^{N}\bigg(\frac{2\rho_{\ell}\mu_{\ell}}{p_\ell}+\frac{c_\ell p_\ell}{\mu_\ell}\bigg) 
	\le \sum_{\ell=1}^{N}\bigg(\frac{2\rho_{\ell}\mu_{\ell}}{ f_\ell^\star}+\frac{c_\ell f_\ell^\star}{\mu_\ell}\bigg).
\end{align}
From \eqref{eq:dummy1} and \eqref{eq:dummy2}, we get 
\begin{align}	\label{eq:gamma-pisr-4}
	\bbE[\Gamma(\pi_{sr})]\le \frac{1}{N}\sum_{\ell=1}^{N}\left(\left(\frac{2\rho_{\ell}\mu_{\ell}}{f_\ell^\star}+\frac{c_\ell f_\ell^\star}{\mu_\ell}\right)+\left(\rho_\ell\gamma_\ell-\frac{\rho_\ell\mu_\ell \theta_\ell}{2}\right)\right).
\end{align}

Since competitive ratio (Definition \ref{def:CR-def}) of $\pi_{sr}$ is $\textsc{cr}_{\pi_{sr}}=\bbE[\Gamma(\pi_{sr})]/\bbE[\Gamma(\pi_{OF}^\star)]$, using Lemma \ref{lemma:Gopt-lb-gen-dist} and Lemma \ref{lemma:gamma-pisr-ub-gen-dist}, we get
\begin{align} 
	\textsc{cr}_{\pi_{sr}}&\le \frac{\frac{1}{N}\sum_{\ell=1}^{N}\left(\frac{\rho_{\ell}\mu_{\ell}}{2f_\ell^\star}\left(4-f_{\ell}^\star\theta_\ell\right)+\rho_\ell \gamma_\ell+\frac{c_\ell f_\ell^\star}{\mu_\ell}\right)}{\frac{1}{N}\sum_{\ell=1}^{N}\left(\frac{\rho_{\ell}\mu_{\ell}}{2f_\ell^\star}+\rho_\ell \gamma_\ell+\frac{c_\ell f_\ell^\star}{\mu_\ell}\right)}, \nonumber \\
	\label{eq:prefinal-cr}
	&\le \max_{\ell}\{4-f_\ell^\star\theta_\ell\}, \\ 
	\label{eq:final-cr}
	&\le \max\left\{4,\ \ 3+\max_{\ell}\{\sigma_\ell^2/\mu_\ell^2\}\right\},
\end{align}
where we get \eqref{eq:final-cr} by substituting $\theta_\ell=1-\sigma_\ell^2/\mu_\ell^2$, and maximizing the R.H.S. of \eqref{eq:prefinal-cr} with respect to $f_\ell^\star\in[0,1]$.
\end{IEEEproof}

\begin{remark} From the proof of Lemma \ref{lemma:Gopt-lb-gen-dist}, it is easy to observe that it holds even for preemptive scheduling policies for which the distribution of the delay seen by the update that is completely transmitted and received by the monitor is the same as the original distribution $\cD_\ell$, e.g. a policy that always preempts if a new update arrives \cite{multisource}, or a greedy policy that chooses to 
preempt an update under transmission if a new update arrives only when the relative cost of transmitting the new update is lower. Since the proposed policy $\pi_{sr}$ is non-preemptive, consequently, the competitive ratio result for the proposed algorithm (Theorem \ref{thm:dist-ind-algo-bound-cr}) also holds against a class of preemptive offline scheduling policies. 
\end{remark}

Proofs of Lemma \ref{lemma:Gopt-lb-gen-dist} and \ref{lemma:gamma-pisr-ub-gen-dist} are provided in Appendix \ref{App:LowerBoundOPT} and \ref{App:SRP-tac-UB}, respectively. Next, we consider the preemptive setting for a G/M/1 system.

\section{Preemptive Setting} \label{sec:preemption} 

In the previous section, we restricted our attention to non-preemptive policies that were allowed to discard packets before making any attempt to transmit them, however, once their transmission began, no preemption was allowed. 
We remove this restriction  in this section, and consider preemptive policies for a G/M/1 system, that can discard any packet or interrupt an ongoing transmission at any time. 
Also, to keep the model general, we allow preemptive policies to retransmit a preempted packet at a later time, either from start (with fresh realization of transmission time), or resume it from the state in which the packet was previously preempted. 
\begin{definition} \label{def:fresh-tx}
	Whenever a source begins to transmit a packet (either a new packet, or a previously preempted packet) from start, i.e., with a fresh realization of transmission time, we call it a fresh transmission. 
\end{definition}
With preemption, counting the transmission cost is bit more tricky. We use the following model. A preemptive policy incurs a cost of $c_\ell$ units for every fresh transmission (Definition \ref{def:fresh-tx}) for source $\ell$. 
There is no cost for resuming the transmission of a preempted packet, from the state in which it was previously preempted. 

As considered earlier, the objective is to minimize a linear combination of the AoI and the average transmission cost of the sources, called the weighted sum cost \eqref{eq:tac}, where the minimization is over the set of causal preemptive policies. Moreover, the competitive ratio definition remains 
similar to \eqref{def:CR-def}, where now both the causal policy and the optimal offline policy is allowed to be preemptive. 


For analyzing preemptive policies, we need the following additional notation.
\begin{enumerate}
\item $\tilde{R}_\ell^\pi(t)$ denotes the number of fresh transmissions (Definition \ref{def:fresh-tx}) by source $\ell$ until time $t$, under policy $\pi$. Thus, the average transmission cost $C_{\ell,\pi}^{av}=c_\ell\tilde{R}_\ell^\pi(t)/t$. Note that $\tilde{R}_\ell^\pi(t)$ is the sum of the number of partially transmitted packets, as well as the number of completely transmitted packets $R_\ell^\pi(t)$.
	Hence, $\tilde{R}_\ell^\pi(t)\ge R_\ell^\pi(t)$.
    \item channel time $\tilde{d}_{\ell i}^\pi$ (for packet $\ell_i^\pi$) denotes the total time in interval $(r_{\ell (i-1)}^\pi,r_{\ell i}^\pi]$ for which source $\ell$ (under policy $\pi$) transmits (keeps the channel busy). As shown in Figure \ref{fig:preemptive}, in interval $(r_{\ell (i-1)}^\pi,r_{\ell i}^\pi]$, source $\ell$ transmits exactly one packet completely (i.e. packet $\ell_i^\pi$), and may transmit multiple other packets partially (before completing the transmission of packet $\ell_i^\pi$). Thus, $\tilde{d}_{\ell i}^\pi$ is equal to the sum of $d_{\ell i}^\pi$, and the time for which source $\ell$ transmits packets in interval $(r_{\ell (i-1)}^\pi,r_{\ell i}^\pi]$ that got preempted before $r_{\ell i}^\pi$. For non-preemptive policies $\pi$, $\tilde{d}_{\ell i}^\pi=d_{\ell i}^\pi$ (as in Figure \ref{fig:nonpreemptive}). 
    Note that the transmission decision of different sources are related by their required channel time to completely transmit a packet.  	
\end{enumerate}

Compared to definitions made for the non-preemptive case, their meanings in the preemptive case are contrasted in Figure  \ref{fig:combined}. In particular, they are as follows.

\begin{enumerate}
	\item Among the sequence of packets generated at source $\ell$ (in increasing order of generation times), let $\ell_i$ and $\ell_i^\pi$, respectively, denote the $i^{th}$ packet generated at source $\ell$, and the $i^{th}$ packet that gets completely transmitted (with or without preemptions) under policy $\pi$. 
	\item $g_{\ell i}^\pi$ and $r_{\ell i}^\pi$, respectively, denotes the generation time and transmission completion time of packet $\ell_i^\pi$.
	\item The interval $\cP_{\ell i}^\pi=(g_{\ell (i-1)}^\pi,g_{\ell i}^\pi]$ is called the $i^{th}$ period of source $\ell$ under policy $\pi$. Also, $T_{\ell i}^\pi=g_{\ell i}^\pi-g_{\ell (i-1)}^\pi$ is called the length of period $\cP_{\ell i}^\pi$. 
	\item $Z_{\ell i}^\pi=r_{\ell i}^\pi-g_{\ell i}^\pi$ denotes the system time of packet $\ell_i^\pi$.
	\item transmission time $d_{\ell i}^\pi$ is the total time for which policy $\pi$ transmits packet $\ell_i^\pi$ in interval $(g_{\ell i}^\pi,r_{\ell i}^\pi]$. Note that for preemptive policies, transmission time of a packet may depend on the policy $\pi$ (unlike non-preemptive policies). Also, $d_{\ell i}^\pi$ may be split over multiple non-contiguous intervals (as shown in Figure \ref{fig:preemptive}).
	\item waiting time $w_{\ell i}^\pi=Z_{\ell i}^\pi-d_{\ell i}^\pi$ is the total time in interval $(g_{\ell i}^\pi,r_{\ell i}^\pi]$ for which packet $\ell_i^\pi$ is not under transmission.
	\item $R_\ell^\pi(t)$ denotes the number of packets of source $\ell$ that are completely transmitted by policy $\pi$ until time $t$.
\end{enumerate}
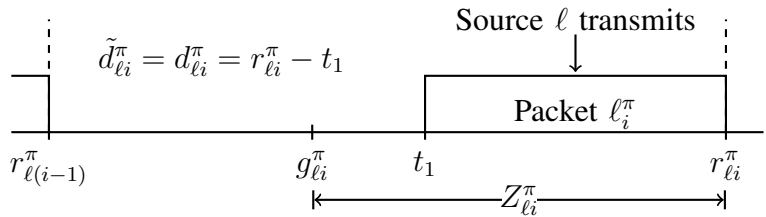
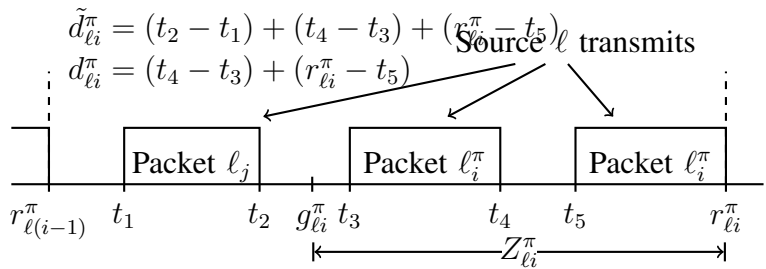
\begin{figure} 
	\begin{subfigure} {\textwidth}
		\begin{center}
			\begin{tikzpicture}[thick,scale=1, every node/.style={scale=1}]
\draw[thick] (0,0) to (10,0); 

\draw (4,-0.1) node[below]{$g_{\ell i}^\pi$} to (4,0.1); 

\draw (0,0.75) to (0.5,0.75) to (0.5,-0.1) node[below]{$r_{\ell (i-1)}^\pi$};
\draw (5.5,-0.1) node[below]{$t_1$} to (5.5,0.75) to (9.5,0.75) to (9.5,-0.1) node[below]{$r_{\ell i}^\pi$}; 
\draw (7.5,0.6) node[below]{Packet $\ell_i^\pi$};

\draw[dashed] (0.5,0) to (0.5,1.5);
\draw[dashed] (9.5,0) to (9.5,1.5);

\draw (7.5,1.5) node{Source $\ell$ transmits};
\draw[->] (7.5,1.3) to (7.5,0.8);


\draw (1,1) node[right]{$\tilde{d}_{\ell i}^\pi=d_{\ell i}^\pi=r_{\ell i}^\pi-t_1$};

\draw[|<->|] (4,-0.9) -- (9.5,-0.9) node[rectangle,inner sep=-1pt,midway,fill=white]{$Z_{\ell i}^{\pi}$};
\end{tikzpicture}
\caption{Illustration of transmission time $d_{\ell i}^\pi$ and channel time $\tilde{d}_{\ell i}^\pi$ for a non-preemptive policy.}   
\label{fig:nonpreemptive}  
		\end{center} 
	\end{subfigure}
\newline
	\begin{subfigure}{\textwidth}
	\begin{center}
		\begin{tikzpicture}[thick,scale=1, every node/.style={scale=1}]
\draw[thick] (0,0) to (10,0); 

\draw (4,-0.1) node[below]{$g_{\ell i}^\pi$} to (4,0.1); 

\draw (0,0.75) to (0.5,0.75) to (0.5,-0.1) node[below]{$r_{\ell (i-1)}^\pi$};
\draw (1.5,-0.1) node[below]{$t_1$} to (1.5,0.75) to (3.3,0.75) to (3.3,-0.1) node[below]{$t_2$}; 
\draw (2.4,0.6) node[below]{Packet $\ell_j$};
\draw (4.5,-0.1) node[below]{$t_3$} to (4.5,0.75) to (6.5,0.75) to (6.5,-0.1) node[below]{$t_4$}; 
\draw (5.5,0.6) node[below]{Packet $\ell_i^\pi$};
\draw (7.5,-0.1) node[below]{$t_5$} to (7.5,0.75) to (9.5,0.75) to (9.5,-0.1) node[below]{$r_{\ell i}^\pi$}; 
\draw (8.5,0.6) node[below]{Packet $\ell_i^\pi$};

\draw[dashed] (0.5,0) to (0.5,1.5);
\draw[dashed] (9.5,0) to (9.5,1.5);

\draw (7.5,1.9) node{Source $\ell$ transmits};
\draw[->] (7.4,1.6) to (8,0.9);
\draw[->] (7.1,1.6) to (5.8,0.9);
\draw[->] (6.7,1.6) to (3.35,0.9); 


\draw (0.6,1.5) node[right]{$d_{\ell i}^\pi=(t_4-t_3)+(r_{\ell i}^\pi-t_5)$};
\draw (0.6,2.1) node[right]{$\tilde{d}_{\ell i}^\pi=(t_2-t_1)+(t_4-t_3)+(r_{\ell i}^\pi-t_5)$};

\draw[|<->|] (4,-0.9) -- (9.5,-0.9) node[rectangle,inner sep=-1pt,midway,fill=white]{$Z_{\ell i}^{\pi}$};
\end{tikzpicture}
\caption{Illustration of transmission time $d_{\ell i}^\pi$ and channel time $\tilde{d}_{\ell i}^\pi$ for a preemptive policy.}   
\label{fig:preemptive}  
		\end{center}
	\end{subfigure}
\caption{Non-preemptive versus Preemptive policy.} 
\label{fig:combined}
\end{figure}

Before we discuss the results for the considered preemptive setting, it is critical to understand the challenges in analyzing preemptive policies, and why we restrict our attention to a G/M/1 system. 
Note that the lower bound \eqref{eq:final-lb-Goff} on the weighted sum cost of an optimal offline non-preemptive policy $\pi_{OF}^\star$ depends on $f_\ell^\star$,  that  is the ratio of the number of packets transmitted by source $\ell$ under $\pi_{OF}^\star$, and the total number of packets generated at source $\ell$. More importantly, with expected service time (equal to the channel time in the non-preemptive case) for each packet of source $\ell$ being $\gamma_\ell$, $f_\ell^\star$'s satisfy the constraint $\sum_{\ell=1}^N\bbE[\tilde{d}_\ell] f_\ell^\star/\mu_\ell\le 1$. This constraint captures the tradeoff 
between different sources as to how many packets can be transmitted by all sources in a given time. 

 
In contrast, for an optimal offline preemptive policy it is difficult to lower bound the expected channel time as a function of $\gamma_\ell$, as described in the next example, and hence an inequality of the type $\sum_{\ell=1}^N\bbE[\tilde{d}_\ell] f_\ell^\star/\mu_\ell\le 1$ is not true in general for the preemptive policies.

\begin{example} \label{ex:preempt-minimize-gamma}
	Consider a single source system ($N=1$), where $\cG$ is some general distribution with finite mean $\mu$, and transmission time $d\sim\cD$ is such that $d=\epsilon$ with probability 0.5, and $\alpha$ otherwise (where $\epsilon\to 0^+$, and $\alpha$ is a large positive constant). Thus, $\gamma=\bbE[d]\approx \alpha/2$.  
	Now, consider a preemptive policy $\pi_\epsilon$ 
	that at any time $t$, if the channel is free and a fresh packet is available, then transmits the packet for at most $\epsilon$ time units. If the transmission of a packet takes more than $\epsilon$ time units to complete, then $\pi_\epsilon$ preempts the packet, 
	and retransmits it (or a newly generated packet if available) with a fresh realization of transmission time $d$. 
	For $\pi_\epsilon$, each fresh transmission completes in $\epsilon$ time units with probability $0.5$. Hence, the expected channel time  for completely transmitted packets under $\pi_\epsilon$ is $2\epsilon\to 0^+$, which can be arbitrarily small compared to $\gamma\approx \alpha/2$. 
\end{example}

Example \ref{ex:preempt-minimize-gamma} is the main reason why we restrict ourselves to a G/M/1 system, where the inter-generation time of packets for source $\ell$ is distributed as $\cG_\ell$, while the service time of any packet of each source is exponentially distributed $\sim \cD_\ell$ with mean $\gamma_\ell$. Next, we show that in a G/M/1 system, the channel time distribution of any preemptive policy for any source $\ell$ is the same as $\cD_\ell$.

%

\begin{lemma} \label{lemma:min-consolidated-time}
   For a G/M/1 system, where the service time $\cD_\ell$ is exponentially distributed with mean $\gamma_\ell<\infty$,  for any preemptive policy $\pi$, the channel times $\tilde{d}_{\ell i}^\pi$ ($\forall i$) are independent and identically distributed with distribution $\cD_\ell$. Thus, from the strong law of large numbers,
	\begin{align} \label{eq:SC-no-preempt}
		\lim_{t\to\infty}\frac{\sum_{i=1}^{R_\ell^\pi(t)}\tilde{d}_{\ell i}^\pi}{R_\ell^\pi(t)}=\bbE[\tilde{d}_{\ell i}^\pi]=\gamma_\ell,
	\end{align} 
with probability 1. 
\end{lemma}
\begin{IEEEproof} 
	See Appendix \ref{appendix:proof-lemma-min-consolidated-time}.
\end{IEEEproof} 

Lemma \ref{lemma:min-consolidated-time} is derived by proving a basic fact about exponential distribution that states that no matter how often you preempt service for a source $\ell$ in hope of finding a smaller service time,  the total time in service is distributed as the original service time distribution $\cD_\ell$.


\begin{remark}
	It is worth noting that Lemma \ref{lemma:min-consolidated-time} \emph{does not} imply that the optimal offline policy is non-preemptive for exponentially distributed $\cD_\ell$'s. In fact, the AoI can be reduced by preempting old packets in service with newly arrived ones, 
	but the total time spent in completely transmitting a single packet has the same distribution as $\cD_\ell$.
\end{remark}


With Lemma \ref{lemma:min-consolidated-time} under our belt, we derive a lower bound on the expected weighted sum cost of an optimal offline preemptive policy.

	\begin{lemma} \label{lemma:lb-OPT-preempt}
	For a G/M/1 system, the expected weighted sum cost for an optimal offline preemptive policy $\tilde{\pi}_{OF}^{\star}$ is 
	\begin{align} \label{eq:lb-OPT-preempt}
		\bbE[\Gamma(\tilde{\pi}_{OF}^{\star})]&\ge 
		\frac{1}{N}\sum_{\ell=1}^{N}\left(\frac{\rho_{\ell}\mu_{\ell}}{2\sfff_\ell^\star}+\frac{c_\ell \sfff_\ell^\star}{\mu_\ell}\right),
	\end{align}
	for some $\sfff_\ell^\star\in[0,1]$ that satisfies $\sum_{\ell=1}^{N}\gamma_\ell \sfff_\ell^\star/\mu_\ell\le 1$.
\end{lemma}
\begin{IEEEproof}
	See Appendix \ref{appendix-lemma-lb-OPT-preempt}.
\end{IEEEproof}

Note that unlike the non-preemptive case \eqref{eq:final-lb-Goff} (lower bound on the weighted sum-cost $\bbE[\Gamma(\pi_{OF}^\star)]$), \eqref{eq:lb-OPT-preempt} does not have the term $\rho_\ell\gamma_\ell$, where $\gamma_\ell$ is equal to the average transmission time of completely transmitted packets. This is because for an optimal offline preemptive policy $\tilde{\pi}_{OF}^\star$, the average transmission time of completely transmitted packets can be arbitrarily close to $0$ (even for a G/M/1 system, which only ensures that the expected channel time for completely transmitted packets of each source $\ell$ is lower bounded by $\gamma_\ell$), as shown in Example \ref{ex:small-tx-time}. 
\begin{example} \label{ex:small-tx-time}

	Consider a single source system with general packet inter-generation time distribution $\cG$ (with mean $\mu<\infty$), and  exponential transmission time distribution $\cD$ (with mean $\gamma<\infty$).  
	Now, consider a preemptive policy $\pi_\epsilon'$, that at any time $t$, if the channel is free, and a fresh packet is available which has never been transmitted in the past (at all), then transmits the packet for at most $\epsilon$ time units. If the packet does not get completely transmitted in $\epsilon$ time units, then $\pi_\epsilon'$ discards the packet (never transmits it again). 
The average transmission time of completely transmitted with $\pi_\epsilon'$ is $\epsilon$ and choosing $\epsilon\to 0^+$, we get that  the average transmission time of completely transmitted packets can be arbitrarily close to $0$ even in a G/M/1 system.

\end{example}

The lower bound \eqref{eq:lb-OPT-preempt} is still `similar' to the lower bound \eqref{eq:final-lb-Goff} for non-preemptive policies. Thus, intuitively a non-preemptive policy similar to Algorithm \ref{algo:distribution-independent-policy} may still have a bounded competitive ratio. In next subsection, we consider a non-preemptive stationary randomized policy $\tilde{\pi}$, and upper-bound its competitive ratio. 

\subsection{Stationary Randomized Policy $\tilde{\pi}_{sr}$} 
Consider a variant of Algorithm \ref{algo:distribution-independent-policy}, denoted by $\tilde{\pi}_{sr}$, which is identical to Algorithm \ref{algo:distribution-independent-policy}, except that the packet marking probability $p_\ell$'s are obtained by solving the following convex optimization problem (instead of \eqref{eq:tx_prob}---\eqref{eq:probability-constraint}):
\begin{align} \label{eq:tx_prob-preempt}
	&\underset{[p_1,...,p_N]}{\arg\min}\sum_{\ell=1}^{N}\left(\frac{3\rho_{\ell}\mu_{\ell}}{p_\ell}+\frac{c_\ell p_\ell}{\mu_\ell}\right), \\
	\label{eq:interference-constraint-preempt}
	& s.t. \ \ \sum_{\ell=1}^{N}\frac{p_\ell \gamma_\ell}{\mu_\ell}\le 1, \\
	\label{eq:probability-constraint-preempt}
	& \hspace{5ex} p_\ell\in [0,1], \ \ \forall \ell\in\{1,\cdots,N\}. 
\end{align} 
\begin{remark}
	Note that the optimization problem \eqref{eq:tx_prob-preempt}---\eqref{eq:probability-constraint-preempt} is identical to the optimization problem \eqref{eq:tx_prob}---\eqref{eq:probability-constraint}, except that in \eqref{eq:tx_prob}, the term $\rho_\ell\mu_\ell/p_\ell$ has coefficient $2$, whereas in \eqref{eq:tx_prob-preempt}, the same term has coefficient $3$. The modification is needed for the proof strategy for Theorem \ref{thm:dist-ind-algo-bound-cr} to work with 
	the lower bound on $\bbE[\Gamma(\tilde{\pi}_{OF}^{\star})]$ in Lemma \ref{lemma:lb-OPT-preempt}, 
	which is different from the lower bound \eqref{eq:final-lb-Goff} for non-preemptive policies. 
\end{remark}

For $\tilde{\pi}_{sr}$, the expected weighted sum cost is upper-bounded as follows.
\begin{lemma} \label{lemma:ub-sr-preempt}
	The expected weighted sum cost for policy $\tilde{\pi}_{sr}$ is 
	\begin{align} \label{eq:ub-sr-preempt}
		\bbE[\Gamma(\tilde{\pi}_{sr})]\le \frac{1}{N}\sum_{\ell=1}^{N}\left(\frac{3\rho_{\ell}\mu_{\ell}}{p_\ell}+\frac{c_\ell p_\ell}{\mu_\ell}-\frac{\rho_\ell\mu_\ell \theta_\ell}{2}\right),
	\end{align}
	where $\theta_\ell=1-\sigma_\ell^2/\mu_\ell^2$ (recall that $\sigma_\ell^2$ and $\mu_\ell$ are respectively the variance and the mean of packet inter-generation times at source $\ell$).
\end{lemma} 
\begin{IEEEproof}
	See Appendix \ref{appendix:lemma-ub-sr-preempt}.
\end{IEEEproof}

 The  second main result of this paper is as follows.
From Lemma \ref{lemma:lb-OPT-preempt} and \ref{lemma:ub-sr-preempt}, we get the following upper-bound on the competitive ratio of $\tilde{\pi}_{sr}$.
\begin{theorem} \label{thm:CR-sr-preempt}
    The stationary randomized policy non-preemptive policy $\tilde{\pi}_{sr}$ has competitive ratio (against an optimal offline preemptive policy)
	\begin{align} \label{eq:CR-sr-preempt}
		\textsc{cr}_{\tilde{\pi}_{sr}}\le\max\{6,\ \ 5+\max_{\ell}\{\sigma_\ell^2/\mu_\ell^2\}\}.
	\end{align}
\end{theorem}


\begin{IEEEproof} 
Computing the ratio of \eqref{eq:ub-sr-preempt} to \eqref{eq:lb-OPT-preempt}, and following the steps in the proof of Theorem \ref{thm:dist-ind-algo-bound-cr}, we get \eqref{eq:CR-sr-preempt}. 
\end{IEEEproof}

Note that the competitive ratio bound \eqref{eq:CR-sr-preempt} for $\tilde{\pi}_{sr}$ exceeds the competitive ratio bound \eqref{eq:CR-for-SR} for $\pi_{sr}$ by only an additive constant $2$. Thus, Theorem \ref{thm:CR-sr-preempt} shows that when distribution $\cD_\ell$'s are exponential, 
then preemption (which may still be used to prioritize newly generated packets, or packets of other source) has limited role in minimizing the weighted sum cost of an optimal offline preemptive policy $\tilde{\pi}_{OF}^\star$. 


\section{Numerical Results} \label{sec:NumericalResults}
In this section, we perform parametric and comparative analysis of the stationary randomized policy $\pi_{sr}$ (Algorithm \ref{algo:distribution-independent-policy}) using numerical simulations. In particular, in Subsection \ref{sec:numerical-parametric}, we simulate $\pi_{sr}$ for different values of the system parameters (such as the number of sources $N$, transmission cost $c_\ell$, and distributions $\cG_\ell$ and $\cD_\ell$), and analyze its corresponding weighted sum cost $\Gamma(\pi_{sr})$. Then, in Subsection \ref{sec:numerical-comparative}, we consider relevant settings for AoI minimization problem from prior work, and compare the performance of $\pi_{sr}$ 
with other state-of-the-art policies.

\subsection{Parametric Analysis} \label{sec:numerical-parametric}

\begin{figure}
	\begin{center}
		\begin{tikzpicture}
			\begin{axis}[xlabel=Number of Nodes ($N$),
				ylabel=Weighted Sum Cost $\Gamma(\pi)$,
				legend cell align=left,
				legend pos=north west,
				grid=major, ymax=15,
				xmin=1,xmax=10,xtick distance=2,
				x tick label style={/pgf/number format/fixed}] 
				\addplot[style={dashed},line width=1pt] table {./NumericalPlots/vsN/NvsUB.dat};
				\addplot[style={solid},line width=1pt] table {./NumericalPlots/vsN/NvsExp_SRP.dat};
				\addplot[style={dashdotted},line width=1pt] table {./NumericalPlots/vsN/NvsExp_MTP.dat};
				\addplot[style={dotted},line width=1pt] table {./NumericalPlots/vsN/NvsLB.dat};
				\legend{$UB_{sr}$, $\pi_{sr}$, $\pi_{th}$, $LB$}
			\end{axis}
		\end{tikzpicture}
		\caption{Effect of number of sources on the weighted sum cost for the system.}
		\label{fig:Cost_vs_N}
	\end{center}
\end{figure}
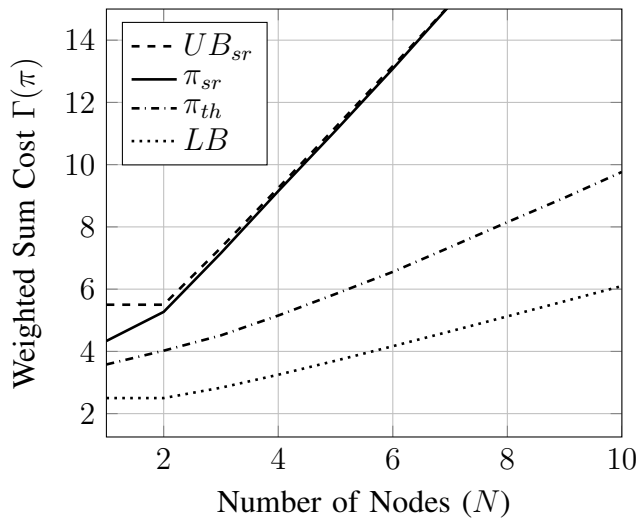

To understand the impact of number of sources on the sum weighted cost \eqref{eq:tac} under policy $\pi_{sr}$, we consider a system consisting of $N$ sources, where for each source $\ell$, $\rho_\ell=1$, $c_\ell=1$, and $\cG_\ell$ and $\cD_\ell$ are exponential distributions with mean $\mu_\ell=2$ and $\gamma_\ell=1$, respectively. 
Figure \ref{fig:Cost_vs_N} shows the plot for the obtained weighted sum cost under policy $\pi_{sr}$ (i.e. $\Gamma(\pi_{sr})$) for different values of $N$, along with the theoretical lower bound \eqref{eq:final-lb-Goff} for an offline optimal policy (henceforth, $LB$), and the theoretical upper bound \eqref{eq:dummy1} on  $\Gamma(\pi_{sr})$ (henceforth, $UB_{sr}$). 

Since $\mu_\ell=2$ and $\gamma_\ell=1$, on average, a packet is generated every two time units, while a transmission takes only one time unit to complete. Hence, when $N=1$, even if $\pi_{sr}$ marks every packet, on average, the channel remains free for half of the time (i.e., the channel is not fully utilized). Thus, addition of one extra source to the system does not increase $\Gamma(\pi_{sr})$ (and $LB$), as shown in Figure \ref{fig:Cost_vs_N}. However, as $N$ increases beyond 2, the overall time for which each source can access the channel, starts decreasing. This leads to an increase in $\Gamma(\pi_{sr})$. Further, in Figure \ref{fig:Cost_vs_N}, since $\Gamma(\pi_{sr})$ and $LB$ both increase linearly with $N$, their ratio is below the 
competitive ratio guarantee \eqref{eq:CR-for-SR}. 

In Figure \ref{fig:Cost_vs_N}, the reason for linear increase in $\Gamma(\pi_{sr})$ with $N\ge 3$ can be explained as follows. 
Note that in the expression for the weighted sum cost \eqref{eq:tac}, for any source $\ell$, the number of periods $R_\ell^{sr}(t)$ decreases linearly with increase in $N$. Whereas, the cost incurred in each period (proportional to $(T_{\ell i}^{sr})^2$) increases quadratically (because period length $T_{\ell i}^{sr}$'s increases linearly with increase in $N$). Hence, $\Gamma(\pi_{sr})$ increases linearly with $N$. The same can be argued for $LB$ as well.


For comparative study, we also consider a multi-source extension of the threshold policy proposed in \cite{saurav2021minimizing}, that is known to be optimal (causal policy) for a single source system with exponentially distributed packet inter-generation times, and zero transmission time. The multi-source adaptation of the threshold policy $\pi_{th}$ in \cite{saurav2021minimizing}, similar to $\pi_{sr}$, consists of two subroutines: $(i)$ TH-PMS, that at each source $\ell$, marks a generated packet if the time elapsed since the generation time of the previously marked packet is greater than the threshold $A^\star=\max\{\sqrt{\sigma_\ell^2+2c_\ell/\rho_\ell}-\mu_\ell,N\gamma\}$, and $(ii)$ TH-NSS, that at any time $t$, if the channel is free, transmits the latest fresh marked packet of the source that has not transmitted for the longest time (despite having a fresh marked packet). 
Figure \ref{fig:Cost_vs_N} shows that the weighted sum cost for $\pi_{th}$ is in fact smaller than that for $\pi_{sr}$, and is close to $LB$. However, the theoretical analysis of this policy is not available and is part of ongoing work. 
 
For the rest of the simulations, we keep $N$ fixed, and unless specified, we use the parameter values from \cite{kadota2019minimizing} as follows. 
We consider a system with $N=4$ sources, with weight $[\rho_1,\rho_2,\rho_3,\rho_4]=[4,4,1,1]$, mean packet inter-generation time $[\mu_1,\mu_2,\mu_3,\mu_4]=\mu\cdot[1,(4/3),2,4]$, mean transmission time for packets $[\gamma_1,\gamma_2,\gamma_3,\gamma_4]=\gamma\cdot[4,2,(4/3),1]$, and cost per transmission $[c_1,c_2,c_3,c_4]=c\cdot[2,1,1,2]$, where for different $\cG_\ell$'s and $\cD_\ell$'s, $\mu$, $\gamma$ and $c$ are parameters that we specify later for each simulation.


To analyze the effect of the ratio $\max_\ell\{\sigma_\ell^2/\mu_\ell^2\}$ on the competitive ratio for $\pi_{sr}$  \eqref{eq:CR-for-SR}, we fix $c=1$, and for each source $\ell$, we begin by choosing $\cG_\ell$ to be the log-normal distribution (with variance $\sigma_\ell^2=\sigma^2$, $\forall \ell$) and $\cD_\ell$ to be the exponential distribution ($\gamma=1$). Then, we simulate the system for different values of $\sigma^2$ and $\mu$. 
\begin{remark}
	Throughout this section, for $\cG_\ell$'s and $\cD_\ell$'s, we either consider the log-normal distribution, or the exponential distribution (as in prior work \cite{sun2017update,saurav2021minimizing,moltafet2019closed}). Since the log-normal distribution is defined using two parameters, its mean and variance can be varied independently (allowing us to vary the ratio $\sigma_\ell^2/\mu_\ell^2$). On the other hand, for the exponential distribution, the ratio of its variance to the square of its mean is constant (equal to 1). Therefore, it is useful for analyzing the effect of change in mean of the distributions $\cG_\ell$ and $\cD_\ell$, when the ratio of their variance to the square of their mean is to be kept constant.
\end{remark}

\begin{figure}
	\begin{center}
		\begin{tikzpicture}
			\begin{axis}[xlabel=Variance of Packet Inter-generation Time ($\sigma^2$),
				ylabel=Weighted Sum Cost $\Gamma(\pi)$,
				legend cell align=left,
				legend pos=north west,
				grid=major, ymax=80,
				xmin=0,xmax=10,xtick distance=2,
				x tick label style={/pgf/number format/fixed}] 
				\addplot[style={dashed},line width=1pt] table {./NumericalPlots/vsRatio/M2_ub.dat};
				\addplot[style={solid},line width=1pt] table {./NumericalPlots/vsRatio/M2_SRP.dat};
				\addplot[style={dashdotted},line width=1pt] table {./NumericalPlots/vsRatio/M3_SRP.dat};
				\addplot[style={dotted},line width=1pt] table {./NumericalPlots/vsRatio/M2_lb.dat};
				\legend{$UB_{sr}; \mu=1$, $\pi_{sr}; \mu=1$, $\pi_{sr}; \mu=2$, $LB; \mu=1$}
			\end{axis}
		\end{tikzpicture}
		\caption{Weighted sum cost as a function of the variance of inter-generation time of packets.}
		\label{fig:Cost_vs_Var}
	\end{center}
\end{figure}
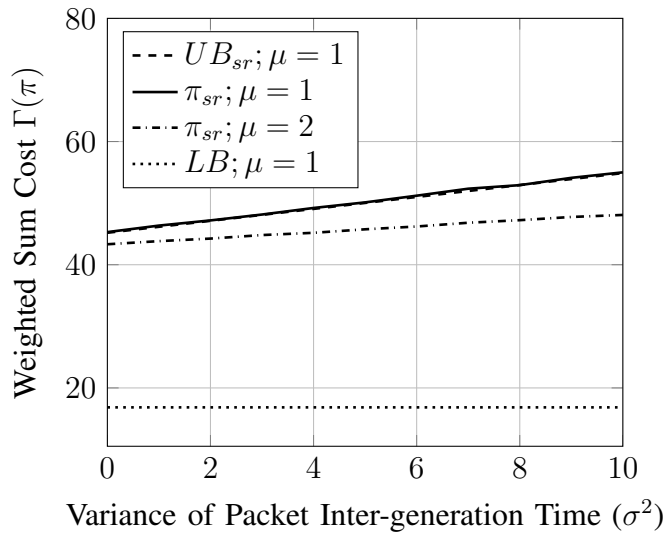

As shown in Figure \ref{fig:Cost_vs_Var}, as $\sigma^2$ increases (i.e., $\sigma^2/\mu^2=\max_\ell\{\sigma_\ell^2/\mu_\ell^2\}$ increases), $\Gamma(\pi_{sr})$ increases linearly (with slope that decreases with increase in $\mu$). Whereas, the lower bound $LB$ remains constant with increase in $\sigma^2$. Hence, the ratio of $\Gamma(\pi_{sr})$ and $LB$ increases linearly with increase in $\sigma^2/\mu^2$, confirming our theoretical result \eqref{eq:CR-for-SR} that the competitive ratio for $\pi_{sr}$ increases linearly with $\max_\ell\{\sigma_\ell^2/\mu_\ell^2\}$. 

Note that in Figure \ref{fig:Cost_vs_Var}, the plot corresponding to $\Gamma(\pi_{sr})$ when $\mu=2$, lies below the plot for $\Gamma(\pi_{sr})$ 
when $\mu=1$ (i.e., $\Gamma(\pi_{sr})$ is smaller when $\mu$ is large). 
Although it may appear counter-intuitive, this is completely justified because of the choice of marking probability $p_\ell$'s. 
Note that for each source $\ell$ in the summand in \eqref{eq:dummy1}, $\mu_\ell$ appears in exactly three terms. In the first two terms, $\mu_\ell$ appears together with $p_\ell$, which by its definition \eqref{eq:tx_prob}, partially compensates for the change in the value of $\mu_\ell$. However, the third term,  $-\rho_\ell\mu_\ell \theta_\ell/2\propto (-\mu_\ell+\sigma_\ell^2/\mu_\ell)$, and hence, for given $\sigma_\ell^2$, $\Gamma(\pi_{sr})$ decreases with increase in $\mu_\ell$.

\begin{figure}
	\begin{center}
		\begin{tikzpicture}
			\begin{axis}[
				xlabel=Mean Packet Inter-generation Time $(\mu)$, 
				ylabel=Weighted Sum Cost $\Gamma(\pi)$,
				legend cell align=left,
				legend pos=north west,
				grid=major, ymax=250,
				xmin=1,xmax=30,xtick distance=3,
				x tick label style={/pgf/number format/fixed}] 
				\addplot[style={dashed},line width=1pt] table {./NumericalPlots/vsArrival/data2_UB.dat};
				\addplot[style={solid},line width=1pt] table {./NumericalPlots/vsArrival/data2.dat};
				\addplot[style={dotted},mark=square*,mark options={style=solid,scale=0.5},line width=1pt] table {./NumericalPlots/vsArrival/data2_NW.dat};
				\addplot[style={dotted},line width=1pt] table {./NumericalPlots/vsArrival/data2_LB.dat};
				\legend{$UB_{sr}$, $\pi_{sr}$, $\pi_{sr}^{wc}$, $LB$}
			\end{axis}
		\end{tikzpicture}
		\caption{Effect of mean packet inter-generation time of sources on the weighted sum cost for the system.}
		\label{fig:Cost_vs_MeanArrival}
	\end{center}
\end{figure}
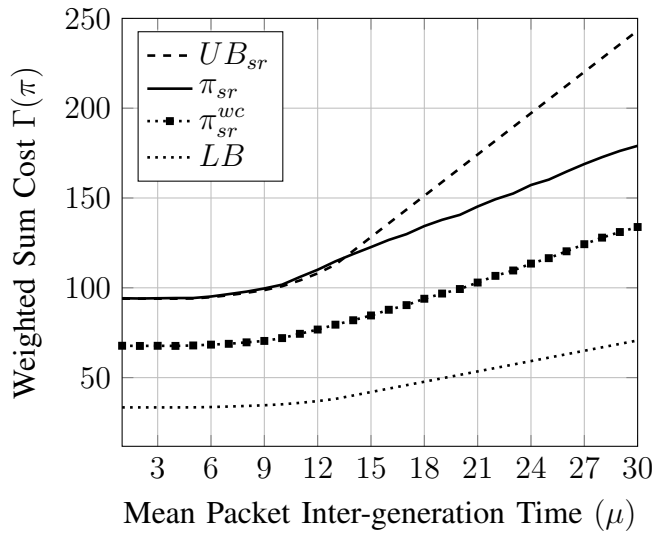

To analyze the effect of the mean packet inter-generation time on $\Gamma(\pi_{sr})$, we simulate the considered system in Figure \ref{fig:Cost_vs_MeanArrival}, by fixing $c=1$, $\gamma=2$, \footnote{$\gamma=2$ ensures that the effect of $\gamma_\ell$'s is visible in the plot of Figure \ref{fig:Cost_vs_MeanArrival}.} and varying $\mu$. Also, we assume $\cG_\ell$'s and $\cD_\ell$'s (for all $\ell$) to be the exponential distribution (because of which  $\sigma_\ell^2/\mu_\ell^2=1$, $\forall \ell$). 

Figure \ref{fig:Cost_vs_MeanArrival} shows that when $\theta_\ell=\sigma_\ell^2/\mu_\ell^2$ are fixed for each source $\ell$, 
$\Gamma(\pi_{sr})$ increases with increase in $\mu$ (i.e. $\mu_\ell$, $\forall \ell$). This is as expected because 
when $\mu_\ell$'s are large, sources need to wait longer for fresh packets to get generated, and hence, they cannot transmit at optimal time instants that would minimize $\Gamma(\pi_{sr})$, even if the channel is free. 
Further, in Figure \ref{fig:Cost_vs_MeanArrival}, note that initially when $\mu$ is small, the rate of increase in $\Gamma(\pi_{sr})$ with respect to $\mu$ is small (almost 0), compared to when $\mu$ is large. 
This is because when $\mu$ is small, the time instants when the sources get to transmit is mainly restricted by their transmission times, and hence, small change in $\mu$ has little effect on $\Gamma(\pi_{sr})$. 

Recall that whenever $\pi_{sr}$ chooses a source $\ell$ to transmit, if the source does not have a fresh marked packet to transmit, then $\pi_{sr}$ waits for random time $d_\ell\sim\cD_\ell$ before it again chooses a source to transmit. As discussed in Remark \ref{remark:sr-non-wc}, the random wait simplifies the theoretical analysis of $\pi_{sr}$. To understand the effect of waiting on the actual performance of $\pi_{sr}$, we consider policy $\pi_{sr}^{wc}$ \footnote{The superscript `$wc$' shows that $\pi_{sr}^{wc}$ is \emph{work conserving}, i.e., it never waits if there is a source with fresh marked packet.} which is identical to $\pi_{sr}$, except that the waiting time for $\pi_{sr}^{wc}$ is $0$ (instead of $d_\ell\sim\cD_\ell$), and whenever $\pi_{sr}^{wc}$ chooses a source which does not have a fresh marked packet, it immediately chooses another source (among all the sources, with probability $\hat{p}_\ell$'s). Using numerical analysis, we compare $\pi_{sr}^{wc}$ with $\pi_{sr}$. 

As shown in Figure \ref{fig:Cost_vs_MeanArrival}, the difference between the weighted sum cost for $\pi_{sr}$ and $\pi_{sr}^{wc}$ does not depend on the mean inter-generation time of packets. This is because of the following two reasons. $(i)$ When $\mu_\ell$'s are small (large), then $p_\ell$'s \eqref{eq:tx_prob} are also small (large), i.e., when packet generation rate is large (small), then packets are marked with smaller (larger) probability. $(ii)$ $\hat{p}_\ell$'s \eqref{eq:sr-p-hat} are inversely proportional to $\mu_\ell$'s, which implies that a source with small packet generation rate is chosen less often to transmit. Thus, the choice of $p_\ell$'s \eqref{eq:tx_prob} and $\hat{p}_\ell$'s \eqref{eq:sr-p-hat} ensure that the probability that $\pi_{sr}$ chooses a source $\ell$ to transmit when it does not have a fresh marked packet, is small. Hence, the number of instances when $\pi_{sr}$ waits is small, independent of the mean packet inter-generation times $\mu_\ell$'s.


\begin{remark}
	From Figures \ref{fig:Cost_vs_N}, \ref{fig:Cost_vs_Var} and \ref{fig:Cost_vs_MeanArrival}, it is evident that under specific settings, the upper bound \eqref{eq:dummy1} on $\bbE[\Gamma(\pi_{sr})]$ is almost tight.
\end{remark}

\begin{figure}
	\begin{center}
		\begin{tikzpicture}
			\begin{axis}[xlabel=Mean Transmission Time $(\gamma)$, ylabel=Weighted Sum Cost $\Gamma(\pi)$,
				legend cell align=left,
				legend pos=north west, xmin=1, xmax=10, x tick label style={/pgf/number format/fixed}, grid=major] 
				\addplot[style={dashed},line width=1pt] table {./NumericalPlots/vsDelay/UB.dat};
				\addplot[style={solid},line width=1pt] table {./NumericalPlots/vsDelay/var0.dat};
				\addplot[style={dashdotted},line width=1pt] table {./NumericalPlots/vsDelay/var4.dat};
				\addplot[style={dotted},mark=square*,mark options={style=solid,scale=0.5},line width=1pt] table {./NumericalPlots/vsDelay/var1_NW.dat};
				\addplot[style={densely dashdotdotted},line width=1pt] table {./NumericalPlots/vsDelay/var10.dat};
				\addplot[style={dotted},line width=1pt] table {./NumericalPlots/vsDelay/LB.dat};
				\legend{$UB_{sr}$, $\pi_{sr}; \nu^2=0$, $\pi_{sr}; \nu^2=4$, $\pi_{sr}^{wc}; \nu^2=4$, $\pi_{sr}; \nu^2=10$, LB}
			\end{axis}
		\end{tikzpicture}
		\caption{Cost Vs Mean Transmission Time}
		\label{fig:Cost_vs_MeanTxTime}
	\end{center}
\end{figure}
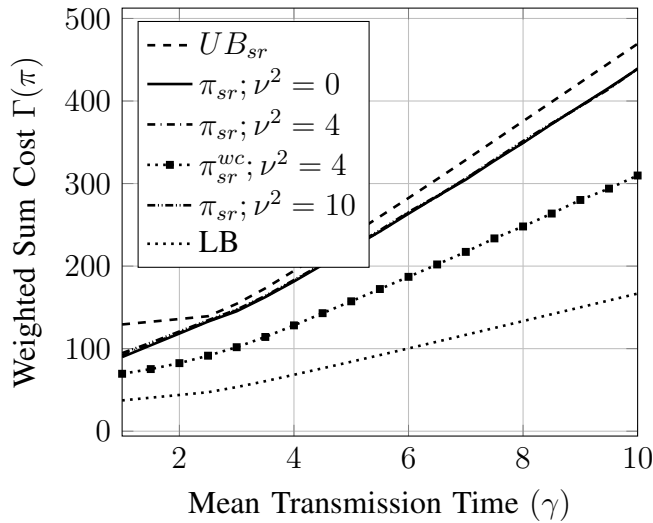

A critical property of $\pi_{sr}$ is that its competitive ratio \eqref{eq:CR-for-SR} is independent of the transmission time distribution $\cD_\ell$ of the sources. To verify this fact, we fix $c=1$ and $\mu=16$,\footnote{$\mu=16$ ensures that the effect of $\mu_\ell$'s is visible in the plot of Figure \ref{fig:Cost_vs_MeanTxTime}.} and for each source $\ell$, we choose $\cG_\ell$ to be the exponential distribution, and $\cD_\ell$ to be the log-normal distribution. For each source $\ell$, defining the variance of $\cD_\ell$ to be $\nu^2$, we simulate the system with different values of parameters $\nu^2$ and $\gamma$. As shown in Figure \ref{fig:Cost_vs_MeanTxTime}, the weighted sum cost $\Gamma(\pi_{sr})$ is less than the upper bound $UB_{sr}$, and $UB_{sr}$ as well as the lower bound $LB$ increase linearly with $\gamma$ (except for small values of $\gamma$, where the effect of $\mu_\ell$'s dominates the effect of $\gamma_\ell$'s). Hence, the ratio of $UB_{sr}$ and $LB$ is a constant, less than the competitive ratio \eqref{eq:CR-for-SR}. Further, similar to the lower bound $LB$ \eqref{eq:final-lb-Goff} and the upper bound $UB_{sr}$ \eqref{eq:dummy1}, the theoretical upper bound on  $\Gamma(\pi_{sr})$ is independent of $\nu^2$ (for different values of $\nu^2$, the plots of $\Gamma(\pi_{sr})$ overlap).

In Figure \ref{fig:Cost_vs_MeanTxTime}, note that the difference in weighted sum cost for $\pi_{sr}$ and $\pi_{sr}^{wc}$ increases (linearly) with increase in $\gamma_\ell$'s. This is because whenever $\pi_{sr}$ chooses a source $\ell$ to transmit which does not have a fresh marked packet to transmit, $\pi_{sr}$ waits for $d_\ell\sim\cD_\ell$ time units, which in expectation, increases linearly with $\gamma_\ell$. 

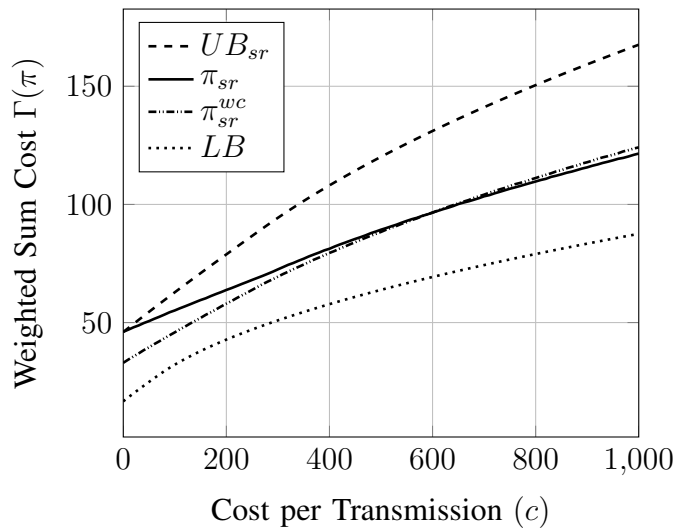
\begin{figure}
	\begin{center}
		\begin{tikzpicture}
			\begin{axis}[xlabel=Cost per Transmission $(c)$, ylabel=Weighted Sum Cost $\Gamma(\pi)$,
				legend cell align=left,
				legend pos=north west, xmin=0, xmax=1000, x tick label style={/pgf/number format/fixed}, grid=major] 
				\addplot[style={dashed},line width=1pt] table {./NumericalPlots/vsCost/SRPvar1UB.dat};
				\addplot[style={solid},line width=1pt] table {./NumericalPlots/vsCost/SRPvar1.dat};
				\addplot[style={densely dashdotdotted},line width=1pt] table {./NumericalPlots/vsCost/SRPvar1NW.dat};
				\addplot[style={dotted},line width=1pt] table {./NumericalPlots/vsCost/LB.dat};
				\legend{$UB_{sr}$, $\pi_{sr}$, $\pi_{sr}^{wc}$, $LB$}
			\end{axis}
		\end{tikzpicture}
		\caption{Effect of cost per transmission of sources on the weighted sum cost of the system.}
		\label{fig:WSC_vs_TxCost}
	\end{center}
\end{figure}

Finally, to understand the impact of cost per transmission $c_\ell$ for the sources on $\Gamma(\pi_{sr})$, we fix $\mu=\gamma=1$, and for each source $\ell$, we choose $\cD_\ell$ to be the exponential distribution, and $\cG_\ell$ to be the log-normal distribution (with variance $\sigma_\ell^2=1$). We simulate the system for different values of $c$, and find that the increase in $\Gamma(\pi_{sr})$ (with increase in $c$) is sub-linear, as shown in Figure \ref{fig:WSC_vs_TxCost}. This suggests that $\pi_{sr}$ compensates for the increase in transmission cost (due to increasing $c$) by appropriately scaling down the transmission frequency of the sources. Also, in Figure \ref{fig:WSC_vs_TxCost}, the plot of $\Gamma(\pi_{sr})$ lies between the plots of $UB_{sr}$ \eqref{eq:dummy1} and $LB$ \eqref{eq:final-lb-Goff},  
which verifies the competitive ratio guarantee \eqref{eq:CR-for-SR}. 

In Figure \ref{fig:WSC_vs_TxCost}, note that when the cost per transmission is large, the weighted sum cost for $\pi_{sr}^{wc}$ exceeds the weighted sum cost for $\pi_{sr}$. This is intuitive because for a source with large cost per transmission, when its AoI is small, waiting is better than transmitting a packet. It highlights the significance of packet marking (SR-PMS) in $\pi_{sr}$, which prevents $\pi_{sr}$ from transmitting `unnecessarily'.

\subsection{Comparative Analysis} \label{sec:numerical-comparative}
In this subsection, we consider some standard settings from prior work, and compare $\pi_{sr}$ with the state-of-the-art policies for those settings.

First, we consider the $N$-source system considered in \cite{kadota2019minimizing}, where time is partitioned into unit length slots, and in each slot, a packet is generated at each source $\ell$ with probability $\mu_\ell^{-1}$. In a slot, at most one source can transmit, and each transmission by source $\ell$ is either successful with probability $\gamma_\ell^{-1}$, or it fails with probability $1-\gamma_\ell^{-1}$. We consider the single packet queue setting, where in any slot, a source only has its latest generated packet. Transmission cost for the sources are assumed to be 0, and the objective is to minimize the weighted sum AoI of the sources. 

For minimizing the weighted sum AoI of the sources, following policies have been proposed in \cite{kadota2019minimizing}: $(i)$ stationary randomized policy for discrete-time setting $\pi_{rd}$, and $(ii)$ Max-Weight Policy $\pi_{mw}$. 
In each slot, among all the sources, $\pi_{rd}$ chooses a source $\ell$ with probability $q_\ell$ (derived in \cite{kadota2019minimizing}), and transmits its latest fresh packet (if the source has fresh packet, else no packet is transmitted in the slot). In \cite{kadota2019minimizing}, $\pi_{rd}$ is shown to have competitive ratio of at most 4. On the other hand, in each slot $t$, $\pi_{mw}$ transmits the latest fresh packet of the source for which the expected weighted\footnote{The weights for the expected reduction in age for the source are defined as part of $\pi_{mw}$, based on the system parameters.} reduction in age upon transmission is maximum. It has been shown that the competitive ratio of $\pi_{mw}$ is no more than that of the proposed policy $\pi_{rd}$. 

For this setting, $\pi_{sr}$ simplifies as follows. 
\begin{enumerate}
	\item Since the transmission cost is 0, and a transmission fails/succeeds in a slot without restricting transmission in future slots, SR-PMS marks every generated packet. Hence, whenever SR-NSS chooses a source to transmit, the source transmits its latest fresh packet (if it has fresh packet, else, the source remains idle in the slot). 
	\item In any slot $t$, SR-NSS chooses a source for transmission, based on the event in slot $t-1$. If there is a failed transmission in slot $t-1$ (packet transmitted but not received at the monitor), then in slot $t$, SR-NSS chooses the same source as in slot $t-1$. Else, in slot $t$, among all the sources, SR-NSS chooses a source $\ell$, with probability $\hat{p}_\ell$ \eqref{eq:sr-p-hat}.
\end{enumerate}

\begin{remark} \label{remark:pisr-notin-PiRD}
	Note that both $\pi_{sr}$ and $\pi_{rd}$ are stationary randomized policies. However, unlike $\pi_{rd}$, if $\pi_{sr}$ picks a source that has a fresh packet, then it picks it repeatedly in each slot, until its transmission is successful. 
	Although it is not obvious which of the two policies is better, 
	it appears intuitively that $\pi_{sr}$ should achieve lower weighted sum AoI than $\pi_{rd}$. This is because once $\pi_{sr}$ picks a source that has a packet to transmit, it never idles in any slot until the source successfully transmits, while $\pi_{rd}$ may again pick a source which does not have packet to transmit, and hence, it may idle in a slot (thus wasting the slot).
\end{remark}

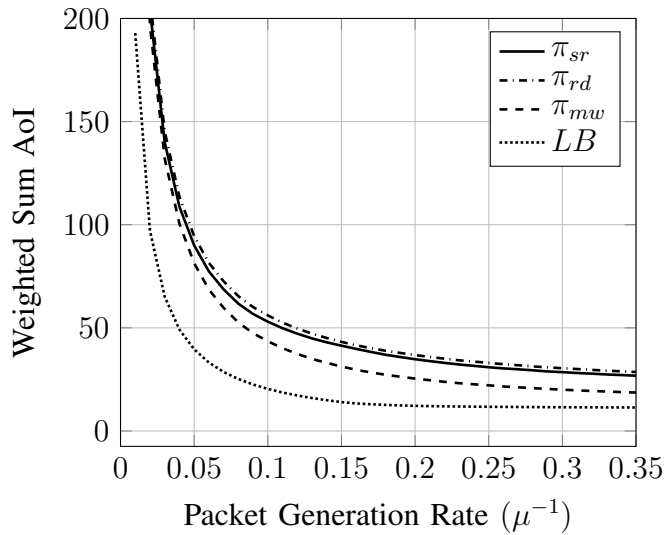
\begin{figure}
	\begin{center}
		\begin{tikzpicture}
			\begin{axis}[
				xlabel=Packet Generation Rate $(\mu^{-1})$, 
				ylabel=Weighted Sum AoI,
				legend cell align=left,
				legend pos=north east,
				grid=major, ymax=200,
				xmin=0,xmax=0.35,xtick distance=0.05,
				x tick label style={/pgf/number format/fixed}] 
				\addplot[style={solid},line width=1pt] table {./NumericalPlots/vsDiscrete/SRP.dat};
				\addplot[style={dashdotted},line width=1pt] table {./NumericalPlots/vsDiscrete/KRP.dat};
				\addplot[style={dashed},line width=1pt] table {./NumericalPlots/vsDiscrete/MWT.dat};
				\addplot[style={densely dotted},line width=1pt] table {./NumericalPlots/vsDiscrete/LB.dat};
				\legend{$\pi_{sr}$, $\pi_{rd}$, $\pi_{mw}$, $LB$}
			\end{axis}
		\end{tikzpicture}
		\caption{Effect of packet generation rate on weighted sum AoI. In this plot, $LB$ denotes the lower bound on weighted sum AoI provided in \cite{kadota2019minimizing}.}
		\label{fig:Comparision_Kadota}
	\end{center}
\end{figure}

For comparing $\pi_{sr}$, $\pi_{rd}$ and $\pi_{mw}$, we simulate the policies for the same parameter values considered in \cite{kadota2019minimizing}, i.e., the number of sources $N=4$, weights $[\rho_1,\rho_2,\rho_3,\rho_4]=[4,4,1,1]$, transmission success probability $[\gamma_1^{-1},\gamma_2^{-1},\gamma_3^{-1},\gamma_4^{-1}]=[0.25,0.5,0.75,1]$, and packet generation rate $[\mu_1^{-1},\mu_2^{-1},\mu_3^{-1},\mu_4^{-1}]=\mu^{-1}\cdot[1,0.75,0.5,0.25]$, where $\mu^{-1}$ is varied in interval $(0,1]$. 
Figure \ref{fig:Comparision_Kadota} shows the plot of weighted sum AoI for the policies $\pi_{sr}$, $\pi_{rd}$ and $\pi_{mw}$, for different values of $\mu^{-1}$ (in Figure \ref{fig:Comparision_Kadota}, $LB$ denotes the lower bound on weighted sum AoI, provided in \cite{kadota2019minimizing}).
As evident from the plot, even though $\pi_{sr}$ is designed for a general setting, its performance in minimizing weighted sum AoI is at par with $\pi_{rd}$, and close to $\pi_{mw}$ (where $\pi_{rd}$ and $\pi_{mw}$ are designed specifically for the considered setting). 
\begin{remark}
	$\pi_{rd}$ and $\pi_{mw}$ assume that for all sources, the packet generation, as well as 
	the transmission success/failure instants are synchronized with the start/end of slots.  
	Since this assumption is not true in the general continuous-time setting of Section \ref{sec:sysModel}, 
	the policies $\pi_{rd}$ and $\pi_{mw}$ (and their corresponding guarantees) do not extend naturally to the setting of Section \ref{sec:sysModel}. 
\end{remark}

  
Next, we consider the single source continuous-time setting of \cite{sun2017update}, where the source can generate a fresh packet at any time (immediately), and the transmission time for every packet is independent and identically distributed as per distribution $\cD$ (with mean $\gamma$). In the considered setting, at any time, at most one packet can be under transmission, and a packet under transmission cannot be preempted. Also, there is zero cost for transmission. The objective is to minimize the AoI of the source.
\begin{remark}
	Since the considered system has only one source, from all the notations, we drop the subscript $\ell$, that is used to index the source in a multi-source system.
\end{remark}

For this setting, \cite{sun2017update} proposed an optimal causal policy $\pi^\star$, that at any time $t$, if the channel is free, and the time elapsed since the generation time of latest transmitted packet is greater than or equal to the threshold $\beta$ (where $\beta\in[0,\infty)$ is computed numerically for each distribution $\cD$), the source generates a fresh packet and begins transmitting it immediately. 

In this setting, since $N=1$, the subroutine SR-NSS of $\pi_{sr}$ (that chooses which source gets to transmit) is redundant. Hence, at any time, if the channel is free, and the source has a marked packet, then $\pi_{sr}$ transmits its latest marked packet. Further, since the source can generate packets at any time, for $\pi_{sr}$, the natural choice for generating and marking packets is when the channel is free (i.e., when the source can transmit), 
and the time elapsed since the generation time of the latest transmitted packet is at least equal to $\gamma$. The rationale for using $\gamma$ as the threshold for $\pi_{sr}$ is as follows. Since the source can generate packets instantaneously at any time, $\mu\to 0^+$. Hence, instead of minimizing \eqref{eq:tx_prob} with respect to $p$, we minimize \eqref{eq:tx_prob} with respect to $\mu/p$ (the expected inter-generation time of marked packets). Since $N=1$, the minimizer of the objective function \eqref{eq:tx_prob} (under constraint \eqref{eq:interference-constraint}) is $\mu/p=\gamma$.   


\begin{remark}
	Note that the threshold-based version of $\pi_{sr}$ proposed for this setting is actually a \emph{stationary deterministic policy} (as against its name, i.e., `stationary randomized policy'). Hence, we denote it by $\pi_{sd}$.
\end{remark}

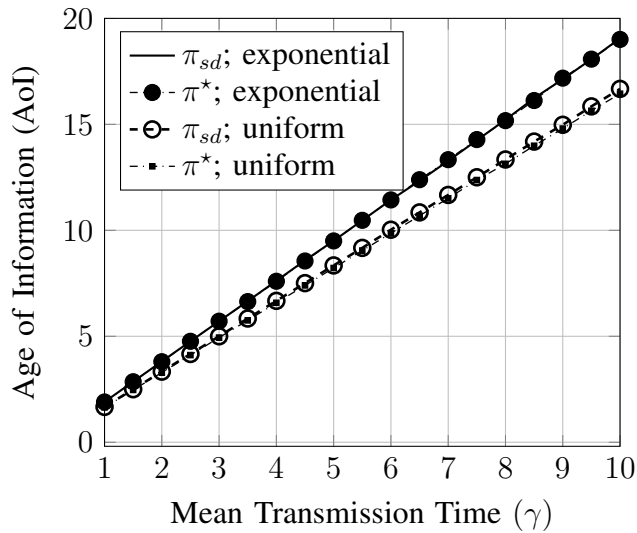
\begin{figure}
	\begin{center}
		\begin{tikzpicture}
			\begin{axis}[
				xlabel=Mean Transmission Time $(\gamma)$, 
				ylabel=Age of Information (AoI),
				legend cell align=left,
				legend pos=north west,
				grid=major, ymax=20,				
				xmin=1,xmax=10,xtick distance=1,				
				x tick label style={/pgf/number format/fixed}] 
				\addplot[style={solid},line width=0.8pt] table {./NumericalPlots/vsUpdateWait/exp_SRP.dat};
				\addplot[style={dashdotted},mark=*,mark options={style=solid,scale=1.5},line width=0.5pt] table {./NumericalPlots/vsUpdateWait/exp_OPT.dat};
				\addplot[style={dashed},mark=o,mark options={style=solid,scale=1.5},line width=1pt] table {./NumericalPlots/vsUpdateWait/unif_SRP.dat};
				\addplot[style={dashdotted},mark=square*,mark options={style=solid,scale=0.5},line width=0.5pt] table {./NumericalPlots/vsUpdateWait/unif_OPT.dat};
				\legend{$\pi_{sd}$; exponential, $\pi^\star$; exponential, $\pi_{sd}$; uniform, $\pi^\star$; uniform}
			\end{axis}
		\end{tikzpicture}
		\caption{AoI as a function of mean transmission time.}
		\label{fig:Comparision_UpdateWait}
	\end{center}
\end{figure}

\begin{figure}
	\begin{center}
		\begin{tikzpicture}
			\begin{axis}[
				xlabel=Mean Transmission Time $(\gamma)$, 
				ylabel=Threshold,
				legend cell align=left,
				legend pos=north west,
				grid=major, ymax=10,				
				xmin=1,xmax=10,xtick distance=1,				
				x tick label style={/pgf/number format/fixed}] 
				\addplot[style={solid},line width=1pt] table {./NumericalPlots/vsUpdateWait/MeanDelay.dat};
				\addplot[style={dashed},line width=1pt] table {./NumericalPlots/vsUpdateWait/beta_exp.dat};
				\addplot[style={dashdotted},line width=1pt] table {./NumericalPlots/vsUpdateWait/beta_unif.dat};
				\legend{$\pi_{sd}$, $\pi^\star$; exponential, $\pi^\star$; uniform, }
			\end{axis}
		\end{tikzpicture}
		\caption{Threshold for $\pi_{sd}$ and $\pi^\star$ as a function of mean transmission time.}
		\label{fig:Comparision_UpdateWait_threshold}
	\end{center}
\end{figure}
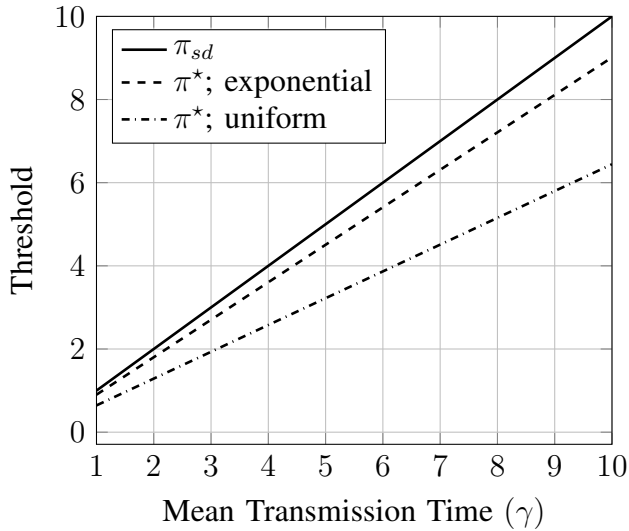

We compare the AoI for $\pi^\star$ and $\pi_{sd}$ (both deterministic threshold policies), by simulating them for different values of $\gamma$. We consider two cases: $(i)$ when $\cD$ is an exponential distribution, and $(ii)$ when $\cD$ is a uniform distribution (over the interval $(0,2\gamma]$, so that the mean is $\gamma$).
Figure \ref{fig:Comparision_UpdateWait} shows the AoI plot for the two policies with respect to $\gamma$. From the plot, it is clear that for both the exponential as well as the uniform distribution (for transmission time), the AoI for $\pi_{sd}$ is very close to the corresponding AoI for $\pi^\star$ (plots almost overlap). To understand the reason for such an observation, Figure \ref{fig:Comparision_UpdateWait_threshold} plots the threshold for $\pi_{sd}$ and $\pi^\star$ for different values of the mean transmission time $\gamma$.

From Figure \ref{fig:Comparision_UpdateWait_threshold}, it is clear that the thresholds for policies $\pi_{sd}$ and $\pi^\star$ are different. However, in Figure \ref{fig:Comparision_UpdateWait_threshold}, since the threshold $\beta$ for $\pi^\star$ is smaller than $\gamma$ (the threshold for $\pi_{sd}$), a likely reason for near optimal AoI for $\pi_{sd}$ is that the transmission times for the packets are close to $\gamma$ (mean transmission time). Hence, under both $\pi^\star$ and $\pi_{sd}$, the source transmits its successive packets close to when the channel becomes free after $\gamma$ time units.



	Note that the threshold $\beta$ (for $\pi^\star$) needs to be computed numerically for each distribution $\cD$. Also, it requires computing an expectation with respect to the distribution $\cD$, which may be a difficult task for certain distributions. Hence, for such distribution $\cD$, policy $\pi_{sd}$ might be a better choice over $\pi^\star$ (however we do not know if $\pi_{sd}$ is near optimal for every $\cD$). 
	This is in addition to the fact that unlike $\pi_{sd}$, there is no known generalization of $\pi^\star$ for the multi-source setup considered in this paper (with stochastic packet generation times, and transmission cost). 
	

\section{Conclusions}
In this paper, in a major departure from prior work, we have considered the scheduling problem of finding the optimal non-preemptive policy to minimize the sum of the AoI and the transmission cost, 
in the presence of multiple sources, and where the inter-generation time of updates and the transmission time/delay for each update follow a general distribution. Mostly prior work has considered fixed scheduling 
policies, and analyzed their AoI distributions. Instead of directly finding the optimal scheduling policy, we propose a randomized scheduling policy 
and upper bound its competitive ratio (by comparing against an offline optimal policy) by the ratio of the variance and the squared mean of the  inter-generation time of updates. Notably the competitive ratio is independent of the transmission time/delay distributions, and is upper bounded by 
$4$ for exponential, uniform, and Rayleigh  inter-generation time distributions. In addition to the upper bound, we also presented 
a tight example to show that the competitive ratio of the considered algorithm has to depend on the ratio of the variance and squared mean of the  inter-generation time of updates. For the preemptive settings, we had to restrict to a G/M/1 system, and showed that a non-preemptive randomized policy has 
a similar competitive ratio as in the non-preemptive setting.
Obvious question that remains open: are there policies that have constant competitive ratios, i.e., independent of the distribution of inter-generation time of updates. 
\bibliographystyle{IEEEtran} 
\bibliography{refs,reflist}

\appendices
\section{General Expression for the Weighted Sum Cost $\Gamma(\pi) \eqref{eq:tac}$} \label{sec:notations}
Next, based on the period lengths, Lemma \ref{lemma:LCFS-finite-eta} identifies a subset of policies $\pi$ that is sufficient to minimize the weighted sum cost $\Gamma(\pi)$ \eqref{eq:tac}. 
\begin{lemma} \label{lemma:LCFS-finite-eta}
	For minimizing the weighted sum cost $\Gamma(\pi)$ \eqref{eq:tac}, it is sufficient to consider only those policies $\pi$ (causal or offline), for which $T_{\ell i}^\pi$ ($\forall i$) and $\eta_\ell^\pi(t)$ are finite with probability 1. 
\end{lemma}
\begin{IEEEproof}
	In the considered model, at each source $\ell$, the inter-generation time of packets $X_\ell$, as well as the transmission time $d_\ell$, is finite with probability 1 (follows from the assumption that both $\bbE[X_\ell]=\mu_\ell$ and $\bbE[d_\ell]=\gamma_\ell$ are finite). Also, the cost per transmission $c_\ell$ is finite. Therefore, under any policy $\pi$ that minimizes the total average cost $\Gamma(\cdot)$, a source will never wait for infinite duration before it transmits a packet of source $\ell$ and allows it to get received at the monitor. Hence, for all such policy $\pi$, $T_{\ell i}^\pi$ and $\eta_\ell^\pi$ must be finite with probability 1.
\end{IEEEproof}

In view of Lemma \ref{lemma:LCFS-finite-eta}, we restrict our attention to policies $\pi$, for which $T_{\ell i}^\pi$ ($\forall i$) and $\eta_\ell^\pi(t)$ are finite with probability 1, and respectively define $\Pi_{S}$ and $\Pi_{OF}$ as the set of all causal and non-causal policies $\pi$, such that $T_{\ell i}^\pi$ ($\forall i$) and $\eta_\ell^\pi(t)$ are finite with probability 1. 
Next, for all policies $\pi\in\Pi_{S}$ or $\pi \in \Pi_{OF}$, Lemma \ref{lemma:actual-Delta} provides a general expression for AoI \eqref{eq:avAge} in terms of the quantities defined in this section so far.
\begin{lemma} \label{lemma:actual-Delta}
	For any policy $\pi\in\Pi_{S}$ or $\pi \in \Pi_{OF}$, the AoI $\Delta_{\ell,\pi}^{av}(t)$ satisfies
	\begin{align} \label{eq:actual-Delta}
        \underset{t\to\infty}{\lim}\Delta_{\ell,\pi}^{av}(t)=\lim_{t\to\infty}\sum_{i=1}^{R_\ell^\pi(t)}\frac{\frac{(T_{\ell i}^\pi)^2}{2}+T_{\ell i}^\pi Z_{\ell i}^\pi}{t}.
	\end{align} 
    Further, for such a policy $\pi$, as $t\to\infty$, $R_\ell^\pi(t)\to\infty$ as well.
\end{lemma}
\begin{IEEEproof} 
	Figure \ref{fig:multi-node-general-age} shows a general age plot for source $\ell$ in terms of the quantities defined in this section so far. Note that in each period $\cP_{\ell i}^\pi$ until time $t$, the age cost is $Q_{\ell i}^\pi=(T_{\ell i}^\pi)^2/2+T_{\ell (i-1)}^\pi Z_{\ell (i-1)}^\pi$. Thus, the AoI for source $\ell$ satisfies 
	\begin{align}
	      \underset{t\to\infty}{\lim}\Delta_{\ell,\pi}^{av}(t)&=\lim_{t\to\infty}\left(\sum_{i=1}^{R_\ell^\pi(t)}\frac{\frac{(T_{\ell i}^\pi)^2}{2}+T_{\ell (i-1)}^\pi Z_{\ell (i-1)}^\pi}{t}+\frac{(\eta_\ell^\pi(t))^2}{2t}\right), \nonumber \\
	      &\stackrel{(a)}{=}\lim_{t\to\infty}\sum_{i=1}^{R_\ell^\pi(t)}\frac{\frac{(T_{\ell i}^\pi)^2}{2}+T_{\ell i}^\pi Z_{\ell i}^\pi}{t}+\lim_{t\to\infty}\frac{(\eta_\ell^\pi(t))^2}{2t}, \nonumber \\
	      &\stackrel{(b)}{=}\lim_{t\to\infty}\sum_{i=1}^{R_\ell^\pi(t)}\frac{\frac{(T_{\ell i}^\pi)^2}{2}+T_{\ell i}^\pi Z_{\ell i}^\pi}{t}, \nonumber
	\end{align} 
	where we get $(a)$ by rearranging the terms in the summation and substituting $T_{\ell 0}^\pi Z_{\ell 0}^\pi=0$ (because initial AoI of all the sources is 0; Assumption \ref{assume:init-AoI}), whereas $(b)$ follows because in the fraction $(\eta_\ell^\pi(t))^2/(2t)$, the numerator $(\eta_\ell^\pi(t))^2$ is finite (with probability 1), but in the denominator, $t\to\infty$.
	
	Further, from \eqref{eq:sumT=t-n}, we have $t=\sum_{i=1}^{R_\ell^{\pi}(t)}T_{\ell i}^\pi+\eta_\ell^\pi(t)$. Since $T_{\ell i}^\pi$ and $\eta_\ell^\pi(t)$ are finite for all $i$, we get that as $t\to\infty$, $R_\ell^\pi(t)\to\infty$ as well.\end{IEEEproof}

Using Lemma \ref{lemma:actual-Delta}, next we derive a 
general expression for the weighted sum cost $\Gamma(\pi)$ \eqref{eq:tac}. 
\begin{lemma} \label{lemma:actual-tac-with-t}
	For any policy $\pi\in\Pi_{S}$ or $\pi \in \Pi_{OF}$, the weighted sum cost
	 \begin{align}
		\label{eq:actual-tac-with-t}	\Gamma(\pi)&=\lim_{t\to\infty}\frac{1}{N}\sum_{\ell=1}^{N}\frac{\sum_{i=1}^{R_\ell^\pi(t)}(\rho_\ell (\frac{(T_{\ell i}^\pi)^2}{2}+T_{\ell i}^\pi Z_{\ell i}^\pi)+c_\ell)}{t},
	\end{align} 
    where $R_\ell^\pi(t)'s$ satisfy
    \begin{align} \label{eq:num-tx}
    	\lim_{t\to\infty}\sum_{\ell=1}^{N}\frac{\gamma_\ell R_\ell^\pi(t)}{t}\le 1, \ \ \text{(with probability 1).}
    \end{align}
\end{lemma}
\begin{IEEEproof}
	Substituting \eqref{eq:avTxCost} and \eqref{eq:actual-Delta} in the expression for weighted sum cost $\Gamma(\pi)$ \eqref{eq:tac}, and rearranging the obtained terms, we get \eqref{eq:actual-tac-with-t}. To obtain \eqref{eq:num-tx}, note that at any time at most one packet can be under transmission, and the transmission of packet $\ell_i^\pi$ takes $d_{\ell i}$ time units, where $d_{\ell i}$'s are independent and identically distributed random variables with mean $\gamma_\ell$. Therefore, $\sum_{\ell=1}^{N}\sum_{i=1}^{R_\ell^\pi(t)}d_{\ell i}\le t$. Dividing both sides by $t$, and taking limit as $t\to\infty$, we get
\begin{align} 
	1&\ge \lim_{t\to\infty}\sum_{\ell=1}^{N}\left(\frac{\sum_{i=1}^{R_\ell^\pi(t)}d_{\ell i}}{R_\ell^\pi(t)}\cdot\frac{R_\ell^\pi(t)}{t}\right), \nonumber \\
	&\stackrel{(a)}{=} \sum_{\ell=1}^{N}\left(\lim_{t\to\infty}\frac{\sum_{i=1}^{R_\ell^\pi(t)}d_{\ell i}}{R_\ell^\pi(t)}\cdot\lim_{t\to\infty}\frac{R_\ell^\pi(t)}{t}\right), \nonumber \\
	&\stackrel{(b)}{=}\lim_{t\to\infty}\sum_{\ell=1}^{N}\frac{\gamma_\ell R_\ell^\pi(t)}{t},\hspace{4ex}\text{(with probability 1)},
\end{align}
where we get $(a)$ because the limit of a product is equal to the product of the limits (when the limits exists, as in the above case), and $(b)$ follows from strong law of large numbers. Note that for $(b)$, we could use strong law of large numbers because $(i)$ $d_{\ell i}$'s (for all $i$) are independent and identically distributed with mean $\gamma_\ell$, and $(ii)$ as $t\to\infty$, $R_\ell^\pi(t)\to\infty$ as well (Lemma \ref{lemma:actual-Delta}).
\end{IEEEproof}


\begin{remark} \label{remark:sumT=t} 
	From \eqref{eq:sumT=t-n}, we get
	$\sum_{i=1}^{R_\ell^\pi(t)}T_{\ell i}^\pi=t-\eta_\ell^\pi(t)$, where $\eta_\ell^\pi(t)$ is finite for all policies $\pi\in\Pi_{S}\cup\Pi_{OF}$ (by definition of $\Pi_S$ and $\Pi_{OF}$). Therefore, for $\pi\in\Pi_{S}\cup\Pi_{OF}$ (i.e., the policies of interest),	
	as $t\to\infty$, we get $\sum_{i=1}^{R_\ell^\pi(t)}T_{\ell i}^\pi=t-\eta_\ell^\pi(t)\approx t$. Hence for simplicity, in the rest of this paper, when $t\to\infty$, we consider $\sum_{i=1}^{R_\ell^\pi(t)}T_{\ell i}^\pi=t$, i.e., any large time $t$ is equal to the sum of the length of periods of source $\ell$ (for any $\ell\in\{1,\cdots,N\}$) until time $t$. 
\end{remark}

\section{Proof of Lemma \ref{lemma:Gopt-lb-gen-dist}} \label{App:LowerBoundOPT}

    From \eqref{eq:system-time}, recall that for any policy $\pi$ (causal or offline), $Z_{\ell i}^\pi\ge d_{\ell i}$ (since $w_{\ell i}^\pi \ge 0$). Therefore, $T_{\ell i}^\pi Z_{\ell i}^\pi\ge T_{\ell i}^\pi d_{\ell i}$. Hence, from \eqref{eq:actual-tac-with-t}, we get that for an offline policy $\pi\in\Pi_{OF}$, 
	\begin{align} \label{eq:Glb-OF-init}
		\Gamma(\pi)&\ge\lim_{t\to\infty}\frac{1}{N}\sum_{\ell=1}^{N}\frac{\sum_{i=1}^{R_\ell^\pi(t)} \rho_\ell(\frac{({T_{\ell i}^\pi})^2}{2}+T_{\ell i}^\pi d_{\ell i})+c_\ell R_\ell^\pi(t)}{t}, \nonumber \\
	    &=\frac{1}{N}\sum_{\ell=1}^{N}\Bigg(\lim_{t\to\infty}\frac{\sum_{i=1}^{R_\ell^\pi(t)} \rho_\ell({T_{\ell i}^\pi})^2}{2t} +\lim_{t\to\infty}\frac{\sum_{i=1}^{R_\ell^\pi(t)}\rho_\ell T_{\ell i}^\pi d_{\ell i}}{t}+\lim_{t\to\infty}\frac{R_\ell^\pi(t) c_\ell}{t}\Bigg).
\end{align}

Let $T_{\ell,\pi}^{av}=\underset{t\to\infty}{\lim}t/R_\ell^\pi(t)=\underset{t\to\infty}{\lim}\sum_{i=1}^{R_\ell^\pi(t)}T_{\ell i}^\pi/R_\ell^\pi(t)$ denote the average period length for a policy $\pi$. Also, define $\delta_{\ell i}^\pi=T_{\ell i}^\pi-T_{\ell,\pi}^{av}$. 
Then, $T_{\ell i}^\pi=\delta_{\ell i}^\pi+T_{\ell,\pi}^{av}$, and $(T_{\ell i}^\pi)^2=(\delta_{\ell i}^\pi)^2+(T_{\ell,\pi}^{av})^2+2\delta_{\ell i}^\pi T_{\ell,\pi}^{av}$. 
Further, since $\sum_{i=1}^{R_\ell^\pi}T_{\ell i}^\pi=t=R_\ell^\pi T_{\ell,\pi}^{av}=\sum_{i=1}^{R_\ell^\pi}T_{\ell,\pi}^{av}$ (follows from the definition of $T_{\ell,\pi}^{av}$), we get $\sum_{i=1}^{R_\ell^\pi(t)}\delta_{\ell i}^\pi=0$.
Therefore, when $t\to\infty$,
\begin{gather} \label{eq:T^2-lb}
\frac{\sum_{i=1}^{R_\ell^\pi(t)}(T_{\ell i}^\pi)^2}{t}=\frac{\sum_{i=1}^{R_\ell^\pi(t)}(\delta_{\ell i}^\pi)^2+R_\ell^\pi(t) (T_{\ell,\pi}^{av})^2+2T_{\ell,\pi}^{av}\sum_{i=1}^{R_\ell^\pi(t)}\delta_{\ell i}^\pi}{R_\ell^\pi(t)T_{\ell,\pi}^{av}}=\frac{\sum_{i=1}^{R_\ell^\pi(t)}(\delta_{\ell i}^\pi)^2}{R_\ell^\pi(t)T_{\ell,\pi}^{av}}+ T_{\ell,\pi}^{av}, \\
\label{eq:Td-lb}
\frac{\sum_{i=1}^{R_\ell^\pi(t)}T_{\ell i}^\pi d_{\ell i}}{t}=\frac{T_{\ell,\pi}^{av}\sum_{i=1}^{R_\ell^\pi(t)}d_{\ell i}+\sum_{i=1}^{R_\ell^\pi(t)}\delta_{\ell i}^\pi d_{\ell i}}{R_\ell^\pi(t)T_{\ell,\pi}^{av}}\stackrel{(a)}{=}\gamma_\ell+\frac{\sum_{i=1}^{R_\ell^\pi(t)}\delta_{\ell i}^\pi d_{\ell i}}{R_\ell^\pi(t)T_{\ell,\pi}^{av}}, \\
\label{eq:Cav-lb}
\frac{c_\ell R_\ell^\pi(t)}{t}=\frac{c_\ell}{T_{\ell,\pi}^{av}},
\end{gather}
where we get $(a)$ (with probability 1), using strong law of large numbers (by definition, $d_{\ell i}$'s are independent and identically distributed, and from Lemma \ref{lemma:actual-Delta}, we know that as $t\to\infty$, $R_\ell^\pi(t)\to\infty$ as well). 

Substituting \eqref{eq:T^2-lb}, \eqref{eq:Td-lb} and \eqref{eq:Cav-lb} in \eqref{eq:Glb-OF-init}, we get
\begin{align} \label{eq:Glb-OF-1}
	\Gamma(\pi)\ge\frac{1}{N}\sum_{\ell=1}^N\left(\frac{\rho_\ell\beta_\ell^\pi}{2T_{\ell,\pi}^{av}}+\frac{\rho_\ell T_{\ell,\pi}^{av}}{2} +\rho_\ell\gamma_\ell+\frac{c_\ell}{T_{\ell,\pi}^{av}}\right),
\end{align}
where $\beta_\ell^\pi=\underset{t\to\infty}{\lim}\sum_{i=1}^{R_\ell^\pi(t)}\delta_{\ell i}^\pi(\delta_{\ell i}^\pi+2d_{\ell i})/R_\ell^\pi(t)$.  
\begin{lemma} \label{lemma:beta-non-negative}
	$\beta_\ell^\pi\ge 0$, $\forall \ell,\pi$, with probability 1.
\end{lemma}
\begin{IEEEproof}
    See Appendix \ref{appendix:lemma-beta-non-negative}.
\end{IEEEproof}
Note that the average period length $T_{\ell,\pi}^{av}$ is always positive. Also, Lemma \ref{lemma:beta-non-negative} shows that $\beta_\ell^\pi$ is non-negative. Therefore, from \eqref{eq:Glb-OF-1} we get 
\begin{align} \label{eq:Glb-OF-2}
	\Gamma(\pi)\ge\frac{1}{N}\sum_{\ell=1}^N\left(\frac{\rho_\ell T_{\ell,\pi}^{av}}{2}+\rho_\ell\gamma_\ell+\frac{c_\ell}{T_{\ell,\pi}^{av}}\right).
\end{align}
\begin{remark}
	For any policy $\pi$, the effect of randomness (variance) in the inter-generation time of packets is captured in $\delta_{\ell i}^\pi=T_{\ell i}^\pi-T_{\ell,\pi}^{av}$, $\forall i$. Therefore, when we lower bound $\beta_\ell^\pi=\underset{t\to\infty}{\lim}\sum_{i=1}^{R_\ell^\pi(t)}\delta_{\ell i}^\pi(\delta_{\ell i}^\pi+2d_{\ell i})/R_\ell^\pi(t)$ in \eqref{eq:Glb-OF-1} by $0$, the $\delta_{\ell i}^\pi$  terms are lost. Hence, the contribution of variance of inter-generation time of packets to the weighted sum cost $\Gamma(\pi)$ \eqref{eq:actual-tac-with-t}) is lower bounded by 0. However, as shown in Example \ref{ex:zero-var}, for an offline optimal policy, this lower bound is tight.
\end{remark}

Next, consider the optimal offline policy $\pi_{OF}^\star$.  
Recall that $h_\ell(t)$ is the number of packets generated at source $\ell$ until time $t$, and $\pi_{OF}^\star$ transmits $R_\ell^\star(t)$ number of these packets. Thus, for  $\pi_{OF}^\star$, the average period length
\begin{align} \label{eq:fl-Tav}
	T_{\ell,\pi^\star_{OF}}^{av}=\lim_{t\to\infty}\frac{t}{ R_\ell^\star(t)}=\lim_{t\to\infty}\frac{t}{h_\ell(t)}\frac{h_\ell(t)}{R_{\ell}^\star(t)}\stackrel{(a)}{=}\frac{\mu_{\ell}}{f_\ell^\star},
\end{align} 
where we get $(a)$ because $(i)$ with probability 1, $\underset{t\to\infty}{\lim}t/h_{\ell}(t)=\mu_\ell$ (using the strong law of large numbers; since $t/h_\ell(t)$ is the average inter-generation time of packets, and $\mu_\ell$ is the expected inter-generation time of packets), and $(ii)$ by definition, $f_\ell^\star=\underset{t\to\infty}{\lim}R_{\ell}^\star(t)/h_{\ell}(t)$. Note that $f_\ell^\star$ is equal to the fraction of total number of packets generated at source $\ell$ until time $t$ that is transmitted by $\pi_{OF}^\star$. Hence, $f_\ell^\star\in[0,1]$. 

Substituting \eqref{eq:fl-Tav} in \eqref{eq:Glb-OF-2}, we get
\begin{equation*} 
	\Gamma(\pi_{OF}^\star)\ge 
	\frac{1}{N}\sum_{\ell=1}^{N}\left(\frac{\rho_{\ell}\mu_{\ell}}{2f_\ell^\star}+\rho_\ell \gamma_\ell+\frac{c_\ell f_\ell^\star}{\mu_\ell}\right),
\end{equation*}
which implies \eqref{eq:final-lb-Goff}. Also, from \eqref{eq:num-tx} and \eqref{eq:fl-Tav}, we get 
\begin{equation*} 
	1\ge \lim_{t\to\infty}\sum_{\ell=1}^{N}\frac{\gamma_\ell R_{\ell}^\star(t)}{t}=\sum_{\ell=1}^{N}\frac{\gamma_\ell}{T_{\ell,\pi^\star_{OF}}^{av}}=\sum_{\ell=1}^{N}\frac{\gamma_\ell f_{\ell}^\star}{\mu_{\ell}}. 
\end{equation*} 

\section{Proof of Lemma \ref{lemma:beta-non-negative}} \label{appendix:lemma-beta-non-negative}

Since $\delta_{\ell i}^\pi=T_{\ell i}^\pi-T_{\ell,\pi}^{av}$, and $T_{\ell i}^\pi$'s and $T_{\ell,\pi}^{av}$ are finite (Lemma \ref{lemma:LCFS-finite-eta}), $\delta_{\ell i}$'s are finite as well. Also, by definition, $\delta_{\ell i}^\pi$'s satisfy $\underset{t\to\infty}{\lim}\sum_{i=1}^{R_\ell^\pi(t)}\delta_{\ell i}^\pi=\underset{t\to\infty}{\lim}\sum_{i=1}^{R_\ell^\pi(t)}(T_{\ell i}^\pi-T_{\ell,\pi}^{av})=0$. 
Now, for some policy $\pi$, let there exists a sequence $\cH=\{\delta_{\ell i}^\pi\}_{i\in\bbN}$, such that the above two conditions are satisfied, and $\beta_\ell^\pi=\underset{t\to\infty}{\lim}\sum_{i=1}^{R_\ell^\pi(t)}\delta_{\ell i}^\pi(\delta_{\ell i}^\pi+2d_{\ell i})/R_\ell^\pi(t)<0$. Then, for any $\omega\in\bbR$,
$\beta_\ell^\pi+\underset{t\to\infty}{\lim}\omega\sum_{i=1}^{R_\ell^\pi(t)}\delta_{\ell i}^\pi/R_\ell^\pi(t)=\underset{t\to\infty}{\lim}\sum_{i=1}^{R_\ell^\pi(t)}\delta_{\ell i}^\pi(\delta_{\ell i}^\pi+\omega+2d_{\ell i})/R_\ell^\pi(t)<0$.

But note that $\delta_{\ell i}^\pi(\delta_{\ell i}^\pi+\omega+2d_{\ell i})<0$, only if $\delta_{\ell i}^\pi\in(-\omega-2d_{\ell i},0)$, i.e., $d_{\ell i}>-(\delta_{\ell i}^\pi+\omega)/2$, which cannot be true when $\omega\to-\infty$ (since $\delta_{\ell i}$ and transmission time $d_{\ell i}$ are finite with probability 1). 
Thus, when $\delta_{\ell i}^\pi\to-\infty$, 
$\delta_{\ell i}^\pi(\delta_{\ell i}^\pi+\omega+2d_{\ell i})>0$ with probability 1. This implies $\beta_\ell^\pi+\underset{t\to\infty}{\lim}\omega\sum_{i=1}^{R_\ell^\pi(t)}\delta_{\ell i}^\pi/R_\ell^\pi(t)>0$ with probability 1, contradicting the existence of sequence $\cH$. Hence, $\beta_\ell^\pi$ must be non-negative with probability 1.

\section{Proof of Lemma \ref{lemma:gamma-pisr-ub-gen-dist}} \label{App:SRP-tac-UB}
Recall that $\pi_{sr}$ (Algorithm \ref{algo:distribution-independent-policy}) transmits only marked packets. Therefore, for $\pi_{sr}$, the period lengths $T_{\ell i}^{sr}$'s (the inter-generation time of completely transmitted packets) can be written as a sum of the inter-generation time of marked packets $T_{\ell j}^m$'s. Also, for each source $\ell$, the number of completely transmitted packets $R_\ell^{sr}(t)$ is upper bounded by the number of marked packets $R_\ell^m(t)$. Moreover, the waiting time $w_{\ell i}^{sr}$ for packets of source $\ell$ is upper bounded by  $\hat{w}_{\ell j}^{sr}$, the difference between the successive time instants when source $\ell$ is chosen to transmit by $\pi_{sr}$. Hence, to prove Lemma \ref{lemma:gamma-pisr-ub-gen-dist}, we upper bound the expectation of the weighted sum cost \eqref{eq:actual-tac-with-t} for $\pi_{sr}$ in terms of $T_{\ell j}^m$'s, $R_\ell^m(t)$'s, $\hat{w}_{\ell j}^{sr}$'s, and the transmission times $d_{\ell i}$'s for completely transmitted packets (under $\pi_{sr}$). Then, using the independence between $T_{\ell j}^{m}$'s and $\hat{w}_{\ell j}^{sr}$'s (which follows because the packet marking by SR-PMS, and the source selection by SR-NSS function independently of each other), we simplify the derived upper bound to get \eqref{eq:dummy1}.

\begin{IEEEproof} 
	Consider the following quantities defined with respect to $\pi_{sr}$:
	$(i)$ $g_{\ell i}^m$ --- the generation time of the $i^{th}$ marked packet of source $\ell$ under $\pi_{sr}$, $(ii)$ $s_{\ell i}^m$ --- the earliest time instant greater than or equal to $g_{\ell i}^m$ when $\pi_{sr}$ chooses source $\ell$ to transmit (at $s_{\ell i}^m$ source $\ell$ has at least one fresh marked packet which got generated at $g_{\ell i}^m$; hence under $\pi_{sr}$, at $s_{\ell i}^m$, source $\ell$ begins to transmit a fresh marked packet), and $(iii)$ $r_{\ell i}^m$ --- the earliest time instant greater than or equal to $s_{\ell i}^m$, when source $\ell$ completes transmitting the packet that it began transmitting at $s_{\ell i}^m$. Also, define $T_{\ell i}^m=g_{\ell i}^m-g_{\ell (i-1)}^m$, and $Z_{\ell i}^m = r_{\ell i}^m-g_{\ell i}^m=w_{\ell i}^m+d_{\ell i}^m$, where $w_{\ell i}^m=s_{\ell i}^m-g_{\ell i}^m$, and $d_{\ell i}^m=r_{\ell i}^m-s_{\ell i}^m$.  
	As shown in Figure \ref{fig:multi-node-age}, the area under the age plot for source $\ell$ under policy $\pi_{sr}$ can be partitioned into trapezoids with area $Q_{\ell 1}^m,Q_{\ell 2}^m,Q_{\ell 3}^m,\cdots$, where $Q_{\ell i}^m=(T_{\ell i}^m)^2/2+T_{\ell i}^m Z_{\ell i}^m$, $\forall i\in\bbN$. Summing the areas of all these trapezoids, dividing it by the time horizon $t$, and following the arguments as in the proof of Lemma \ref{lemma:actual-Delta} (which uses Property 3 in Lemma \ref{lemma:finite-T-sr} below), we get the long-term AoI under $\pi_{sr}$ to be
	\begin{align} \label{eq:actual-Delta-sr}
		\underset{t\to\infty}{\lim}\Delta_{\ell,sr}^{av}(t)=\lim_{t\to\infty}\frac{\sum_{i=1}^{R_\ell^m(t)}\left(\frac{(T_{\ell i}^m)^2}{2}+T_{\ell i}^m Z_{\ell i}^m\right)}{t},
	\end{align} 
	where $R_\ell^m(t)$ denotes the number of packets marked by $\pi_{sr}$ until time $t$, and as $t\to\infty$, $R_\ell^m(t)\to\infty$ as well. Also, as in Remark \ref{remark:sumT=t}, we get $\sum_{i=1}^{R_\ell^m(t)}T_{\ell i}^m=t$.


	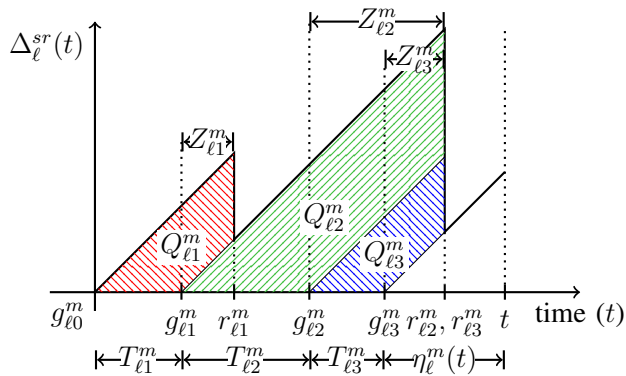
\begin{figure}
	\begin{center}
		\begin{tikzpicture}[thick,scale=1, every node/.style={scale=0.9}]
		\draw[->] (-0.25,0) to (6.8,0) node[below]{time ($t$)};
		\draw[->] (0.35,-0.25) to (0.35,3.6) node[below left]{$\Delta_{\ell}^{sr}(t)$};
		\draw (0.35,0) node[below left]{$g_{\ell 0}^{m}$} to (2.2,1.85) to (2.2,0.7) to (5,3.5) to (5,0.8) to (5.8,1.6); 
		
		\fill[pattern=north west lines, pattern color=red] (0.35,0) to (2.2,1.85) to (2.2,0.7) to (1.5,0) to (0.35,0);
		\fill[pattern=north east lines, pattern color=green!50!gray] (1.5,0) to (5,3.5) to (5,1.8) to (3.2,0) to (1.5,0);
		
		\fill[pattern=north west lines, pattern color=blue] (5,1.8) to (3.2,0) to (4.2,0) to (5,0.8) to (5,1.8);

		\draw (1.5,-0.1) node[below]{$g_{\ell 1}^{m}$} to (1.5,0.1);
		\draw (3.2,-0.1) node[below]{$g_{\ell 2}^{m}$} to (3.2,0.1);
		\draw (4.2,-0.1) node[below]{$g_{\ell 3}^{m}$} to (4.2,0.1); 
		
		\draw (2.2,-0.1) node[below]{$r_{\ell 1}^{m}$} to (2.2,0.1);
		\draw (5,-0.1) to (5,0.1);
		\draw (5,-0.1) node[below]{$r_{\ell 2}^{m},r_{\ell 3}^{m}$};
		\draw (5.8,-0.1) node[below]{$t$} to (5.8,0.1);
		
		\draw[dotted] (1.5,0.1) to (1.5,2);
		\draw[dotted] (3.2,0.1) to (3.2,3.6);
		\draw[dotted] (4.2,0.1) to (4.2,3);
		\draw[dotted] (2.2,0.7) to (2.2,0.1);
		\draw[dotted] (5,0.8) to (5,0.1);
		\draw[dotted] (5.8,0.1) to (5.8,3.5);
		
		\draw[thin] (1.5,0) to (2.2,0.7); 
		\draw[thin] (3.2,0) to (5,1.8); 
		\draw[thin] (4.2,0) to (5,0.8); 
		
        \draw[|<->] (0.35,-0.9) -- (1.5,-0.9) node[rectangle,inner sep=-1pt,midway,fill=white]{$T_{\ell 1}^{m}$}; 
        \draw[|<->] (1.5,-0.9) -- (3.2,-0.9) node[rectangle,inner sep=-1pt,midway,fill=white]{$T_{\ell 2}^{m}$};
        \draw[|<->|] (3.2,-0.9) -- (4.2,-0.9) node[rectangle,inner sep=-1pt,midway,fill=white]{$T_{\ell 3}^{m}$};
        \draw[<->|] (4.2,-0.9) -- (5.8,-0.9) node[rectangle,inner sep=-1pt,midway,fill=white]{$\eta_{\ell}^{m}(t)$};
		
        \draw[|<->|] (1.5,2) -- (2.2,2) node[rectangle,inner sep=-1pt,midway,fill=white]{$Z_{\ell 1}^{m}$}; 
        \draw[|<->|] (3.2,3.6) -- (5,3.6) node[rectangle,inner sep=-1pt,midway,fill=white]{$Z_{\ell 2}^{m}$};
        \draw[|<->|] (4.2,3.1) -- (5,3.1) node[rectangle,inner sep=-1pt,midway,fill=white]{$Z_{\ell 3}^{m}$};
		
		\draw (1.3,0.6)-- (1.7,0.6) node[rectangle,inner sep=1.5pt,midway,fill=white]{$Q_{\ell 1}^{m}$};
		\draw (3.2,1)-- (3.6,1) node[rectangle,inner sep=1.5pt,midway,fill=white]{$Q_{\ell 2}^{m}$};
		\draw (4,0.5)-- (4.4,0.5) node[rectangle,inner sep=1.5pt,midway,fill=white]{$Q_{\ell 3}^{m}$};		
		\end{tikzpicture}
		\caption{Sample age plot of source $\ell$ under stationary randomized policy $\pi_{sr}$ (Algorithm \ref{algo:distribution-independent-policy}).} 
		\label{fig:multi-node-age} 
	\end{center}
\end{figure}

 \begin{lemma} \label{lemma:finite-T-sr}
 	For each source $\ell\in\{1,\cdots,N\}$ and $i\in\bbN$,
 	\begin{enumerate}
 		\item $T_{\ell i}^m$'s are independent and identically distributed,
 		\item $\bbE[T_{\ell i}^m]=\mu_{\ell}/p_{\ell}$ and $\bbE[(T_{\ell i}^m)^2]=(\sigma_{\ell}^2/p_\ell)+(2-p_{\ell})\mu_\ell^2/p_\ell^2$, and 
 		\item $T_{\ell i}^m$ and $\eta_\ell^m(t)$  (where, $\forall t\ge0$, $\eta_\ell^m(t)$ denotes the difference between the generation time of the latest marked packet until time $t$, and time $t$) are finite with probability 1. 
 	\end{enumerate}
 \end{lemma}
\begin{IEEEproof}
	See Appendix \ref{appendix:lemma-finite-T-sr}.
\end{IEEEproof}

Further, recall that under $\pi_{sr}$, whenever a source $\ell$ is chosen for transmission, it either transmits a fresh marked packet, or does not transmit at all.
Therefore, the packets transmitted by a source under $\pi_{sr}$ is a subset of its marked packets, implying that the average transmission cost \eqref{eq:avTxCost} for $\pi_{sr}$ is $C_{\ell,sr}^{av}\le c_\ell R_\ell^{m}(t)/t$. Hence, the weighted sum cost \eqref{eq:tac} for $\pi_{sr}$ is 
\begin{align} \label{eq:tac-sr}
	\Gamma(\pi_{sr})&\le \lim_{t\to\infty}\frac{1}{N}\sum_{\ell=1}^N\left(\frac{\rho_\ell\sum_{i=1}^{R_\ell^m(t)}\left(\frac{(T_{\ell i}^m)^2}{2}+T_{\ell i}^m Z_{\ell i}^m\right)}{t}+\frac{c_\ell R_\ell^m(t)}{t}\right), \nonumber \\
	&\stackrel{(a)}{=}\frac{1}{N}\sum_{\ell=1}^N \left(\lim_{t\to\infty}\frac{\sum_{i=1}^{R_\ell^m(t)}\left(\rho_\ell\frac{(T_{\ell i}^m)^2}{2}+c_\ell\right)}{\sum_{i=1}^{R_\ell^m(t)}T_{\ell i}^m}\right)+\frac{1}{N}\sum_{\ell=1}^N\left(\lim_{t\to\infty}\frac{\rho_\ell \sum_{i=1}^{R_\ell^m(t)}T_{\ell i}^m Z_{\ell i}^m}{\sum_{i=1}^{R_\ell^m(t)}T_{\ell i}^m}\right),
\end{align}
 where we get $(a)$ by rearranging the terms, and substituting $t=\sum_{i=1}^{R_\ell^m(t)}T_{\ell i}^m$.
 From Lemma \ref{lemma:finite-T-sr}, it is known that $T_{\ell i}^m$'s are independent and identically distributed. Also, as $t\to\infty$, $R_\ell^m(t)\to\infty$ as well. Therefore, applying the renewal reward theorem \cite{ross2014introduction} to each term in the first summation on the R.H.S. of \eqref{eq:tac-sr}, we get
 \begin{align} \label{eq:sr-tac-term1}
 	\lim_{t\to\infty}\frac{\sum_{i=1}^{R_\ell^m(t)}\left(\rho_\ell\frac{(T_{\ell i}^m)^2}{2}+c_\ell\right)}{\sum_{i=1}^{R_\ell^m(t)}T_{\ell i}^m}=\frac{\frac{\rho_\ell}{2}\bbE[(T_{\ell 1}^{m})^2]+c_\ell}{\bbE[T_{\ell 1}^{m}]}\stackrel{(a)}{=}\frac{\rho_\ell}{2}\left(\frac{\sigma_\ell^2}{\mu_\ell}+\frac{\mu_\ell}{p_\ell}(2-p_\ell)\right)+\frac{c_\ell p_\ell}{\mu_\ell},
 \end{align}
where $(a)$ follows by substituting $\bbE[T_{\ell 1}^m]=\mu_\ell/p_\ell$, and $\bbE[(T_{\ell i}^m)^2]=(\sigma_{\ell}^2/p_\ell)+(2-p_{\ell})\mu_\ell^2/p_\ell^2$ (from Lemma \ref{lemma:finite-T-sr}). Thus, substituting \eqref{eq:sr-tac-term1} back into \eqref{eq:tac-sr}, rearranging its terms, and taking expectation (jointly with respect to the distributions $\cG_\ell$ and $\cD_\ell$, for each source $\ell$) on both sides, we get
\begin{align} \label{eq:exp-tac-sr}
	\bbE[\Gamma(\pi_{sr})]&\le 
	\frac{1}{N}\sum_{\ell=1}^N \left(\frac{\rho_\ell \mu_\ell}{2 p_\ell}(2-p_\ell\theta_\ell)+\frac{c_\ell p_\ell}{\mu_\ell}\right)+\frac{1}{N}\sum_{\ell=1}^N\bbE\left[\lim_{t\to\infty}\frac{\rho_\ell \sum_{i=1}^{R_\ell^m(t)}T_{\ell i}^m Z_{\ell i}^m}{\sum_{i=1}^{R_\ell^m(t)}T_{\ell i}^m}\right],
\end{align}
where $\theta_\ell=1-\sigma_\ell^2/\mu_\ell^2$.
    
\begin{figure} 
\begin{center}
\begin{tikzpicture}[thick,scale=1, every node/.style={scale=1}]
\draw[thick] (-0.1,0.1) to (11.1,0.1); 
        		
\draw (0,-0.1) node[below]{$0$} to (0,0.05); 
        		
\draw (4.7,0.1) to (4.7,-0.05) node[below]{$g_{\ell 3}^m$}; 
        		
\draw (1.5,-0.05) node[below]{$g_{\ell 1}^m$} to (1.5,0.1);
\draw (3,-0.05) node[below]{$g_{\ell 2}^m$} to (3,0.1); 
\draw (6,-0.05) node[below]{$g_{\ell 4}^m$} to (6,0.1); 
\draw (9.3,-0.05) node[below]{$g_{\ell 5}^m$} to (9.3,0.1); 

\draw (0,0.1) to (0,0.25) node[above]{$s_{\ell 0}^{m}$};
\draw (3.7,0.1) to (3.7,0.25) node[above]{$s_{\ell 1}^{m},s_{\ell 2}^{m}$};
\draw (5.4,0.1) to (5.4,0.25) node[above]{$r_{\ell 1}^{m},r_{\ell 2}^{m}$}; 
\draw (7,0.1) to (7,0.25) node[above]{$s_{\ell 3}^{m},s_{\ell 4}^{m}$}; 
\draw (8.5,0.1) to (8.5,0.25) node[above]{$r_{\ell 3}^{m},r_{\ell 4}^{m}$}; 
\draw (10,0.1) to (10,0.25) node[above]{$s_{\ell 5}^{m}$}; 
\draw (11,0.1) to (11,0.25) node[above]{$r_{\ell 5}^{m}$};



\draw[|<->] (4.7,1.8) -- (7,1.8) node[rectangle,inner sep=-1pt,midway,fill=white]{$w_{\ell 3}^{m}$};
\draw[|<->] (6,1.2) -- (7,1.2) node[rectangle,inner sep=-1pt,midway,fill=white]{$w_{\ell 4}^{m}$};
\draw[|<->|] (7,1.8) -- (8.5,1.8) node[rectangle,inner sep=-1pt,midway,fill=white]{$d_{\ell 3}^{m}$};
\draw[|<->|] (7,1.2) -- (8.5,1.2) node[rectangle,inner sep=-1pt,midway,fill=white]{$d_{\ell 4}^{m}$};


\draw[|<->] (0,2.5) -- (3.7,2.5) node[rectangle,inner sep=-1pt,midway,fill=white]{$\hat{w}_{\ell 1}^{sr}$};
\draw[|<->|] (3.7,2.5) -- (7,2.5) node[rectangle,inner sep=-1pt,midway,fill=white]{$\hat{w}_{\ell 2}^{sr}$};
\draw[<->|] (7,2.5) -- (8.5,2.5) node[rectangle,inner sep=-1pt,midway,fill=white]{$d_{\ell 2}$};

\draw[|<->] (0,-0.8) -- (1.5,-0.8) node[rectangle,inner sep=-1pt,midway,fill=white]{$T_{\ell 1}^m$}; 
\draw[|<->] (1.5,-0.8) -- (3,-0.8) node[rectangle,inner sep=-1pt,midway,fill=white]{$T_{\ell 2}^m$};
\draw[|<->] (3,-0.8) -- (4.7,-0.8) node[rectangle,inner sep=-1pt,midway,fill=white]{$T_{\ell 3}^m$};
\draw[|<->] (4.7,-0.8) -- (6,-0.8) node[rectangle,inner sep=-1pt,midway,fill=white]{$T_{\ell 4}^m$};
\draw[|<->|] (6,-0.8) -- (9.3,-0.8) node[rectangle,inner sep=-1pt,midway,fill=white]{$T_{\ell 5}^m$};
\draw[<-] (9.3,-0.8) to (10,-0.8);
\draw[loosely dotted] (10,-0.8) to (11,-0.8);

\draw[|<->] (0,-1.3) -- (3,-1.3) node[rectangle,inner sep=-1pt,midway,fill=white]{$T_{\ell 1}^{sr}$};
\draw[|<->] (3,-1.3) -- (6,-1.3) node[rectangle,inner sep=-1pt,midway,fill=white]{$T_{\ell 2}^{sr}$};
\draw[|<->|] (6,-1.3) -- (9.3,-1.3) node[rectangle,inner sep=-1pt,midway,fill=white]{$T_{\ell 3}^{sr}$};
\draw[<-] (9.3,-1.3) to (10,-1.3);
\draw[loosely dotted] (10,-1.3) to (11,-1.3);

\draw[dotted] (4.7,0.1) to (4.7,1.7);
\draw[dotted] (6,0.1) to (6,1.1);
\draw[dotted] (3.7,0.8) to (3.7,2.4);
\draw[dotted] (8.5,0.8) to (8.5,2.4);
\draw[dotted] (7,0.8) to (7,2.4);
\draw[dotted] (0,0.8) to (0,2.4);
\end{tikzpicture}
\caption{Relation between different quantities defined with respect to $\pi_{sr}$. Note that $s_{\ell 3}^m=s_{\ell 4}^m$, because it is the earliest time instant after both $g_{\ell 3}^m$ and $g_{\ell 4}^m$ when source $\ell$ gets the opportunity to transmit.}   
\label{fig:PMS-NSS}  
\end{center} 
\end{figure}
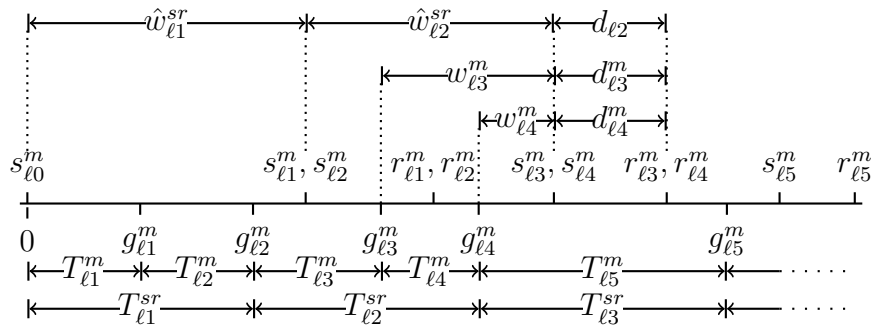
    
Next, we upper bound the second summation on the R.H.S. of \eqref{eq:exp-tac-sr}. Recall that $Z_{\ell i}^m=r_{\ell i}^m-g_{\ell i}^m=w_{\ell i}^m+d_{\ell i}^m$, where $w_{\ell i}^m=s_{\ell i}^m-g_{\ell i}^m$ is the time duration (since the generation time of $i^{th}$ marked packet of source $\ell$), after which $\pi_{sr}$ chooses source $\ell$ to transmit, and $d_{\ell i}^m=r_{\ell i}^m-s_{\ell i}^m$ is equal to the total transmission time for the packet that source $\ell$ begins to transmit at time $s_{\ell i}^m$. Also, as shown in Figure \ref{fig:PMS-NSS}, $w_{\ell i}^m$'s are upper-bounded by the difference between the successive time instants (before and after $g_{\ell i}^m$) when source $\ell$  is chosen to transmit (let this difference be denoted by $\hat{w}_{\ell i}^{sr}$'s). Therefore, $T_{\ell i}^m Z_{\ell i}^m=T_{\ell i}^m (w_{\ell i}^m+d_{\ell i}^m)\le T_{\ell i}^m(\hat{w}_{\ell i}^{sr}+d_{\ell i}^m)$, which implies 
\begin{align} \label{eq:sr-tac-term2}
	\bbE\left[\lim_{t\to\infty}\frac{\rho_\ell \sum_{i=1}^{R_\ell^m(t)}T_{\ell i}^m Z_{\ell i}^m}{\sum_{i=1}^{R_\ell^m(t)}T_{\ell i}^m}\right]\le \bbE\left[\lim_{t\to\infty}\frac{\rho_\ell \sum_{i=1}^{R_\ell^m(t)}T_{\ell i}^m \hat{w}_{\ell i}^{sr}}{\sum_{i=1}^{R_\ell^m(t)}T_{\ell i}^m}\right]+\bbE\left[\lim_{t\to\infty}\frac{\rho_\ell \sum_{i=1}^{R_\ell^m(t)}T_{\ell i}^m d_{\ell i}^m}{\sum_{i=1}^{R_\ell^m(t)}T_{\ell i}^m}\right].
\end{align}

Note that the first term on the R.H.S. of \eqref{eq:sr-tac-term2} is 
\begin{align} \label{eq:sr-tac-term2-a}
	\bbE\left[\lim_{t\to\infty}\frac{\rho_\ell \sum_{i=1}^{R_\ell^m(t)}T_{\ell i}^m \hat{w}_{\ell i}^{sr}}{\sum_{i=1}^{R_\ell^m(t)}T_{\ell i}^m}\right]\stackrel{(a)}{=}\bbE\left[\frac{\rho_\ell \bbE[T_{\ell i}^m \hat{w}_{\ell i}^{sr}]}{\bbE[T_{\ell i}^m]}\right]\stackrel{(b)}{=}\frac{\rho_\ell \bbE[T_{\ell i}^m]\bbE[\hat{w}_{\ell i}^{sr}]}{\bbE[T_{\ell i}^m]}\stackrel{(c)}{\le}\frac{\rho_\ell\mu_\ell}{p_\ell},
\end{align}
where we get $(a)$ by applying renewal reward theorem (renewal properties follow from Lemma \ref{lemma:iid-w-d} below), $(b)$ follows because $T_{\ell i}^m$ and $\hat{w}_{\ell i}^{sr}$ are mutually independent (Lemma \ref{lemma:iid-w-d}), and $(c)$ follows because $\bbE[\hat{w}_{\ell i}^{sr}]\le\mu_\ell/p_\ell$ (Lemma \ref{lemma:iid-w-d}).

\begin{lemma} \label{lemma:iid-w-d}
	$\hat{w}_{\ell i}^{sr}$'s are independent and identically distributed ($\forall i\in\bbN$), with mean $\bbE[\hat{w}_{\ell i}^{sr}]\le\mu_\ell/p_\ell$. Moreover, $\hat{w}_{\ell i}^{sr}$'s are independent of $T_{\ell j}^m$'s ($\forall i,j\in\bbN$). 
\end{lemma}
\begin{IEEEproof}
	See Appendix \ref{appendix:lemma-iid-w-d}.
\end{IEEEproof}

In the second term on the R.H.S. of \eqref{eq:sr-tac-term2}, note that $d_{\ell i}^m$'s are not independent for every $i$. 
In fact, if the $\kappa_{j-1}^{th}$ and $\kappa_{j}^{th}$ marked packets of source $\ell$ are respectively its $(j-1)^{th}$ and $j^{th}$ transmitted packets, then as shown in Figure \ref{fig:PMS-NSS}, $\forall i\in\{\kappa_{j-1}+1,\cdots,\kappa_j\}$, $d_{\ell i}^m=d_{\ell j}$, where $d_{\ell j}$ denotes the transmission time for the $j^{th}$ transmitted packet. Additionally, $\sum_{i=\kappa_{j-1}+1}^{\kappa_{j}}T_{\ell i}^m=T_{\ell j}^{sr}$, where $T_{\ell j}^{sr}$ is the period length under $\pi_{sr}$, which by definition, is equal to the inter-generation time of marked packets that get transmitted. Therefore, $\sum_{i=\kappa_{j-1}+1}^{\kappa_{j}}T_{\ell i}^m d_{\ell i}^m=T_{\ell j}^{sr}d_{\ell j}$, which implies
\begin{align} \label{eq:sr-tac-term2-b1}
	\bbE\left[\lim_{t\to\infty}\frac{\rho_\ell \sum_{i=1}^{R_\ell^m(t)}T_{\ell i}^m d_{\ell i}^m}{\sum_{i=1}^{R_\ell^m(t)}T_{\ell i}^m}\right]=\bbE\left[\lim_{t\to\infty}\frac{\rho_\ell \sum_{j=1}^{R_\ell^{sr}(t)}T_{\ell j}^{sr} d_{\ell j}}{\sum_{j=1}^{R_\ell^{sr}(t)}T_{\ell j}^{sr}}\right],
\end{align}
where $R_\ell^{sr}(t)$ denotes the number of packets transmitted by source $\ell$ under $\pi_{sr}$ until time $t$. In the simplified expression on the R.H.S. of \eqref{eq:sr-tac-term2-b1}, it is not obvious if $T_{\ell j}^{sr}d_{\ell j}$ and $T_{\ell (j+1)}^{sr}d_{\ell (j+1)}$ are independent. Hence, unlike \eqref{eq:sr-tac-term2-a}, we cannot apply the renewal reward theorem directly to simplify \eqref{eq:sr-tac-term2-b1}. Therefore, we take an alternate approach as follows. 

Note that when packet $\ell_j^{sr}$ is transmitted, $T_{\ell j}^{sr}$ gets fixed, whereas $d_{\ell j}$ is realized after that, from distribution $\cD_\ell$, independent of $T_{\ell j}^{sr}$. Therefore, $T_{\ell j}^{sr}$ and $d_{\ell j}$ are mutually independent. Hence, if we define $\varphi_{\ell j}=T_{\ell j}^{sr}/t$, then $\varphi_{\ell j}$ and $d_{\ell j}$ are mutually independent as well. 
Also, since $t=\sum_{j=1}^{R_\ell^{sr}(t)}T_{\ell i}^{sr}$ (Remark \ref{remark:sumT=t}), $\varphi_{\ell j}=T_{\ell j}^{sr}/t\in[0,1]$, and $\sum_{i=1}^{R_\ell^{sr}(t)}\varphi_{\ell i}=1$. Therefore, $\bbE\left[\underset{t\to\infty}{\lim}\frac{\rho_\ell \sum_{j=1}^{R_\ell^{sr}(t)}T_{\ell j}^{sr} d_{\ell j}}{\sum_{j=1}^{R_\ell^{sr}(t)}T_{\ell j}^{sr}}\right]$
\begin{align} \label{eq:sr-tac-term2-b2}
	\stackrel{(a)}{=}\rho_\ell\bbE\left[\sum_{j=1}^{\infty}\varphi_{\ell j} d_{\ell j}\right]
	\stackrel{(b)}{=}\rho_{\ell}\sum_{j=1}^{\infty}\bbE\left[\varphi_{\ell j} d_{\ell j}\right]
	\stackrel{(c)}{=}\rho_{\ell}\sum_{j=1}^{\infty}\bbE\left[\varphi_{\ell j}\right] \bbE\left[d_{\ell j}\right]
	\stackrel{(d)}{=} \rho_\ell\gamma_\ell\bbE\left[\sum_{j=1}^{\infty}\varphi_{\ell j}\right]\stackrel{(e)}{=}\rho_\ell\gamma_\ell,
\end{align}
where we get $(a)$ because as $t\to\infty$, $R_\ell^{sr}(t)\to\infty$ as well (Lemma \ref{lemma:actual-Delta}), we get $(b)$ as an application of Tonelli's theorem \cite{patrick1995probability} ($\varphi_{\ell j} d_{\ell j}$ is non-negative, and finite with probability 1; by definition), $(c)$ follows because $\varphi_{\ell j}$ and $d_{\ell j}$ are mutually independent, $(d)$ follows because $d_{\ell i}$'s are independent and identically distributed with mean $\gamma_\ell$, and we get $(e)$ because $\sum_{j=1}^{\infty}\varphi_{\ell j}=\underset{t\to\infty}{\lim}\sum_{j=1}^{R_\ell^{sr}(t)}\varphi_{\ell j}=1$. 

From \eqref{eq:sr-tac-term2-b1} and \eqref{eq:sr-tac-term2-b2}, we get 
\begin{align} \label{eq:sr-tac-term2-b}
	\bbE\left[\lim_{t\to\infty}\frac{\rho_\ell \sum_{i=1}^{R_\ell^m(t)}T_{\ell i}^m d_{\ell i}^m}{\sum_{i=1}^{R_\ell^m(t)}T_{\ell i}^m}\right]=\rho_\ell\gamma_\ell.
\end{align}
Further, substituting \eqref{eq:sr-tac-term2-a} and \eqref{eq:sr-tac-term2-b} into \eqref{eq:sr-tac-term2}, we get
\begin{align} \label{eq:sr-tac-term2-final}
	\bbE\left[\lim_{t\to\infty}\frac{\rho_\ell \sum_{i=1}^{R_\ell^m(t)}T_{\ell i}^m Z_{\ell i}^m}{\sum_{i=1}^{R_\ell^m(t)}T_{\ell i}^m}\right]\le \frac{\rho_\ell\mu_\ell}{p_\ell}+\rho_\ell\gamma_\ell.
\end{align}
Finally, substituting \eqref{eq:sr-tac-term2-final} into \eqref{eq:exp-tac-sr}, and rearranging terms, we get Lemma \ref{lemma:gamma-pisr-ub-gen-dist}. 
\end{IEEEproof}

\section{Proof of Lemma \ref{lemma:finite-T-sr}} \label{appendix:lemma-finite-T-sr}

Since $\pi_{sr}$ marks/discards every packet generated at source $\ell$ with fixed probability $p_\ell$, and inter-generation time of packets at source $\ell$ are independent and identically distributed, we conclude that $T_{\ell i}^m$'s (inter-generation time of marked packets) are independent and identically distributed ($\forall i\in\bbN$). 
Also,  
\begin{align} \label{eq:T-sr}
	T_{\ell i}^m=\sum_{j=1}^{K_{\ell i}^{m}}X_{\ell i j},
\end{align}
where $K_{\ell i}^{m}$ is a geometrically distributed random variable (with success probability $p_\ell$), that denotes the number of packets generated at source $\ell$ (after its $(i-1)^{th}$ marked packet), until a packet gets marked under $\pi_{sr}$ (at time $g_{\ell i}^m$).

In \eqref{eq:T-sr}, since $K_{\ell i}^{m}$ is independent of  $X_{\ell i j}$'s, i.e., the inter-generation time of packets at source $\ell$, 
using the Wald's equation \cite{mckay2019probability}, we get
\begin{align} \label{eq:mean-T-sr}
	\bbE[T_{\ell i}^m]=\bbE[K_{\ell i}^{m}]\bbE[X_{\ell i j}]=\mu_{\ell}/p_{\ell}. 
\end{align}

Similarly, squaring both sides of \eqref{eq:T-sr}, and using the Wald's equation \cite{mckay2019probability}, we get
\begin{align} \label{eq:mean-T2-sr}
	\bbE[(T_{\ell i}^m)^2]&=\bbE\left[\sum_{j=1}^{K_{\ell i}^m}X_{\ell i j}^2+\sum_{j=1}^{K_{\ell i}^m}\sum_{k=1,k\ne j}^{K_{\ell i}^m}X_{\ell i j}X_{\ell i k}\right], \nonumber \\
	&=\bbE[K_{\ell i}^{m}]\bbE[X_{\ell i j}^2]+\bbE[(K_{\ell i}^m)^2-K_{\ell i}^m]\bbE[X_{\ell i j}]^2, \nonumber \\ 
	&=\bbE[K_{\ell i j}^{m}]\sigma_{\ell}^2+\bbE[(K_{\ell i j}^{m})^2]\mu_\ell^2, \nonumber \\
	&=\frac{\sigma_{\ell}^2}{p_\ell}+\frac{(2-p_{\ell})\mu_\ell^2}{p_\ell^2}. 
\end{align}

Recall that for each source $\ell$, $c_\ell<\infty$ and $0<\rho_\ell<\infty$ (Remark \ref{remark:finite-tx-cost}), and $\mu_\ell,\gamma_\ell<\infty$ (by definition). Hence, for $p_\ell=\mu_\ell/(\vartheta N \gamma_\ell)$ (where $\vartheta=\max\{1,\sum_{\ell=1}^N(\mu_\ell/\gamma_\ell)\}$), \eqref{eq:interference-constraint} and \eqref{eq:probability-constraint} are satisfied, and the objective \eqref{eq:tx_prob} is finite. Therefore, the minimum value of the objective \eqref{eq:tx_prob} must be finite, which is possible only if $\mu_\ell/p_\ell$ is finite (for $p_\ell$ that minimizes \eqref{eq:tx_prob}). Hence, $\bbE[T_{\ell i}^m]$ (equal to $\mu_\ell/p_\ell$) must be finite, which implies $T_{\ell i}^m$'s are finite with probability 1. 

Note that by definition, $\eta_\ell^m(t)$ is the length of interval in which no packet is marked. Since $T_{\ell i}^m$'s (inter-generation time of marked packets) are finite (with probability 1), $\eta_\ell^m(t)$ must be finite with probability 1 as well.



\section{Proof of Lemma \ref{lemma:iid-w-d}} \label{appendix:lemma-iid-w-d} 

When the channel is free, among all the sources, SR-NSS (in $\pi_{sr}$) chooses a source $j$ with probability $\hat{p}_j$ \eqref{eq:sr-p-hat}.  
Also, each time a source $j$ is chosen, the channel remains busy for $d_{j k}$ time units, where $d_{j k}$'s are independent and identically distributed random variables ($\forall k\in\bbN$) with distribution $\cD_j$ (and mean $\gamma_j$). 
Hence, $\hat{w}_{\ell i}^{sr}=\sum_{j=1}^{N}\sum_{k=1}^{\tilde{\kappa}_{\ell j}}d_{j k}$, where $\tilde{\kappa}_{\ell j}$ are independent and identically distributed random variables ($\forall j\in\{1,\cdots,N\}$) that denote the number of times source $j$ is chosen between two successive instants when source $\ell$ gets chosen (naturally, $\hat{\kappa}_{\ell \ell}=1$).  
Therefore, $\hat{w}_{\ell i}^{sr}$'s are independent and identically distributed ($\forall i\in\bbN$).
    
Also, under $\pi_{sr}$, sources are chosen with fixed probability, independent of $d_{j k}$'s. Therefore, $\hat{\kappa}_{\ell j}$ and $d_{j k}$'s are mutually independent ($\forall j\in\{1,\cdots,N\}, k\in\bbN$). Hence, using the Wald's equation \cite{mckay2019probability}, we get
\begin{align} \label{eq:sr-exp-hat-w}
	\bbE[\hat{w}_{\ell i}^{sr}]
	=\sum_{j=1}^N \bbE[\hat{\kappa}_{\ell j}]\bbE[d_{j k}]=\sum_{j=1}^N \frac{\hat{p}_j}{\hat{p}_\ell}\gamma_j 
	\stackrel{(a)}{=}\frac{\mu_\ell}{p_\ell}\sum_{j=1}^N \gamma_j\frac{p_j}{\mu_j} 
	\stackrel{(b)}{\le}\frac{\mu_\ell}{p_\ell},
\end{align}
where $(a)$ follows from \eqref{eq:sr-p-hat}, and $(b)$ follows from \eqref{eq:interference-constraint}. 

Further, since $\hat{\kappa}_{\ell j}$'s and $d_{j k}$'s are independent of packet generation instants $g_{\ell i}$'s (for $\ell\in\{1,\cdots,N\}$ and $i\in\bbN$) and SR-PMS (which marks packets at each source $\ell$ with fixed probability $p_\ell$), mutual independence of $\hat{w}_{\ell i}^{sr}$'s and $T_{\ell j}^m$'s ($\forall i,j\in\bbN$) follows.

\section{Proof of Lemma \ref{lemma:min-consolidated-time}} \label{appendix:proof-lemma-min-consolidated-time} 

Lemma \ref{lemma:min-consolidated-time} follows from a basic fact (Lemma \ref{lemma:exp-preempt}) about exponential random variables, which may be well known. However, in absence of a readily available reference, we first prove it using first principles, and then use it to prove Lemma \ref{lemma:min-consolidated-time}.
 
Let $Y_i > 0 $ for $i\in \mathbb{N}$ be an arbitrary sequence. 
Also, let $d_i$ for $i\in \mathbb{N}$ be a sequence of independent exponentially distributed random variables, each with mean $\gamma$. 
Consider the following random variables $i^\star = \min_{ d_i < Y_i}  i$, and
$\tilde{d} = \sum_{i=1}^{i^\star-1} Y_i + d_{i^\star}$.  
\begin{lemma} \label{lemma:exp-preempt}
$\tilde{d}$ is exponentially distributed with mean $\gamma$, independent of $Y_i$'s and $i^\star$. 
\end{lemma}
\begin{IEEEproof}
For any $t$, let $\tau(t) = \min_{\sum_{i=1}^k Y_i > t} k$. Then, we can write $t=(\sum_{i=1}^{\tau(t)-1} Y_i)+\delta_t$, 
for some $\delta_t\ge 0$. Consider a sequence of mutually disjoint intervals $I_j=[\sum_{i=1}^{j-1}Y_i, \  \sum_{i=1}^{j}Y_i)$ for $j\in\bbN$. From definition of $\tau(t)$ and $i^\star$, we get that $t\in I_{\tau(t)}$, while $\tilde{d}\in I_{i^\star}$.

Using the law of total probability, we get 
\begin{align} \label{eq:law-total-prob}
	\bbP(\tilde{d}>t)=&\bbP(\tilde{d}>t|i^\star>\tau(t))\bbP(i^\star>\tau(t)) \nonumber \\
	&+\bbP(\tilde{d}>t|i^\star=\tau(t))\bbP(i^\star=\tau(t))\nonumber \\
	&+\bbP(\tilde{d}>t|i^\star<\tau(t))\bbP(i^\star<\tau(t)).
\end{align}
Since, $\bbP(\tilde{d}>t|i^\star>\tau(t))=1$ and $\bbP(\tilde{d}>t|i^\star<\tau(t))=0$, \eqref{eq:law-total-prob} simplifies to
\begin{align}
\label{eq:law-total-prob-2}
\bbP(\tilde{d}>t)=\bbP(i^\star>\tau(t)) +\bbP(\tilde{d}>t|i^\star=\tau(t))\bbP(i^\star=\tau(t)).
\end{align}

Note that $i^\star>\tau(t)$ is equivalent to the event where $d_i>Y_i$, $\forall i\le\tau(t)$.
Since $d_i$'s are independent and exponentially distributed with mean $\gamma$, we get  $\bbP(i^\star>\tau(t))=\prod_{i=1}^{\tau(t)}\exp(-Y_i/\gamma)=\exp(-\sum_{i=1}^{\tau(t)}Y_i/\gamma)$, and $\bbP(i^\star=\tau(t))$
$$=\prod_{i=1}^{\tau(t)-1}\exp(-Y_i/\gamma)\cdot (1-\exp(-Y_{\tau(t)}/\gamma))=\exp\left(-\sum_{i=1}^{\tau(t)-1}Y_i/\gamma\right)\cdot (1-\exp(-Y_{\tau(t)}/\gamma)).$$ Also, $$\bbP(\tilde{d}>t|i^\star=\tau(t))=\frac{\bbP(d_{i^\star}\in (\delta_t,Y_{\tau(t)}])}{\bbP(\tilde{d}<Y_{\tau(t)})}=\frac{\exp(-\delta_t/\gamma)-\exp(-Y_{\tau(t)}/\gamma)}{1-\exp(-Y_{\tau(t)}/\gamma)}.$$ 
Hence, from \eqref{eq:law-total-prob-2}, we get $\bbP(\tilde{d}>t)$
\begin{align}
	&=\exp\left(-\frac{\sum_{i=1}^{\tau(t)}Y_i}{\gamma}\right) +\frac{\exp\left(-\frac{\delta_t}{\gamma}\right)-\exp\left(-\frac{Y_{\tau(t)}}{\gamma}\right)}{1-\exp\left(-\frac{Y_{\tau(t)}}{\gamma}\right)}\cdot\exp\left(-\frac{\sum_{i=1}^{\tau(t)-1}Y_i}{\gamma}\right)\cdot \left(1-\exp\left(-\frac{Y_{\tau(t)}}{\gamma}\right)\right), \nonumber \\
	&=\exp\left(-\frac{\sum_{i=1}^{\tau(t)}Y_i}{\gamma}\right)+\exp\left(-\frac{\sum_{i=1}^{\tau(t)-1}Y_i+\delta_t}{\gamma}\right)+\exp\left(-\frac{\sum_{i=1}^{\tau(t)-1}Y_i+Y_{\tau(t)}}{\gamma}\right), \nonumber \\
	&=\exp\left(-\frac{t}{\gamma}\right).
\end{align}
Hence, $\tilde{d}$ is exponentially distributed with mean $\gamma$, independent of $Y_i$'s and $i^\star$.
\end{IEEEproof}

Note that any preemptive policy $\pi$ 
transmits packets of source $\ell$ over a sequence of intervals $I_{\ell 1}^\pi,I_{\ell 2}^\pi,\cdots$ (end of each interval is marked by a preemption), until one of its packet gets completely transmitted. Let the length of these intervals after time $r_{\ell (i-1)}^\pi$ (when the $i-1^{st}$ packet of source $\ell$ got completely transmitted) be $Y_{\ell 1}^\pi,Y_{\ell 2}^\pi,\cdots$. Because of the memoryless property of the exponential distribution, at the start of each interval $I_{\ell j}^\pi$, the remaining transmission time $d_{\ell j}$ of the packet under transmission is exponentially distributed with mean $\gamma_\ell$, irrespective of whether the packet was under transmission in any of the previous intervals. Hence, the required channel time for completely transmitting the $i^{th}$ packet of source $\ell$ is $\tilde{d}_{\ell i}^\pi=\sum_{j=1}^{j^\star-1}Y_{\ell j}^\pi+d_{\ell j^\star}$, where $j^\star=\min{d_{\ell j}<Y_{\ell j}^\pi}j$, and $d_{\ell j}$'s are independent and exponentially distributed with mean $\gamma_\ell$. Therefore, from Lemma \ref{lemma:exp-preempt}, we get that $\tilde{d}_{\ell i}^\pi$ is exponentially distributed with mean $\gamma_\ell$, and independent of the preemptive policy $\pi$ (since $\pi$ controls $Y_{\ell j}^\pi$'s, and $\tilde{d}_{\ell i}^\pi$ is independent of $Y_{\ell j}$'s).

Further, because the channel times $\tilde{d}_{\ell i}^\pi$'s ($\forall i$) are independent of the interval lengths $Y_{\ell j}^\pi$'s and the random variable $j^\star$, $\tilde{d}_{\ell i}^\pi$'s only depend on the transmission times $d_{\ell j}$'s of packets transmitted in successive intervals $I_{\ell 1}^\pi,I_{\ell 2}^\pi,\cdots$. Since $d_{\ell j}$'s are independent across $j\in\bbN$, this implies that $\tilde{d}_{\ell i}^\pi$'s are also independent for different $i$'s. 
Combining this with the previous result, 
we get Lemma \ref{lemma:min-consolidated-time}. 

\section{Proof of Lemma \ref{lemma:lb-OPT-preempt}} \label{appendix-lemma-lb-OPT-preempt}

By definition, the AoI of a source only depends on the packets that it transmits completely. Hence, AoI of each source $\ell$ under a preemptive policy $\pi$ can be written in terms of $T_{\ell i}^\pi$, $Z_{\ell i}^\pi$ and $R_{\ell i}^\pi(t)$ as in \eqref{eq:actual-Delta}. Also, the average transmission cost of source $\ell$ is $c_\ell \tilde{R}_\ell^\pi(t)/t$. Hence, the weighted sum cost for any preemptive policy $\pi$ is 
\begin{align} \label{eq:tac-preemptive}
	\Gamma(\pi)&=\lim_{t\to\infty}\frac{1}{N}\sum_{\ell=1}^{N}\frac{\sum_{i=1}^{R_\ell^\pi(t)}\rho_\ell (\frac{(T_{\ell i}^\pi)^2}{2}+T_{\ell i}^\pi Z_{\ell i}^\pi)+c_\ell\tilde{R}_\ell^\pi(t)}{t}.
\end{align}

Since period length $T_{\ell i}^\pi$ and system time $Z_{\ell i}^\pi$ are non-negative (for any causal/offline preemptive policy $\pi$), substituting $T_{\ell i}^\pi Z_{\ell i}^\pi= 0$ in \eqref{eq:tac-preemptive},\footnote{We lower bound $T_{\ell i}^\pi Z_{\ell i}^\pi$ by 0 (instead of $T_{\ell i}^\pi d_{\ell i}^\pi$ for non-preemptive policies) because when preemption is allowed, 
in certain settings, transmission times $d_{\ell i}^\pi$ for completely transmitted packets can be arbitrarily small (Example \ref{ex:small-tx-time}). 
} 
 and using the fact that $\tilde{R}_\ell^\pi(t)\ge R_\ell^\pi(t)$, for any offline preemptive policy $\pi$, we get 
\begin{align} \label{eq:gpi-preempt-1}
	\Gamma(\pi)&\ge
	\frac{1}{N}\sum_{\ell=1}^{N}\Bigg(\lim_{t\to\infty}\frac{\sum_{i=1}^{R_\ell^\pi(t)} \rho_\ell({T_{\ell i}^\pi})^2}{2t} +\lim_{t\to\infty}\frac{c_\ell R_\ell^\pi(t)}{t}\Bigg).
\end{align}

Note that $T_{\ell,\pi}^{av}=\underset{t\to\infty}{\lim}t/R_\ell^\pi(t)=\underset{t\to\infty}{\lim}\sum_{i=1}^{R_\ell^\pi(t)}T_{\ell i}^\pi/R_\ell^\pi(t)$ is the average period length for preemptive policy $\pi$. Defining $\delta_{\ell i}^\pi=T_{\ell i}^\pi-T_{\ell,\pi}^{av}$, and following the steps as in \eqref{eq:T^2-lb} and \eqref{eq:Cav-lb} to simplify  \eqref{eq:gpi-preempt-1}, we get 
\begin{align} \label{eq:preempt-Glb-OF-1}
	\Gamma(\pi)\ge\frac{1}{N}\sum_{\ell=1}^N\left(\frac{\rho_\ell\beta_\ell^\pi}{2T_{\ell,\pi}^{av}}+\frac{\rho_\ell T_{\ell,\pi}^{av}}{2} +\frac{c_\ell}{T_{\ell,\pi}^{av}}\right),
\end{align}
where $\beta_\ell^\pi=\underset{t\to\infty}{\lim}\sum_{i=1}^{R_\ell^\pi(t)}(\delta_{\ell i}^\pi)^2/R_\ell^\pi(t)\ge 0$. Thus, substituting $\beta_\ell^\pi=0$ in \eqref{eq:preempt-Glb-OF-1}, we get

\begin{align} \label{eq:tempeq}
	\Gamma(\pi)\ge\frac{1}{N}\sum_{\ell=1}^N\left(\frac{\rho_\ell T_{\ell,\pi}^{av}}{2}+\frac{c_\ell}{T_{\ell,\pi}^{av}}\right)\implies\Gamma(\tilde{\pi}_{OF}^\star)\ge\frac{1}{N}\sum_{\ell=1}^N\left(\frac{\rho_\ell T_{\ell,\tilde{\pi}_{OF}^\star}^{av}}{2}+\frac{c_\ell}{T_{\ell,\tilde{\pi}_{OF}^\star}^{av}}\right),
\end{align}
where $\tilde{\pi}_{OF}^\star$ denotes an optimal offline preemptive policy. Also, as shown in \eqref{eq:fl-Tav}, $T_{\ell,\tilde{\pi}_{OF}^\star}^{av}=\mu_\ell/\sfff_\ell^\star$, where $\sfff_\ell^\star\in[0,1]$ denotes the ratio of the number of packets of source $\ell$ that are completely transmitted by $\tilde{\pi}_{OF}^\star$, to the total number of packets generated at it (until the time horizon $t\to\infty$). Hence, substituting $T_{\ell,\tilde{\pi}_{OF}^\star}^{av}=\mu_\ell/\sfff_\ell^\star$ in \eqref{eq:tempeq}, we get \eqref{eq:lb-OPT-preempt}.
	

Further, note that between two successive transmission completion instants $r_{\ell (i-1)}^\pi$ and $r_{\ell i}^\pi$, source $\ell$ transmits for time equal to channel time $\tilde{d}_{\ell i}^\pi$. Also, at any time, at most one source can transmit. Therefore, for any preemptive policy $\pi$, 
$\sum_{\ell=1}^N\sum_{i=1}^{R_\ell^\pi(t)}\tilde{d}_{\ell i}^\pi\le t$. Dividing both sides by $t$, and taking limit as $t\to\infty$, for an optimal offline preemptive policy $\tilde{\pi}_{OF}^\star$, we get
\begin{align} \label{eq:proof-interference-constraint-1}
	1&\stackrel{(a)}{\ge} \lim_{t\to\infty}\sum_{\ell=1}^{N}\left(\frac{\sum_{i=1}^{R_\ell^{\star}(t)}\tilde{d}_{\ell i}^\star}{R_\ell^{\star}(t)}\cdot\frac{R_\ell^{\star}(t)}{h_\ell(t)}\cdot\frac{h_\ell(t)}{t}\right), \nonumber \\
	&\stackrel{(b)}{=} \sum_{\ell=1}^{N}\left(\lim_{t\to\infty}\frac{\sum_{i=1}^{R_\ell^{\star}(t)}\tilde{d}_{\ell i}^\star}{R_\ell^{\star}(t)}\cdot\lim_{t\to\infty}\frac{R_\ell^{\star}(t)}{h_\ell(t)}\cdot \lim_{t\to\infty}\frac{h_\ell(t)}{t}\right), \nonumber \\
	&\stackrel{(c)}{\ge}\sum_{\ell=1}^N\frac{\gamma_\ell \sfff_\ell^{\star}}{\mu_\ell}, 
\end{align}
where in $(a)$, $R_\ell^\star(t)$ denotes the number of completely transmitted packets of source $\ell$ under $\tilde{\pi}_{OF}^\star$ (until time $t$), and $\tilde{d}_{\ell i}^\star$'s denote the channel time for completely transmitted packets of source $\ell$ under $\tilde{\pi}_{OF}^\star$.
We get $(b)$ because limit of product is equal to product of limits (when limits exists), and $(c)$ follows because $\underset{t\to\infty}{\lim}\sum_{i=1}^{R_\ell^\star(t)}\tilde{d}_{\ell i}^\star/R_\ell^\star(t)\ge \gamma_\ell$ (Lemma \ref{lemma:min-consolidated-time}),
$\underset{t\to\infty}{\lim}R_\ell^\star(t)/h_\ell(t)=\sfff_\ell^\star$ (by definition), 
and from strong law of large numbers, $\underset{t\to\infty}{\lim}h_\ell(t)/t=1/\bbE[X_\ell]=1/\mu_\ell$. 

\section{Proof of Lemma \ref{lemma:ub-sr-preempt}} \label{appendix:lemma-ub-sr-preempt}
Recall that $\tilde{\pi}_{sr}$ differs from $\pi_{sr}$ only in the choice of $p_\ell$'s. Therefore, following the same steps as in the proof of Lemma \ref{lemma:gamma-pisr-ub-gen-dist}, we get the following upper bound on $\bbE[\Gamma(\tilde{\pi}_{sr})]$ (in terms of $p_\ell$'s):  
\begin{align} \label{eq:ub-sr-preempt-1}
	\bbE[\Gamma(\tilde{\pi}_{sr})]&\le \frac{1}{N}\sum_{\ell=1}^{N}\left(\left(\frac{2\rho_{\ell}\mu_{\ell}}{p_\ell}+\frac{c_\ell p_\ell}{\mu_\ell}\right)+\left( \rho_\ell\gamma_\ell-\frac{\rho_\ell\mu_\ell \theta_\ell}{2}\right)\right). 
\end{align}

Since $p_\ell$'s satisfy \eqref{eq:interference-constraint-preempt}, we get $1\ge \sum_{i=1}^N \gamma_i p_i/\mu_i\ge \gamma_\ell p_\ell/\mu_\ell$ (since $\gamma_i,p_i,\mu_i\ge0$, $\forall i\in\{1,\cdots,N\}$), which implies $\gamma_\ell\le \mu_\ell/p_\ell$. Therefore, substituting $\gamma_\ell= \mu_\ell/p_\ell$ in \eqref{eq:ub-sr-preempt-1}, we get
\begin{align} \label{eq:ub-sr-preempt-2}
	\bbE[\Gamma(\tilde{\pi}_{sr})]&\le \frac{1}{N}\sum_{\ell=1}^{N}\left(\left(\frac{2\rho_{\ell}\mu_{\ell}}{p_\ell}+\frac{c_\ell p_\ell}{\mu_\ell}\right)+\left( \frac{\rho_{\ell}\mu_{\ell}}{p_\ell}-\frac{\rho_\ell\mu_\ell \theta_\ell}{2}\right)\right), \nonumber \\
	&= \frac{1}{N}\sum_{\ell=1}^{N}\left(\frac{3\rho_{\ell}\mu_{\ell}}{p_\ell}+\frac{c_\ell p_\ell}{\mu_\ell}-\frac{\rho_\ell\mu_\ell \theta_\ell}{2}\right). 
\end{align}

\end{document}